\documentclass[aip,apl,longbibliography,reprint]{revtex4-1}

\usepackage{amssymb,amsmath}

\usepackage{graphicx}

\usepackage{color}

\newcommand{\vicente}[1]{{ #1}}

\newcommand\beq{\begin{equation}}
\newcommand\eeq{\end{equation}}
\newcommand\beqa{\begin{eqnarray}}
\newcommand\eeqa{\end{eqnarray}}
\newcommand{\dd}{\text{d}}

\newcommand{\al}{\alpha}
\begin{document}
\title{Transport properties in a model of confined granular mixtures at moderate densities}
\author{David Gonz\'alez M\'endez}
\email{dgonzalezm@unex.es}
\affiliation{Departamento de F\'{\i}sica, Universidad de Extremadura, E-06071 Badajoz, Spain}
\author{Vicente Garz\'o}
\email{vicenteg@unex.es} \homepage{https://fisteor.cms.unex.es/investigadores/vicente-garzo-puertos/}
\affiliation{Departamento de
F\'{\i}sica and Instituto de Computaci\'on Cient\'{\i}fica Avanzada (ICCAEx), Universidad de Extremadura, E-06071 Badajoz, Spain}

\begin{abstract}

This work derives the Navier--Stokes hydrodynamic equations for a model of a confined, quasi-two-dimensional, $s$-component mixture of inelastic, smooth, hard spheres. Using the inelastic version of the revised Enskog theory, macroscopic balance equations for mass, momentum, and energy are obtained, and constitutive equations for the fluxes are determined
through a first-order Chapman--Enskog expansion. As for elastic collisions, the transport coefficients are given in terms of the solutions of a set of coupled linear integral equations. Approximate solutions to these equations for diffusion transport coefficients and shear viscosity are achieved by assuming steady-state conditions and considering leading terms in a Sonine polynomial expansion. These transport coefficients are expressed in terms of the coefficients of restitution, concentration, the masses and diameters of the mixture's components, and the system's density. The results apply to moderate densities and are not limited to particular values of the coefficients of restitution, concentration, mass, and/or diameter ratios. As an application, the thermal diffusion factor is evaluated to analyze segregation driven by temperature gradients and gravity, providing criteria that distinguish whether larger particles accumulate near the hotter or colder boundaries.

\end{abstract}

%\draft
%\pacs{PACS number(s): 45.70.Mg, 05.20.Dd, 51.10.+y}

%\bigskip %\narrowtext

\date{\today}
\maketitle

\section{Introduction}
\label{sec1}

A paradigmatic situation for studying transport properties in confined granular systems corresponds to a quasi-two-dimensional vibrating box.\cite{OU98,LCG99,PMEU04,CMS12,CMS15,GS18,CMSSGS19} In this setup, particles are confined in a box in which the vertical $z$-direction is slightly larger than the diameter of a particle. The system is driven by vertical vibrations of the box; thus, energy is injected into the particles' vertical degrees of freedom as they collide with the bottom plate of the box. This kinetic energy is subsequently dissipated and redistributed to the particles' horizontal degrees of freedom by inter-particle collisions.

\begin{figure}
%[hbtp]
\begin{center}
\begin{tabular}{lr}
\resizebox{7.4cm}{!}{\includegraphics{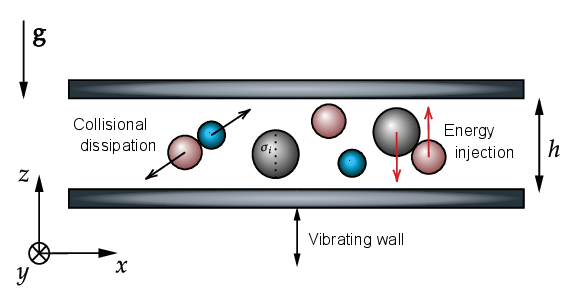}}
%\resizebox{7.4cm}{!}{\includegraphics{diagram-20260213-3.pdf}}
%&\resizebox{6.5cm}{!}{\includegraphics{Dpdil.pdf}}
\end{tabular}
\end{center}
\caption{Schematic illustration of the confined quasi-two-dimensional system. Vertical vibration is imposed on the box to inject external energy into the system. During nonplanar collisions between grains, the vertical energy gained from the vibration of the plate is transferred to the $xy$ components of the velocities. This kinetic energy is then dissipated and redistributed among the particles' horizontal degrees of freedom due to their collisions. It is important to note that, although our study aims to capture the phenomenology of confined systems, we are modeling an unconfined two-dimensional system where collisions are described by the $\Delta$-model.
\label{fig0}}
\end{figure}

Nevertheless, describing the confined system above using a kinetic theory is quite intricate, primarily due to the restrictions imposed by confinement on the corresponding Boltzmann/Enskog collisional operator. Although some recent progress has been made by considering this approach,\cite{MBGM22, MGB22, MPGM23} it is common in granular literature to start with a coarse-grained model that more effectively accounts for the effect of confinement on grain dynamics. This was the objective of the collisional model (referred to as the $\Delta$-model) proposed in Ref. \onlinecite{BRS13}. Apart from the coefficient of restitution, which characterizes the inelasticity of collisions, the $\Delta$-model introduces the factor $\Delta >0$ (which is responsible for velocity injection) to mimic the transfer of kinetic energy from the vertical degrees of freedom of granular particles to the horizontal ones. \vicente{A schematic illustration of the studied system is provided in Fig.\ \ref{fig0}.}

In the context of kinetic theory, the $\Delta$-model has primarily been used to determine the dynamic properties of monocomponent confined granular gases. The results have been specially devoted to the low-density regime, in which the inelastic Boltzmann equation is used as a starting point.\cite{BGMB13,BMGB14,SRB14,BBMG15,BBGM16} These results were then extended to moderate densities by considering the inelastic Enskog equation.\cite{GBS18,GBS20} Apart from results in kinetic theory, the $\Delta$-model has been also employed in recent years in the study of systems with long-range interactions,\cite{JMV16} absorbing phase transitions in driven granular systems,\cite{MPSTSF24,MPSF25a} the formation of quasi-long-range ordered phases,\cite{PMFBRSF24,MP24} the non-equilibrium coexistence between a fluid and a crystal of granular hard disks\cite{MPSF25} and the study of hyperuniformity.\cite{MCh25}

Within the context of the $\Delta$-model, work focusing on confined granular mixtures has been more scarce. This is primarily because determining transport properties in multicomponent granular mixtures is more complicated than in monodisperse granular systems. Not only is the number of transport coefficients larger, but they also depend on parameters such as diameters, masses, concentrations, and coefficients of restitution of each one of the components (or species) of the mixture. Thus, most of studies have been restricted to the \emph{dilute} regime. In this regime, Brito \emph{et al.} \cite{BSG20} used theoretical and computational tools to analyze the lack of energy equipartition of the homogeneous steady state (HSS) in granular binary mixtures. They then took the local version of the homogeneous state as the reference state to solve the inelastic Boltzmann equation using the Chapman--Enskog method,\cite{CC70} conveniently adapted to dissipative dynamics. Explicit expressions for the Navier--Stokes transport coefficients of the binary mixture were obtained in terms of the parameter space of the system.\cite{GBS21} As an application of the results reported in Ref.\ \onlinecite{GBS21}, the stability analysis of the HSS and the particle segregation problem induced by a thermal gradient and gravity were analyzed.\cite{GBS24,GBS24a}

An interesting and challenging problem is to extend previous works on \emph{dilute} granular mixtures (which have been carried in the context of the inelastic Boltzmann kinetic equation)\cite{GBS21,GBS24} to the (inelastic) Enskog kinetic theory for a description of hydrodynamics and transport of confined granular mixtures at higher densities. Since this theory applies to moderate densities (e.g., solid volume fraction of hard spheres, $\phi \lesssim 0.25$), a comparison between the kinetic theory and molecular dynamics (MD) simulations is feasible. Due to the intricacies of the general problem, it is common to first consider a binary mixture in which one species is present at a tracer concentration, meaning its concentration is very small. In the tracer limit, the pressure tensor and heat flux of the mixture are the same as those of the excess species. Thus, the relevant flux of the problem is the mass transport of tracer particles. This flux has been recently obtained in Ref.\ \onlinecite{GGBS24}, and the corresponding forms of the diffusion transport coefficients have been explicitly determined by considering the lowest Sonine polynomial approximation.

The objective of this paper is to surpass the tracer limit\cite{GGBS24} and examine a confined $s$-component mixture with an arbitrary concentration. Here, mass, momentum, and heat fluxes are calculated up to first order in the spatial gradients of the hydrodynamic fields. As with elastic collisions,\cite{CC70,FK72} the corresponding Navier-Stokes transport coefficients are precisely expressed in terms of a set of coupled linear integral equations. However, explicitly determining the complete set of Navier-Stokes transport coefficients is quite cumbersome. For example, in a binary system ($s=2$), there are 12 relevant transport coefficients (10 transport coefficients and two first-order contributions to partial temperatures and cooling rates), and one must solve 10 coupled integral equations. This is a lengthy task, even for a binary mixture. For this reason, in this work, although we have derived the complete set of integral equations verifying the transport coefficients, we will only solve the integral equations explicitly for the four diffusion coefficients associated with the mass flux of a binary mixture, as well as the shear and bulk viscosity coefficients.

As usual, the expressions derived for the transport coefficients are approximate because they are obtained by solving the above integral equations by considering the leading terms in a Sonine polynomial expansion of the first-order distribution function. Here, for the sake of simplicity, we consider the relevant state of confined dense granular mixtures with \emph{steady} temperature, as previously discussed for dilute granular mixtures.\cite{GBS21} This allows us to provide analytic expressions for the transport properties in terms of the mixture's parameters.

Since the expressions of the diffusion transport coefficients are available, as an application we have also obtained the so-called thermal diffusion factor $\Lambda$.\cite{KCL87} Knowledge of this quantity provides a segregation criterion showing the transition between regions of the parameter space where $\Lambda>0$ (where larger particles tend to accumulate near the cooler plate) to regions where $\Lambda<0$ (where larger particles tend to accumulate near the hotter plate).
The first situation ($\Lambda>0$) is usually referred to in granular literature as the Brazil nut Effect (BNE), while the second situation ($\Lambda<0$) is referred to as the Reverse Brazil nut Effect (BNE). Since the origin of segregation is due to both gravity and the thermal gradient, we will analyze limiting cases in which the influence of each on segregation can be disentangled.

The plan of the paper is as follows. The $\Delta$-model and the inelastic Enskog equation for an $s$-component granular mixture are introduced in Sec.\ \ref{sec2}. The balance equations for mass, momentum, and energy are also derived in this section and the kinetic and collisional contributions to the fluxes are defined in terms of the one-particle velocity distribution functions $f_i(\mathbf{r}, \mathbf{v};t)$. The application of the Chapman--Enskog method to solve the Enskog kinetic equation to first order in spatial gradients is described in Sec.\ \ref{sec3}. In Sec.\ \ref{sec4}, the first-order distribution functions $f_i$ are expressed in terms of the quantities $\boldsymbol{\mathcal{A}}_i$, $\boldsymbol{\mathcal{B}}_i$, $\mathcal{C}_{i,\lambda \beta}$, and $\mathcal{D}_{i}$. These quantities obey a set of coupled linear integral equations which are approximately solved in Sec.\ \ref{sec5} for the diffusion transport coefficients and the shear and bulk viscosities by considering the leading terms in a Sonine polynomial expansion. Some technical details on the above calculations are provided in Appendixes \ref{appA}, \ref{appB}, and \ref{appC}. The case of a binary granular mixture is considered in Sec.\ \ref{sec6} where the dependence of the above transport coefficients on the inelasticity in collisions is widely analyzed for dilute and dense mixtures. Thermal diffusion segregation is studied in Sec.\ \ref{sec7}, while a brief discussion on the results reported in this paper is offered in Sec.\ \ref{sec8}.

\section{Enskog kinetic equation for a model of a confined quasi-two-dimensional multicomponent granular mixture}
\label{sec2}

\subsection{Collision rules for a granular mixture in the $\Delta$-model}

We consider an $s$-multicomponent granular mixture of inelastic, smooth hard disks ($d=2$) or spheres ($d=3$) of masses $m_i$ and diameters $\sigma_i$. In the context of the $\Delta$-model, the relationship between the pre-collisional velocities $(\mathbf{v}_1, \mathbf{v}_2)$ of two spherical particles of species $i$ and $j$, respectively, and their corresponding post-collisional velocities $(\mathbf{v}_1',\mathbf{v}_2')$ is
\beq
\label{2.1}
\mathbf{v}_1'=\mathbf{v}_1-\mu_{ji}\left(1+\alpha_{ij}\right)(\widehat{{\boldsymbol {\sigma }}}\cdot \mathbf{g}_{12})\widehat{{\boldsymbol {\sigma }}}-2\mu_{ji}\Delta_{ij} \widehat{{\boldsymbol {\sigma }}},
\eeq
\beq
\label{2.2}
{\bf v}_{2}'=\mathbf{v}_{2}+\mu_{ij}\left(1+\alpha_{ij}\right)(\widehat{{\boldsymbol {\sigma }}}\cdot \mathbf{g}_{12})\widehat{{\boldsymbol {\sigma }}}+2\mu_{ij}\Delta_{ij} \widehat{{\boldsymbol {\sigma }}},
\eeq
where $\mu_{ij}=m_i/(m_i+m_j)$, $\mathbf{g}_{12}=\mathbf{v}_1-\mathbf{v}_2$ is the relative velocity, and $\widehat{{\boldsymbol {\sigma}}}$ is the unit collision vector joining the centers of the two colliding spheres and pointing from particle 1 of species $i$ to particle 2 of species $j$. Particles are approaching if $\widehat{{\boldsymbol {\sigma}}}\cdot \mathbf{g}>0$. In Eqs.\ \eqref{2.1} and \eqref{2.2}, $0<\al_{ij}\leq 1$ is the (constant) coefficient of normal restitution for collisions $i$-$j$, and $\Delta_{ij}$ is an extra velocity added to the relative motion. This extra velocity points outward in the normal direction $\widehat{\boldsymbol {\sigma}}$, as required by the conservation of angular momentum. \cite{L04bis} The relative velocity after collision is
\beq
\label{2.3}
\mathbf{g}_{12}'=\mathbf{v}_1'-\mathbf{v}_2'=\mathbf{g}_{12}-(1+\al_{ij})(\widehat{{\boldsymbol {\sigma}}}\cdot \mathbf{g}_{12})
\widehat{\boldsymbol {\sigma}}-2\Delta_{ij} \widehat{{\boldsymbol {\sigma }}}.
\eeq
According to Eq.\ \eqref{2.3} one easily gets the relation
\beq
\label{2.4}
(\widehat{{\boldsymbol {\sigma}}}\cdot \mathbf{g}_{12}')=-\al_{ij} (\widehat{{\boldsymbol {\sigma}}}\cdot \mathbf{g}_{12})-2\Delta_{ij}.
\eeq

Similarly, the collision rules for the so-called restituting collisions $\left(\mathbf{v}_1'',\mathbf{v}_2''\right)\to \left(\mathbf{v}_1,\mathbf{v}_2\right)$ with the same collision vector $\widehat{{\boldsymbol {\sigma }}}$ are defined as
\beq
\label{2.4a}
\mathbf{v}_1''=\mathbf{v}_1-\mu_{ji}\left(1+\alpha_{ij}^{-1}\right)(\widehat{{\boldsymbol {\sigma }}}\cdot \mathbf{g}_{12})\widehat{{\boldsymbol {\sigma }}}-2\mu_{ji}\Delta_{ij}\al_{ij}^{-1} \widehat{{\boldsymbol {\sigma }}},
\eeq
\beq
\label{2.5a}
\mathbf{v}_2''=\mathbf{v}_2+\mu_{ij}\left(1+\alpha_{ij}^{-1}\right)(\widehat{{\boldsymbol {\sigma }}}\cdot \mathbf{g}_{12})\widehat{{\boldsymbol {\sigma}}}+2\mu_{ij}\Delta_{ij}\al_{ij}^{-1} \widehat{{\boldsymbol {\sigma }}}.
\eeq
Equations \eqref{2.4a}--\eqref{2.5a} lead to
\beq
\label{2.6}
(\widehat{{\boldsymbol {\sigma}}}\cdot \mathbf{g}_{12}'')=-\al_{ij}^{-1}(\widehat{{\boldsymbol {\sigma}}}\cdot \mathbf{g}_{12})-2\Delta_{ij}\al_{ij}^{-1}.
\eeq

For practical purposes, it is also convenient to know the volume transformation for a direct and a restituting collision. In the first case, one has $\dd\mathbf{v}_1' \dd\mathbf{v}_2'=\al_{ij}\dd\mathbf{v}_1 \dd\mathbf{v}_2$ while in the second case one has the relation $\dd\mathbf{v}_1'' \dd\mathbf{v}_2''=\al_{ij}^{-1}\dd\mathbf{v}_1 \dd\mathbf{v}_2$.

\subsection{Enskog kinetic equation for granular mixtures}

At a kinetic level, the knowledge of the velocity distribution functions $f_i(\mathbf{r}, \mathbf{v};t)$ ($i=1,2,\cdots ,s$) of the species $i$ provides all the relevant information on the state of the granular mixture. For moderate densities, in the presence of the gravity field $m_i \mathbf{g}$, the set of Enskog kinetic equations are
\beq
\label{2.7}
\frac{\partial}{\partial t}f_i+\mathbf{v}\cdot \nabla f_i+\mathbf{g}\cdot \frac{\partial f_i}{\partial \mathbf{v}}
=\sum_{j=1}^s\; J_{ij}[\mathbf{r},\mathbf{v}|f_i,f_j],
\eeq
where the Enskog collision operators $J_{ij}$ for collisions between particles of species $i$ and $j$ in the $\Delta$-model read \cite{BSG20}
%\beqa
%\label{2.8}
%& & J_{ij}[\mathbf{r},\mathbf{v}_1|f_i,f_j]\equiv\nonumber\\
%& & \sigma_{ij}^{d-1} \int \mathrm{d}{\bf v}_{2}\int \mathrm{d} %\widehat{\boldsymbol{\sigma}}\;
%\Theta (-\widehat{{\boldsymbol {\sigma }}}\cdot {\bf %g}_{12}-2\Delta_{ij})\nonumber\\
%& & \times
%(-\widehat{\boldsymbol {\sigma }}\cdot {\bf g}_{12}-2\Delta_{ij})
%\al_{ij}^{-2}\chi_{ij}(\mathbf{r},\mathbf{r}+{\boldsymbol %{\sigma}}_{ij})f_i(\mathbf{r},\mathbf{v}_1'';t)
%\nonumber\\
%& & \times
%f_j(\mathbf{r}+{\boldsymbol {\sigma}}_{ij},\mathbf{v}_2'';t)-\sigma_{ij}^{d-1}\int \mathrm{d} {\bf v}_{2}\int \mathrm{d}\widehat{\boldsymbol{\sigma}}\;
%\Theta (\widehat{{\boldsymbol {\sigma }}}\cdot {\bf %g}_{12})\nonumber\\
%& & (\widehat{\boldsymbol {\sigma }}\cdot {\bf g}_{12})\chi_{ij}(\mathbf{r},\mathbf{r}+{\boldsymbol {\sigma}}_{ij})f_i(\mathbf{r},\mathbf{v}_1;t)
%f_j(\mathbf{r}+{\boldsymbol {\sigma}}_{ij},\mathbf{v}_2;t).
%\nonumber\\
%\eeqa
\begin{widetext}
\beqa
\label{2.8}
& & J_{ij}[\mathbf{r},\mathbf{v}_1|f_i,f_j]=\sigma_{ij}^{d-1} \int \mathrm{d}{\bf v}_{2}\int \mathrm{d} \widehat{\boldsymbol{\sigma}}\;
\Theta (-\widehat{{\boldsymbol {\sigma }}}\cdot {\bf g}_{12}-2\Delta_{ij})(-\widehat{\boldsymbol {\sigma }}\cdot {\bf g}_{12}-2\Delta_{ij})
\al_{ij}^{-2}\chi_{ij}(\mathbf{r},\mathbf{r}+{\boldsymbol {\sigma}}_{ij})f_i(\mathbf{r},\mathbf{v}_1'';t)
\nonumber\\
& & \times f_j(\mathbf{r}+{\boldsymbol {\sigma}}_{ij},\mathbf{v}_2'';t)-\sigma_{ij}^{d-1}\int \mathrm{d} {\bf v}_{2}\int \mathrm{d}\widehat{\boldsymbol{\sigma}}\;
\Theta (\widehat{{\boldsymbol {\sigma }}}\cdot {\bf g}_{12})(\widehat{\boldsymbol {\sigma }}\cdot {\bf g}_{12})\chi_{ij}(\mathbf{r},\mathbf{r}+{\boldsymbol {\sigma}}_{ij})f_i(\mathbf{r},\mathbf{v}_1;t)
f_j(\mathbf{r}+{\boldsymbol {\sigma}}_{ij},\mathbf{v}_2;t).
\eeqa
\end{widetext}
Here, $\Theta(x)$ is the Heaviside step function, $\boldsymbol{\sigma}_{ij}=\sigma_{ij}\widehat{\boldsymbol{\sigma}}$ and $\sigma_{ij}=(\sigma_i+\sigma_j)/2$. Moreover, $\chi_{ij}(\mathbf{r},\mathbf{r}+{\boldsymbol {\sigma}}_{ij})$ is the pair correlation function of two hard spheres, one of the species $i$ and the other of species $j$, at contact, i.e., when the distance between their centers is $\sigma_{ij}$. The quantities $\chi_{ij}$ accounts for volume excluded effects and spatial correlations not present in the Boltzmann equation. As noted in our previous papers, \cite{GBS18,BSG20,GBS21,GGBS24} although the $\Delta$-model was mainly proposed to describe quasi-two dimensional systems, the calculations worked out in this paper will be performed for an arbitrary number of dimensions $d$.

It should be noted that a modification of the revised Enskog equation for molecular fluids has been recently proposed.\cite{TT25} This modification transforms the correlation functions from functions of densities to functionals of densities in a simple form, eliminating the series structure. Here, we consider the series structure of the functions $\chi_{ij}$ in the same manner as in the revised Enskog theory developed in Ref.\ \onlinecite{LCK83}.

\begin{widetext}
As in the conventional inelastic hard sphere (IHS) model (where $\Delta_{ij}=0$),\cite{G19} an important property of the
Enskog collision operators for an arbitrary function $\psi_i(\mathbf{v}_1)$ is \cite{BGMB13,SRB14}
%\beqa
%\label{2.9}
%I_{\psi_i}&\equiv& \int\; d \mathbf{v}_1\; \psi_i(\mathbf{v}_1) J_{ij}[\mathbf{r},\mathbf{v}_1|f_i,f_j]\nonumber\\
%&=&\sigma_{ij}^{d-1}\int \mathrm{d} \, \mathbf{v}_1\int\ d {\bf v}_{2}\int \mathrm{d} \widehat{\boldsymbol{\sigma}}\,
%\Theta (\widehat{{\boldsymbol {\sigma }}}\cdot {\bf g}_{12})(\widehat{\boldsymbol {\sigma}}\cdot {\bf g}_{12})
%\nonumber\\
%& & \times  \chi_{ij}(\mathbf{r},\mathbf{r}+{\boldsymbol {\sigma}}_{ij})f_i(\mathbf{r},\mathbf{v}_1;t)
%f_j(\mathbf{r}+{\boldsymbol {\sigma}}_{ij},\mathbf{v}_2;t)\nonumber\\
%& & \times \left[\psi_i(\mathbf{v}_1')-\psi_i(\mathbf{v}_1)\right],
%\eeqa
%where $\mathbf{v}_1'$ is defined by Eq.\ \eqref{2.1}.
%\begin{widetext}
\beqa
\label{2.9}
I_{\psi_i}&\equiv&\int\; \dd \mathbf{v}_1\; \psi_i(\mathbf{v}_1) J_{ij}[\mathbf{r},\mathbf{v}_1|f_i,f_j]\nonumber\\
&=&\sigma_{ij}^{d-1}\int \mathrm{d}\mathbf{v}_1\int\ \dd {\bf v}_{2}\int \mathrm{d} \widehat{\boldsymbol{\sigma}}\,
\Theta (\widehat{{\boldsymbol {\sigma }}}\cdot {\bf g}_{12})(\widehat{\boldsymbol {\sigma}}\cdot {\bf g}_{12})
\chi_{ij}(\mathbf{r},\mathbf{r}+{\boldsymbol {\sigma}}_{ij})f_i(\mathbf{r},\mathbf{v}_1;t)
f_j(\mathbf{r}+{\boldsymbol {\sigma}}_{ij},\mathbf{v}_2;t)\left[\psi_i(\mathbf{v}_1')-\psi_i(\mathbf{v}_1)\right],\nonumber\\
\eeqa
\end{widetext}
where $\mathbf{v}_1'$ is defined by Eq.\ \eqref{2.1}. A consequence of the property \eqref{2.9} is that the balance equations for the densities of mass, momentum, and energy can be derived by following similar mathematical steps as those made for the IHS model.\cite{G19} They are given by
\begin{equation}
\text{D}_tn_{i}+n_{i}\nabla \cdot {\bf U}+\frac{\nabla \cdot {\bf j}_{i}}{m_{i}}
=0,  \label{2.10}
\end{equation}
\begin{equation}
\text{D}_t{\bf U}+\rho ^{-1}\nabla \cdot \mathsf{P}=\mathbf{g},  \label{2.11}
\end{equation}
\begin{equation}
\text{D}_tT-\frac{T}{n}\sum_{i=1}^s\frac{\nabla \cdot {\bf j}_{i}}{m_{i}}+\frac{2}{dn}
\left( \nabla \cdot {\bf q}+\mathsf{P}:\nabla {\bf U}\right) =-\zeta \,T.
\label{2.12}
\end{equation}
In Eqs.\ \eqref{2.10}--\eqref{2.12},
\begin{equation}
n_{i}=\int \mathrm{d}{\bf v}f_{i}({\bf v})
\label{2.13}
\end{equation}
is the number density of species $i$,
\beq
\label{2.14}
\mathbf{U}=\rho^{-1}\sum_{i=1}^s m_{i}\int \mathrm{d} {\bf v}{\bf v}f_{i}({\bf v})
\eeq
is the mean flow velocity, and
\begin{equation}
T=\frac{1}{d n}\sum_{i=1}^s m_{i}\int \mathrm{d}{\bf
v}V^{2}f_{i}({\bf v}) \label{2.15}
\end{equation}
is the (global) granular temperature. In addition, $\text{D}_t=\partial_t+\mathbf{U}\cdot \nabla$ is the material derivative, $\rho=\sum_i \rho_i=\sum_i m_i n_i$ is the total mass density, and ${\bf V}={\bf v}-{\bf U}$ is the peculiar velocity. Apart from the granular temperature $T$, at a kinetic level it is convenient to introduce the partial temperatures $T_i$ for each species; they measure their mean kinetic energies. They are defined as
\beq
\label{2.16}
n_i T_i=\frac{m_{i}}{d}\int \mathrm{d}{\bf v}V^{2}f_{i}({\bf v}).
\eeq

In the balance equations \eqref{2.10}--\eqref{2.12},
\begin{equation}
{\bf j}_{i}=m_{i}\int \mathrm{d}{\bf v}\,{\bf V}\,f_{i}({\bf v}),
\label{2.17}
\end{equation}
is the mass flux for the species $i$ relative to the local flow, $\mathsf{P}$ is the pressure tensor, and $\mathbf{q}$ is the heat flux. While the mass flux $\mathbf{j}_i$ has only \emph{kinetic} contributions, the pressure tensor and the heat flux have both \emph{kinetic} and \emph{collisional} transfer contributions, i.e., $\mathsf{P}=\mathsf{P}^\text
{k}+\mathsf{P}^\text{c}$ and $\mathbf{q}=\mathbf{q}^\text{k}+\mathbf{q}^\text{c}$. The kinetic contributions are given as usual by
\begin{equation}
\mathsf{P}^\text
{k}=\sum_{i=1}^s\,\int \mathrm{d}{\bf v}\,m_{i}{\bf V}{\bf V}\,f_{i}({\bf v}),
\label{2.18}
\end{equation}
\begin{equation}
{\bf q}^\text{k}=\sum_{i=1}^s\,\int \mathrm{d}{\bf v}\,\frac{1}{2}m_{i}V^{2}{\bf V}\,f_{i}({\bf v}).
\label{2.19}
\end{equation}

The collisional transfer contributions for the pressure tensor and the heat flux can be derived by following similar steps as those made in the $\Delta$-model for monocomponent granular gases. \cite{GBS18} They are given, respectively, by
\beqa
\label{2.20}
\mathsf{P}^{\text{c}}&=&\sum_{i,j}\frac{1+\alpha_{ij}}{2}m_{ij}\sigma_{ij}^{d}
\int \mathrm{d}\mathbf{v}_1\int \mathrm{d}\mathbf{v}_2
\int \mathrm{d}\widehat{\boldsymbol {\sigma}}\,\Theta (\widehat{{\boldsymbol {\sigma}}}
\cdot \mathbf{g}_{12})\nonumber\\
& &
\times (\widehat{\boldsymbol {\sigma }}\cdot {\bf g}_{12})
\widehat{\boldsymbol {\sigma}}
\widehat{\boldsymbol {\sigma }}\left[(\widehat{\boldsymbol {\sigma}}\cdot {\bf g}_{12})+\frac{2\Delta_{ij}}{1+\al_{ij}}\right]\int_{0}^{1}\; d\lambda
\nonumber \\
& & \times f_{ij}\left[\mathbf{r}-\lambda \boldsymbol{\sigma}_{ij},\mathbf{v}_1,
\mathbf{r}+(1-\lambda)\boldsymbol{\sigma}_{ij},\mathbf{v}_2,t\right],
\eeqa
\beqa
\label{2.21}
{\bf q}^\text{c}&=&\sum_{i,j}\frac{1+\alpha_{ij}}{8}m_{ij} \sigma_{ij}^{d}
\int \mathrm{d}\mathbf{v}_{1}\int \mathrm{d}\mathbf{v}_{2}\int
\dd\widehat{\boldsymbol {\sigma}}\,\Theta (\widehat{\boldsymbol{\sigma}}\cdot
\mathbf{g}_{12})\nonumber\\
& & \times
(\widehat{\boldsymbol {\sigma}}\cdot \mathbf{g}_{12})^{2}\widehat{\boldsymbol {\sigma}}\Big[4
(\widehat{\boldsymbol {\sigma}}\cdot {\bf G}_{ij})+(\mu_{ji}-\mu_{ij})(1-\al_{ij})\nonumber\\
& &\times
(\widehat{\boldsymbol{\sigma}}\cdot
\mathbf{g}_{12})\Big]\int_{0}^{1}d\lambda \nonumber\\
& & \times
f_{ij}\left[\mathbf{r}-
\lambda{\boldsymbol{\sigma}}_{ij},\mathbf{v}_{1},\mathbf{r}+(1-\lambda)
{\boldsymbol {\sigma}}_{ij},\mathbf{v}_{2},t\right]
\nonumber\\
& &-\sum_{i,j} \frac{m_i}{4} \sigma_{ij}^d \Delta_{ij}
\int \mathrm{d}\mathbf{v}_{1}\int \mathrm{d}\mathbf{v}_{2}\int
d\widehat{\boldsymbol {\sigma}}\Theta (\widehat{\boldsymbol{\sigma}}\cdot
\mathbf{g}_{12})\nonumber\\
& & \times (\widehat{\boldsymbol {\sigma}}\cdot \mathbf{g}_{12})
\widehat{\boldsymbol {\sigma}}
\Big[4\mu_{ji}^2\Delta_{ij} +4\mu_{ji}^2\al_{ij} (\widehat{\boldsymbol {\sigma}}\cdot \mathbf{g}_{12})\nonumber\\
& & -
4\mu_{ji} (\widehat{\boldsymbol {\sigma}}\cdot \mathbf{G}_{ij})\Big]
\int_{0}^{1}\dd\lambda \nonumber\\
& & \times
f_{ij}\left[\mathbf{r}-\lambda {\boldsymbol{\sigma}}_{ij},\mathbf{v}_{1},\mathbf{r}+(1-\lambda)
{\boldsymbol {\sigma}}_{ij},\mathbf{v}_{2},t\right].
\eeqa
The cooling rate $\zeta$ is
\beqa
\label{2.22}
\zeta&=&-\frac{2}{d n T}\sum_{i,j}\sigma_{ij}^{d-1}m_{ij}\int \dd \mathbf{v}_1\int \dd \mathbf{v}_2
\int \mathrm{d}\widehat{\boldsymbol {\sigma }}\,\Theta (\widehat{{\boldsymbol {\sigma}}}
\cdot \mathbf{g}_{12})\nonumber\\
& & \times (\widehat{\boldsymbol {\sigma}}\cdot {\bf g}_{12})\Big[\Delta_{ij}^2+\al_{ij}\Delta_{ij}(\widehat{{\boldsymbol {\sigma}}}\cdot \mathbf{g}_{12})-\frac{1-\al_{ij}^2}{4}(\widehat{{\boldsymbol {\sigma}}}\cdot \mathbf{g}_{12})^2\Big]
\nonumber\\
& & \times
f_{ij}(\mathbf{r},\mathbf{v}_1,\mathbf{r}+\boldsymbol{\sigma}_{ij},\mathbf{v}_2,t).
\eeqa
In Eqs.\ \eqref{2.20}--\eqref{2.22}, $m_{ij}=m_im_j/(m_i+m_j)$ is the reduced mass, $\mathbf{G}_{ij}=\mu_{ij}\mathbf{V}_1+\mu_{ji}\mathbf{V}_2$ is the center-of-mass velocity, and $f_{ij}$ is defined as
\beq
\label{2.23}
f_{ij}(\mathbf{r}_1, \mathbf{v}_1, \mathbf{r}_2, \mathbf{v}_2)=\chi_{ij}(\mathbf{r}_1, \mathbf{r}_2)f_i(\mathbf{r}_1, \mathbf{v}_1;t)f_j(\mathbf{r}_2, \mathbf{v}_2; t).
\eeq
It is important to remark that, in contrast to the IHS model, the cooling rate (which is due to dissipative collisions) can take negative values. This property allows the granular mixture in the $\Delta$-model to reach a steady temperature in the homogeneous state.

As usual, the balance equations \eqref{2.10}--\eqref{2.12} do not constitute a closed set of hydrodynamic equations for the
fields $n_{i}$, ${\bf U}$ and $T$. These equations become a closed set once the fluxes $\mathbf{j}_i$, $\mathsf{P}$, $\mathbf{q}$ and the cooling rate $\zeta$ are expressed in terms of the above hydrodynamic fields and their spatial gradients. To obtain this functional dependence one has to solve the set of Enskog equations \eqref{2.7} by means of the Chapman--Enskog method \cite{CC70} conveniently modified to account for the inelasticity of collisions.

\section{Chapman--Enskog method}
\label{sec3}

The Chapman--Enskog method \cite{CC70} is applied in this section to solve the Enskog kinetic equation \eqref{2.7} up to first order in spatial gradients. As widely discussed in many textbooks, \cite{CC70,FK72} two stages are present in the relaxation of a molecular gas (elastic collisions) toward equilibrium. In the first stage (\textit{kinetic} regime), the main effect of collisions is to relax the distribution function toward the so-called \textit{local} equilibrium distribution function. During this stage, the system's evolution depends on its initial state. Then, a second, slower stage (the \textit{hydrodynamic} regime) is achieved, in which the system has completely forgotten its initial preparation, and the microscopic state of the gas is described in terms of hydrodynamic fields.

One also expects the existence of the above two stages for granular gases. However, in the kinetic stage, the distribution function generally relaxes toward a time-dependent, nonequilibrium distribution (the homogeneous cooling state in the conventional IHS model) rather than a local equilibrium distribution. A crucial point is that, although the kinetic energy is not conserved (since collisions between particles are inelastic), the granular temperature $T$  can still be considered a slow field, as in conventional fluids. This assumption is clearly supported by the good agreement found between granular hydrodynamics and computer simulations in several non-equilibrium situations.\cite{DHGD02,G19} More details on applying the Chapman--Enskog method to granular mixtures can be found in Ref.\ \onlinecite{G19}, for example.

Based on the above arguments, in the hydrodynamic regime, the set of Enskog equations \eqref{2.7} admits a \emph{normal} (or hydrodynamic) solution where all the space and time dependence of the distributions $f_i$ only occurs through a functional dependence on the hydrodynamic fields $n_i$, $\mathbf{U}$, and $T$. As usual, \cite{CC70} this functional dependence can be made explicit by assuming small spatial gradients. In this case, $f_i(\mathbf{r}, \mathbf{v};t)$ can be written as a series expansion in powers of the spatial gradients of the hydrodynamic fields:
\beq
\label{3.1}
f_i=f_i^{(0)}+f_i^{(1)}+\cdots,
\eeq
where the approximation $f_i^{(k)}$ is of order $k$ in the spatial gradients.

The expansion \eqref{3.1} yields similar expansions for the fluxes and the cooling rate when substituted into Eqs.\ \eqref{2.17}--\eqref{2.22}:
\beq
\label{3.2}
\mathbf{j}_i=\mathbf{j}_i^{(0)}+\mathbf{j}_i^{(1)}+\cdots, \quad \mathsf{P}=\mathsf{P}^{(0)}+\mathsf{P}^{(1)}+\cdots,
\eeq
\beq
\label{3.3}
\mathbf{q}=\mathbf{q}^{(0)}+\mathbf{q}^{(1)}+\cdots, \quad \zeta=\zeta^{(0)}+\zeta^{(1)}+\cdots.
\eeq
Although the partial temperatures $T_i$ are not hydrodynamic quantities, they are also involved in the evaluation of the bulk viscosity\cite{KS79,GGG19b,CHGG22} and the cooling rate.\cite{GGG19b} Its expansion is
\beq
\label{3.4}
T_i=T_i^{(0)}+T_i^{(1)}+\cdots
\eeq
Finally, the time derivatives $\partial_t$ must be also expanded as
\begin{equation}
\label{3.5}
\partial_t=\partial_t^{(0)}+\partial_t^{(1)}+\cdots
\end{equation}
The action of the time derivatives $\partial_t^{(k)}$ on $n_i$, $\mathbf{U}$, and $T$ can be obtained from the balance equations \eqref{2.10}--\eqref{2.12} after taking into account the expansions \eqref{3.1}--\eqref{3.3} and collecting terms of the same order in the spatial gradients. Moreover, in the presence of the gravitational field, one has to characterize the magnitude of the force relative to spatial gradients. As for elastic collisions, \cite{CC70} we assume here that $\mathbf{g}$ must be considered to be at least of first order in spatial gradients.

As usual in the Chapman--Enskog method, \cite{CC70} the hydrodynamic fields $n_i$, $\mathbf U$, and $T$ are defined in terms of the zeroth-order distributions $f_i^{(0)}$:
\beq
\label{3.6}
\int \mathrm{d}\mathbf{v}\left(f_i-f_i^{(0)}\right)=0, \quad i=1,\cdots,s,
\eeq
\beq
\label{3.7}
\sum_{i=1}^s\int \mathrm{d}\mathbf{v}\; \left\{m_i\mathbf{v}, \frac{m_i}{2}V^2\right\}\left(f_i-f_i^{(0)}\right)=\left\{\mathbf{0},0\right\}.
\eeq
As a consequence, the remainder distributions of the expansion \eqref{3.1} must obey the orthogonality conditions:
\beq
\label{3.8}
\int \mathrm{d}\mathbf{v}f_i^{(k)}=0,
\eeq
and
\beq
\label{3.9}
\sum_{i=1}^s\int \mathrm{d}\mathbf{v}\; \left\{m_i\mathbf{v}, \frac{m_i}{2}V^2\right\}f_i^{(k)}=\left\{\mathbf{0},0\right\},
\eeq
for $k\geq 1$. The identities \eqref{3.9} lead to the constraints
\begin{equation}
\label{3.10}
\sum_{i=1}^s\mathbf{j}_i^{(k)}=\mathbf{0}, \quad \sum_{i=1}^s n_i T_i^{(k)}=0,
\end{equation}
for $k\geq 1$. As expected, the second condition in Eq.~\eqref{3.10} prevents that the (global) granular temperature $T$ is affected by the spatial gradients.

\vicente{What follows uses Latin indices to label the particle species (running from 1 to $s$) and Greek indices to label the spatial dimensions ($d=2$ for disks and $d=3$ for spheres). In addition, Einstein summation convention over repeated Greek indices is assumed in this paper.}

\subsection{Zeroth-order approximation}

In the absence of spatial gradients, the Enskog equation \eqref{2.7} reads
\beq
\label{3.11}
\partial_t^{(0)} f_i^{(0)}=\sum_{j=1}^s\; J_{ij}^{(0)}[\mathbf{v}|f_i^{(0)},f_j^{(0)}],
\eeq
where
%\beqa
%\label{3.12}
%& & J_{ij}^{(0)}[\mathbf{v}_1|f_i^{(0)},f_j^{(0)}]\equiv
%\sigma_{ij}^{d-1}\chi_{ij} \int \mathrm{d}{\bf v}_{2}\int \mathrm{d} \widehat{\boldsymbol{\sigma}}\;
%\nonumber\\
%& & \times
%\Theta (-\widehat{{\boldsymbol {\sigma }}}\cdot {\bf g}_{12}-2\Delta_{ij})(-\widehat{\boldsymbol {\sigma }}\cdot {\bf g}_{12}-2\Delta_{ij})
%\al_{ij}^{-2}f_i^{(0)}(\mathbf{v}_1'')\nonumber\\
%& & \times
%f_j^{(0)}(\mathbf{v}_2'')-\sigma_{ij}^{d-1}\chi_{ij}\int \mathrm{d} {\bf v}_{2}\int \mathrm{d}\widehat{\boldsymbol{\sigma}}\;
%\Theta (\widehat{{\boldsymbol {\sigma }}}\cdot {\bf g}_{12})\nonumber\\
%& & (\widehat{\boldsymbol {\sigma }}\cdot {\bf g}_{12})f_i^{(0)}(\mathbf{v}_1)
%f_j^{(0)}(\mathbf{v}_2).
%\eeqa
\begin{widetext}
\beqa
\label{3.12}
J_{ij}^{(0)}[\mathbf{v}_1|f_i^{(0)},f_j^{(0)}]&\equiv&
\sigma_{ij}^{d-1}\chi_{ij} \int \mathrm{d}{\bf v}_{2}\int \mathrm{d} \widehat{\boldsymbol{\sigma}}\;
\Theta (-\widehat{{\boldsymbol {\sigma }}}\cdot {\bf g}_{12}-2\Delta_{ij})(-\widehat{\boldsymbol {\sigma }}\cdot {\bf g}_{12}-2\Delta_{ij})
\al_{ij}^{-2}f_i^{(0)}(\mathbf{v}_1'')
f_j^{(0)}(\mathbf{v}_2'')
\nonumber\\
& &
-\sigma_{ij}^{d-1}\chi_{ij}\int \mathrm{d} {\bf v}_{2}\int \mathrm{d}\widehat{\boldsymbol{\sigma}}\;
\Theta (\widehat{{\boldsymbol {\sigma }}}\cdot {\bf g}_{12})(\widehat{\boldsymbol {\sigma }}\cdot {\bf g}_{12})f_i^{(0)}(\mathbf{v}_1)
f_j^{(0)}(\mathbf{v}_2).
\eeqa
\end{widetext}
The balance equations to this order give
\begin{equation}
\partial_{t}^{(0)}n_{i}=\partial_t^{(0)}U_\lambda=0,\quad
T^{-1}\partial_{t}^{(0)}T=-\zeta^{(0)}.
\label{3.13}
\end{equation}
Here, the cooling rate $\zeta^{(0)}$ is determined by Eq.\ \eqref{2.22} to
zeroth order. After performing the angular integrals, $\zeta^{(0)}$ can be written as
\beqa
\label{3.14}
\zeta^{(0)}&=&-\frac{2\pi^{\frac{d-1}{2}}}{d n T}\sum_{i=1}^s\sum_{j=1}^s\sigma_{ij}^{d-1}m_{ij}\chi_{ij}\int \dd \mathbf{v}_1\int \dd \mathbf{v}_2
\nonumber\\
& & \times \Bigg[\frac{\Delta_{ij}^2}{\Gamma\left(\frac{d+1}{2}\right)}g_{12}+\frac{\sqrt{\pi}}{d\Gamma\left(\frac{d}{2}\right)}g_{12}^2
\al_{ij}\Delta_{ij}
\nonumber\\
& & -\frac{1-\al_{ij}^2}{4}\frac{g_{12}^3}{\Gamma\left(\frac{d+3}{2}\right)}\Bigg]
f_{i}^{(0)}(\mathbf{v}_1)f_{j}^{(0)}(\mathbf{v}_2).
\eeqa
Upon obtaining Eq.\ \eqref{3.14} use has been of the result \cite{NE98}
\beq
\label{3.15}
B_k\equiv\int \mathrm{d} \widehat{\boldsymbol{\sigma}}\;\Theta (\widehat{{\boldsymbol {\sigma}}}\cdot {\bf g}_{12})(\widehat{{\boldsymbol {\sigma}}}\cdot \widehat{{\bf g}}_{12})^k=\pi^{\frac{d-1}{2}}\frac{\Gamma\left(\frac{k+1}{2}\right)}{\Gamma\left(\frac{k+d}{2}\right)},
\eeq
for positive integers $k$. Here, $\widehat{{\bf g}}_{12}=\mathbf{g}_{12}/g_{12}$.

The time evolution equation for the partial temperatures $T_i^{(0)}$ can be easily derived from Eq.\ \eqref{3.11} and the definition \eqref{2.16}:
\beq
\label{3.15.1}
\partial_{t}^{(0)}\ln T_i^{(0)}= -\zeta_i^{(0)},
\eeq
where the partial cooling rates $\zeta_i^{(0)}$ are given by
\beq
\label{3.15.2}
\zeta_i^{(0)}=-\frac{1}{d n_i T_i^{(0)}}\sum_{j=1}^s \int \mathrm{d}\mathbf{v} m_i v^2 J_{ij}^{(0)}[f_i^{(0)},f_j^{(0)}].
\eeq
The relationship between the cooling rates $\zeta^{(0)}$ and $\zeta_i^{(0)}$ is
\beq
\label{3.13.1}
\zeta^{(0)}=\sum_{i=1}^s\; x_i \gamma_i  \zeta_i^{(0)},
\eeq
where $x_i=n_i/n$ and $\gamma_i=T_i^{(0)}/T$ are the concentration (or mole fraction) and temperature ratio of species $i$, respectively. Combining Eqs.\ \eqref{3.13} and \eqref{3.15.1} one gets the time evolution of the temperature ratio $\gamma_i(t)=T_i^{(0)}(t)/T(t)$ as
\beq
\label{3.15.2bis}
\partial_{t}^{(0)} \ln \gamma_i=\zeta^{(0)}-\zeta_i^{(0)}.
\eeq

According to Eq.\ \eqref{3.13}, the Enskog equation \eqref{3.11} can be rewritten as
\beq
\label{3.16}
-\zeta^{(0)} T \frac{\partial f_i^{(0)}}{\partial T}=\sum_{j=1}^s\; J_{ij}^{(0)}[\mathbf{v}|f_i^{(0)},f_j^{(0)}].
\eeq
As for the conventional IHS model, the exact solution to Eq.\ \eqref{3.16} is not known to date. However, in the hydrodynamic regime, dimensional analysis and symmetry considerations yield the scaled solution
\beq
\label{3.17}
f_i^{(0)}(\mathbf{V};t)=n_i v_\text{th}(t)^{-d} \varphi_i(\mathbf{c}; \Delta_{\ell j}^*), \quad \ell, j=1,\cdots,s,
\eeq
where $\mathbf{c}=\mathbf{V}/v_\text{th}$ and $v_\text{th}(t)=\sqrt{2T(t)/\overline{m}}$ is a thermal velocity of the mixture defined in terms of the granular temperature $T(t)$. In addition, $\Delta_{ij}^*=\Delta_{ij}/v_\text{th}$ and $\overline{m}=\sum_i m_i/s$ is the average mass. It is important to remark that the consistency of the scaled solution \eqref{3.17} has been confirmed by computer simulations performed for the $\Delta$-model for monocomponent \cite{BGMB13,BMGB14} and multicomponent \cite{BSG20} granular gases. The Enskog equation for the scaled distributions $\varphi_i(\mathbf{c})$ can be easily obtained from Eq.\ \eqref{3.17} as
\beq
\label{3.18}
\frac{1}{2}\zeta_0^*\left(\frac{\partial}{\partial \mathbf{c}}\cdot \left(\mathbf{c}\varphi_i\right)
+\sum_{\ell, j=1}^s\; \Delta_{\ell j}^*\frac{\partial \varphi_i}{\partial \Delta_{\ell j}^*}\right)=\sum_{j=1}^s\; J_{ij}^{(0)*}[\mathbf{c}|\varphi_i,\varphi_j],
\eeq
where $\zeta_0^*=\zeta^{(0)}/\nu$, $J_{ij}^{(0)*}=v_\text{th}^d J_{ij}^{(0)}/(n_i \nu)$, $\nu=n \overline{\sigma}^{d-1}v_\text{th}$ is an effective collision frequency, and $\overline{\sigma}=\sum_i \sigma_i/s$.

Since in this approximation the distribution functions $f_i^{(0)}$ are isotropic in velocity space, then  $\mathbf{j}_i^{(0)}=\mathbf{q}^{(0)}=\mathbf{0}$ and $P_{\lambda\beta}^{(0)}=p\delta_{\lambda\beta}$. The hydrostatic pressure $p=nTp^*$, where
\beqa
\label{3.19}
p^*&=&1+\frac{\pi^{d/2}}{d\Gamma\left(\frac{d}{2}\right)}\sum_{i,j}\mu_{ji}n\sigma_{ij}^d\chi_{ij}x_ix_j\Bigg[(1+\al_{ij})
\gamma_i
\nonumber\\
& &+
\frac{2}{\sqrt{\pi}}
\frac{\Gamma\left(\frac{d}{2}\right)}{\Gamma\left(\frac{d+1}{2}\right)}
\frac{m_i}{\overline{m}}\Delta_{ij}^*\int \dd\mathbf{c}_1 \int \mathrm{d}\mathbf{c}_2\;
g_{12}^*\nonumber\\
& & \times \varphi_i(\mathbf{c_1})\varphi_j(\mathbf{c_2})\Bigg].
\eeqa

\subsubsection{Steady solution: Maxwellian approximation}

An interesting particular case corresponds to the steady state solution to Eq.\ \eqref{3.16}. In this situation, $\partial_t^{(0)}T=\partial_t^{(0)}T_i^{(0)}=0$ and hence, according to Eqs.\ \eqref{3.13} and \eqref{3.15.1}, the cooling rates vanish:
\beq
\label{3.21.1}
\zeta^{(0)}=\zeta_1^{(0)}=\cdots=\zeta_s^{(0)}=0.
\eeq

\vicente{According to Eq.\ \eqref{3.15.2}, determining the zeroth-order contributions to the cooling rates requires knowledge of the distribution functions $f_i^{(0)}(\mathbf{V})$. In the steady state and for elastic collisions, $\gamma_i=1$ and $\Delta_{ij}=0$ and the solution to Eq.\ \eqref{3.18} is the Gaussian or Maxwellian distribution $\varphi_1(\mathbf{c})=\varphi_2(\mathbf{c})=\cdots=\varphi_s(\mathbf{c})=\pi^{-d/2} e^{-c^2}$.
However,  an exact solution to Eq.\ \eqref{3.18} has not yet been obtained for inelastic collisions. A systematic approximation for the isotropic distributions $\varphi_i(\mathbf{c})$ can be found by expanding them into a complete set of orthogonal polynomials with a Gaussian measure. In practice, generalized Laguerre or Sonine polynomials $S_p^{(i)}(c^2)$ are used in kinetic theory.\cite{CC70,FK72} Thus, $\varphi_i(\mathbf{c})$ can be written as
\beq
\label{3.21.2}
\varphi_i(\mathbf{c})=\varphi_{i,\text{M}}(\mathbf{c})\Big[1+\sum_{p=1}^\infty a_p^{(i)} S_p^{(i)}(c^2)\Big],
\eeq
where the cumulants $a_p^{(i)}$ are given in terms of the velocity moments of $\varphi_i$. However, results obtained for binary granular mixtures\cite{BSG20} have clearly shown that in general the magnitude of the coefficients $a_p^{(i)}$ is very small. For practical purposes, one can therefore take the Maxwellian approximations
\beq
\label{4.19.1}
\varphi_{i,\text{M}}(\mathbf{c})=\pi^{-d/2} \theta_i^{d/2} e^{-\theta_i c^2},
\eeq
or equivalently
\beq
\label{3.20}
f_{i,\text{M}}(\mathbf{V})=n_i\left(\frac{m_i}{2\pi T_i^{(0)}}\right)^{d/2} \exp \left(-\frac{m_iV^2}{2 T_i^{(0)}}\right)
\eeq
to estimate the zeroth-order contributions to the cooling rates. Here, $\theta_i=m_i/(\overline{m}\gamma_i)$}.
In this approximation, the (dimensionless) cooling rate $\zeta_0^*=\zeta^{(0)}/\nu=\sum_i x_i \gamma_i \zeta_i^*$ where the (dimensionless) partial cooling rates $\zeta_i^*=\zeta_i^{(0)}/\nu$ are given by \cite{BSG20}
\begin{widetext}
\beqa
\label{3.21}
\zeta_{i}^*&=&\frac{4\pi^{(d-1)/2}}{d\Gamma\left(\frac{d}{2}\right)}\sum_{j=1}^s
x_j\chi_{ij}
\left(\frac{\sigma_{ij}}{\overline{\sigma}}\right)^{d-1}\mu_{ji}(1+\al_{ij})\theta_i^{-1/2}
\left(1+\theta_{ij}\right)^{1/2}
\left[1-\frac{1}{2}\mu_{ji}(1+\alpha_{ij})(1+\theta_{ij}) \right]\nonumber\\
& &-\frac{4\pi^{d/2}}{d\Gamma\left(\frac{d}{2}\right)}\sum_{j=1}^s x_j\chi_{ij}
\left(\frac{\sigma_{ij}}{\overline{\sigma}}\right)^{d-1}\mu_{ji}\Delta_{ij}^*\left[
\frac{2\mu_{ji}\Delta_{ij}^*}{\sqrt{\pi}}\theta_i^{1/2}\left(1+\theta_{ij}\right)^{1/2}
-1+\mu_{ji}(1+\al_{ij})\left(1+\theta_{ij}\right)\right],
\eeqa
\end{widetext}
where $\theta_{ij}=m_iT_j^{(0)}/m_jT_i^{(0)}$ gives the ratio between the mean-square
velocity of the particles of the species $j$ relative to that of the particles of the species $i$. In addition, the (reduced) pressure $p^*$ can be finally written as
\beqa
\label{3.19.1}
p^*&=&1+\frac{\pi^{d/2}}{d\Gamma\left(\frac{d}{2}\right)}\sum_{i,j}\mu_{ji}n\sigma_{ij}^d\chi_{ij}x_ix_j\Bigg[(1+\al_{ij})
\gamma_i
\nonumber\\
& &
+
\frac{2}{\sqrt{\pi}}\frac{m_i}{\overline{m}}\Delta_{ij}^*\left(\frac{\theta_i+\theta_j}{\theta_i\theta_j}\right)^{1/2}
\Bigg].
\eeqa

The theoretical predictions
of the dependence of the temperature ratio $T_1^{(0)}/T_2^{(0)}$ on the parameter space of a binary granular mixture has been recently \cite{BSG20} compared against computer simulations. Since the comparison shows in general a quite good agreement, one can conclude that the estimate \eqref{3.21} is quite reliable for not quite strong values of inelasticity and/or for moderate densities.

\section{First-order approximation. Mass flux and pressure tensor}
\label{sec4}

The implementation of the Chapman--Enskog method to first order in the spatial gradients follows similar steps as those made in the conventional IHS model for dense granular mixtures \cite{GDH07,GHD07} and more recently in the $\Delta$-model for dilute granular mixtures. \cite{BSG20,GBS21} Some mathematical details are provided in the Appendix \ref{appA} and only the final results for the integral equations verifying the kinetic transport coefficients are displayed in this section.

\begin{widetext}
As expected, the first-order velocity distribution function $f_i^{(1)}(\mathbf{r}, \mathbf{v};t)$ is given by
\beqa
\label{4.1}
f_i^{(1)}(\mathbf{V})&=&\boldsymbol{\mathcal{A}}_i\left(\mathbf{V}\right)\cdot  \nabla \ln
T+\sum_{j=1}^s\boldsymbol{\mathcal{B}}_{ij}\left(\mathbf{V}\right) \cdot \nabla \ln n_j+\mathcal{C}_{i,\lambda \beta}(\mathbf{V})\frac{1}{2}\left(\partial_\beta U_\lambda+\partial_\lambda U_\beta-\frac{2}{d}\delta_{\lambda\beta}\nabla \cdot \mathbf{U}\right)\nonumber\\
& & +\mathcal{D}_i
\left(\mathbf{V}\right) \nabla \cdot \mathbf{U},
\eeqa
The unknowns $\boldsymbol{\mathcal{A}}_i$, $\boldsymbol{\mathcal{B}}_{ij}$, $\mathcal{C}_{i,\lambda \beta}$, and $\mathcal{D}_i$ are the solutions of the following set of coupled linear integral equations:
\begin{equation}
\label{4.2}
-\zeta^{(0)}T \frac{\partial \boldsymbol{\mathcal{A}}_i}{\partial T}-
\frac{1}{2}\zeta^{(0)}\Bigg(1-\Delta^* \frac{\partial \ln \zeta_0^*}{\partial \Delta^*}\Bigg)
\boldsymbol{\mathcal{A}}_i-
\sum_{j=1}^s\left(J_{ij}^{(0)}[\boldsymbol{\mathcal{A}}_i,f_j^{(0)}]+
J_{ij}^{(0)}[f_i^{(0)},\boldsymbol{\mathcal{A}}_j]\right)=\mathbf{A}_i,
\end{equation}
\begin{equation}
\label{4.3}
-\zeta^{(0)}T \frac{\partial \boldsymbol{\mathcal{B}}_{ij}}{\partial T}
-\sum_{\ell=1}^s\left(J_{i\ell}^{(0)}[\boldsymbol{\mathcal{B}}_{ij},f_\ell^{(0)}]+
J_{i\ell}^{(0)}[f_i^{(0)},\boldsymbol{\mathcal{B}}_{\ell j}]\right)
=\mathbf{B}_{ij}+n_j \frac{\partial \zeta^{(0)}}{\partial n_j}
\boldsymbol{\mathcal{A}}_i,
\end{equation}
\begin{equation}
\label{4.4}
-\zeta^{(0)}T \frac{\partial \mathcal{C}_{i,\lambda \beta}}{\partial T}
-\sum_{j=1}^s\left(J_{ij}^{(0)}[\mathcal{C}_{i,\lambda\beta},f_j^{(0)}]+
J_{ij}^{(0)}[f_i^{(0)},\mathcal{C}_{j,\lambda\beta}]\right)=C_{i,\lambda\beta},
\end{equation}
\begin{equation}
\label{4.5}
-\zeta^{(0)}T \frac{\partial \mathcal{D}_{i}}{\partial T}-\zeta^{(1,1)}T \frac{\partial f_i^{(0)}}{\partial T}
-\sum_{j=1}^s\left(J_{ij}^{(0)}[\mathcal{D}_{i},f_j^{(0)}]+
J_{ij}^{(0)}[f_i^{(0)},\mathcal{D}_{j}]\right)=D_{i}',
\end{equation}
\end{widetext}
where the quantities $\mathbf{A}_i$, $\mathbf{B}_{ij}$, $C_{i,\lambda\beta}$, and $D_i'$ are defined in terms of the zeroth-order distributions $f_i^{(0)}$. Their explicit forms are provided in the Appendix \ref{appA}. The first-order contribution to the cooling rate  $\zeta^{(1,1)}$ is given in terms of the unknown $\mathcal{D}_{i}$ and is defined in Eq.\ \eqref{a16}. Moreover,
in Eq.\ \eqref{4.2} we have introduced the shorthand notation
\beq
\label{4.24}
\Delta^*\frac{\partial X}{\partial \Delta^*}
\equiv \sum_{i=1}^s\sum_{j=1}^s\;\Delta_{ij}^*\frac{\partial X}{\partial \Delta_{ij}^*}.
\eeq
In the particular case $\Delta_{ij}^*=\Delta^*$, only one of the $s(s+1)/2$ terms of the identity \eqref{4.24} must be considered. In the low-density limit ($n_i\sigma_{i}^d\to 0$), Eqs.\ \eqref{4.2}--\eqref{4.5} are consistent with those obtained in Ref.\ \onlinecite{GBS21} from the Boltzmann kinetic equation.

As mentioned in Sec.\ \ref{sec1}, although the solution to Eqs.\ \eqref{4.2}--\eqref{4.5} allows us to obtain the complete set of Navier--Stokes transport coefficients of the mixture, in this paper we will focus on the explicit determination of the diffusion transport coefficients and the shear and bulk viscosities. This will be carried out in Sec.\ \ref{sec5}.

\subsection{Diffusion transport coefficients}

The constitutive equation for the mass flux $\mathbf{j}_i^{(1)}$ to first order in spatial gradients can be written  using simple symmetry arguments. As for the IHS model, \cite{GDH07,G19} the mass flux is given by
\beq
\label{4.6}
\mathbf{j}_i^{(1)}=-\sum_{j=1}^s \frac{m_im_j n_j}{\rho}D_{ij}\nabla \ln n_j-\rho D_i^T \nabla \ln T.
\eeq
According to the constraint \eqref{3.10}, $\sum_{i=1}^s \mathbf{j}_i^{(1)}=\mathbf{0}$. In Eq.\ \eqref{4.6}, $D_{ij}$ are the mutual diffusion coefficients and $D_i^T$ are the thermal diffusion coefficients. The mass flux has only kinetic contributions. The transport coefficients $D_i^T$ and $D_{ij}$ can be easily expressed in terms of the solutions of the integral equations \eqref{4.2} and\eqref{4.3}, respectively. Since the first-order contribution to mass flux is defined as
\beq
\label{4.7}
\mathbf{j}_i^{(1)}=\int \mathrm{d}\mathbf{v}\; m_i \mathbf{V} f_i^{(1)}(\mathbf{V}),
\eeq
then the diffusion transport coefficients can be identified as
\beq
\label{4.8}
D_{i}^{T}=-\frac{m_i}{d \rho}\int \dd\mathbf{v} \mathbf{V} \cdot
\boldsymbol{\mathcal{A}}_{i}\left( \mathbf{V}\right),
\eeq
\beq
\label{4.9}
D_{ij}=-\frac{\rho }{d m_{j}n_{j}}\int \dd\mathbf{v} \mathbf{V}\cdot
\boldsymbol{\mathcal{B}}_{ij}\left( \mathbf{V}\right).
\eeq

\subsection{Pressure tensor}

The constitutive equation for the first-order contribution $\mathsf{P}^{(1)}$ to the pressure tensor is \cite{GDH07,G19}
\beq
\label{4.10}
P_{\lambda\beta}^{(1)}=-\eta\left(\partial_\lambda U_\beta+\partial_\beta U_\lambda-\frac{2}{d}\delta_{\lambda\beta}\nabla \cdot \mathbf{U}\right) -
\delta_{\lambda\beta} \eta_b  \nabla \cdot \mathbf{U},
\eeq
where $\eta$ and $\eta_b$ are the shear and bulk viscosity coefficients, respectively. While $\eta$ has kinetic and collisional contributions, $\eta_b$ has only collisional contributions. As in the case of the diffusion transport coefficients, since
\beq
\label{4.11}
P_{\lambda\beta}^{\text{k}(1)}=\sum_{i=1}^s\int \mathrm{d}\mathbf{v}\; m_i V_\lambda V_\beta f_i^{(1)}(\mathbf{V}),
\eeq
then the kinetic contribution $\eta_k$ to the shear viscosity $\eta$ can be written as
\beq
\label{4.12}
\eta_k=\sum_{i=1}^s\eta_{i}^k,
\eeq
where the partial kinetic coefficients $\eta_i^k$ are defined as
\beq
\label{4.13}
\eta_{i}^k=-\frac{1}{(d+2)(d-1)}\int \mathrm{d}\mathbf{v} R_{i,\lambda \beta}(\mathbf{V})\mathcal{C}_{i,\lambda\beta}(\mathbf{V}).
\eeq
In Eq. \eqref{4.13} we have introduced the polynomial
\beq
\label{4.14}
{R}_{i,\lambda\beta}(\mathbf{V})=m_i \left(V_\lambda V_\beta-\frac{1}{d}V^2 \delta_{\lambda \beta} \right).
\eeq
Note that upon writing Eq.\  \eqref{4.13} we have accounted for that $\mathcal{C}_{i,\lambda\beta}(\mathbf{V})$ is a traceless tensor.

The collisional contributions to the pressure tensor $P_{\lambda\beta}^{(1)}$ have been worked out in the Appendix \ref{appB}. From these contributions one can identify the collisional contribution $\eta_\text{c}$ to $\eta$ and the bulk viscosity coefficient $\eta_{b}$. The coefficient $\eta_{b}$ can be written as
\beq
\label{4.14.1}
\eta_{b}=\eta_{b}^{(\text{I})}+\eta_{b}^{(\text{II})},
\eeq
where
\begin{widetext}
\beq
\label{4.15}
\eta_{b}^{(\text{I})}=\frac{\pi^{d/2}}{d^2\Gamma\left(\frac{d}{2}\right)}\sum_{i=1}^s\sum_{j=1}^sn_i n_j \sigma_{ij}^{d+1}\chi_{ij} m_{ij}v_\text{th}\Bigg[\frac{(d+1)}{2\sqrt{\pi}}\frac{\Gamma\left(\frac{d}{2}\right)}{\Gamma\left(\frac{d+3}{2}\right)}
(1+\al_{ij})
I_{\eta_b}'+\Delta_{ij}^*\Bigg],
\eeq
\beq
\label{4.16}
\eta_{b}^{(\text{II})}=-\frac{\pi^{d/2}}{d\Gamma\left(\frac{d}{2}\right)}\sum_{i=1}^s\sum_{j=1}^sn_i n_j \sigma_{ij}^{d}\chi_{ij}\mu_{ji}
\Bigg[1+\al_{ij}+\frac{4\theta_i}{\sqrt{\pi}}\frac{\Gamma\left(\frac{d}{2}\right)}{\Gamma\left(\frac{d+1}{2}\right)}I_{\eta_b}''\Delta_{ij}^*
\Bigg]\varpi_i.
\eeq
\end{widetext}
Here, $I_{\eta_b}'$ and $I_{\eta_b}''$ are the dimensionless integrals
\beq
\label{4.17}
I_{\eta_b}'=\int \mathrm{d}\mathbf{c}_1\int \mathrm{d}\mathbf{c}_2 \,g_{12}^*\varphi_i(\mathbf{c}_1)\varphi_j(\mathbf{c}_2)
\eeq
and
\beq
I_{\eta_b}'' = \int \dd \mathbf{c}_1 \int \dd \mathbf{c}_2\, g_{12}^* \left( \theta_i c_1^2 - \frac{d}{2} \right) \varphi_{i,M}(\textbf{c}_1)\varphi_{j}(\textbf{c}_2),
\eeq
where $g_{12}^*=g_{12}/v_\text{th}$ and the coefficients $\varpi_i$ define the first-order contributions $T_i^{(1)}$ to  the partial temperatures \cite{GGG19b} as $T_i^{(1)}=\varpi_i \nabla \cdot \mathbf{U}$.

The collisional shear viscosity coefficient $\eta_c$ is
\begin{widetext}
\beq
\label{4.18}
\eta_{c}=\frac{2\pi^{d/2}}{d\Gamma\left(\frac{d}{2}\right)}\sum_{i=1}^s\sum_{j=1}^s n_j  \sigma_{ij}^{d}\chi_{ij}
\Bigg[\frac{\mu_{ji}}{(d+2)}(1+\al_{ij})+\frac{4d}{\sqrt{\pi}(d+1)}
\frac{\Gamma\left(\frac{d}{2}\right)}{\Gamma\left(\frac{d+1}{2}\right)}
\frac{m_{ij}m_i}{\overline{m}^2}\frac{I_{\eta_c}}{\gamma_i^2}\left(\frac{\theta_j}{\theta_i+\theta_j}\right)^2\Delta_{ij}^*\Bigg]\eta_i^\text{k}+
\frac{d}{d+2}\eta_{b}^{(\text{I})},
\eeq
\end{widetext}
where the dimensionless integral $I_{\eta_c}$ is
\beq
\label{4.19}
I_{\eta_c}=
\int \mathrm{d}\mathbf{c}_1\int \mathrm{d}\mathbf{c}_2 \,
g_{12}^{*-1}g_{12,x}^{*2}g_{12,y}^{*2}\varphi_{i,\text{M}}(\mathbf{c}_1)\varphi_{j,\text{M}}(\mathbf{c}_2).
\eeq
%Here,
%\beq
%\label{4.19.1}
%\varphi_{i,\text{M}}(\mathbf{c})=\pi^{-d/2}\theta_i^{d/2}e^{-\theta_i c^2}
%\eeq
%is the Maxwellian approximation to the scaled distribution $\varphi_{i}(\mathbf{c})$ of the species $i$.

In the limit of mechanically equivalent particles ($m_i=m$, $\sigma_i=\sigma$, and $\al_{ij}=\al$), $\gamma_i=1$, $\varpi_i =0$, and Eqs. \eqref{4.14.1}--\eqref{4.19} agree with previous results \cite{GBS18} derived from the Enskog equation for monocomponent granular gases. In addition, when $\Delta_{ij}^*=0$, the results are consistent with those obtained in the conventional IHS model. \cite{GDH07,GHD07}

\subsection{Steady state conditions}

As is the case with dilute granular mixtures, \cite{GBS21} determining the kinetic contributions to the diffusion transport coefficients and shear viscosity requires numerically solving first-order differential equations in the dimensionless parameters $\Delta_{ij}^*$. Years ago, this type of study was carried out within the $\Delta$-model for a monocomponent dilute granular gas. \cite{BBMG15} However, the goal of the present work is to obtain simple analytical expressions for the transport coefficients. Thus, the relevant state of a two-dimensional, confined granular mixture with a stationary temperature is considered. In this case, the constraints \eqref{3.21.1} apply and hence, the integral equations \eqref{4.2}--\eqref{4.4} associated with the transport coefficients $D_i^T$, $D_{ij}$, and $\eta$, respectively, reduce to
\begin{equation}
\label{4.20}
\frac{1}{2}\nu \Delta^*\frac{\partial \zeta_0^*}{\partial \Delta^*}
\boldsymbol{\mathcal{A}}_i-
\sum_{j=1}^s\left(J_{ij}^{(0)}[\boldsymbol{\mathcal{A}}_i,f_j^{(0)}]+
J_{ij}^{(0)}[f_i^{(0)},\boldsymbol{\mathcal{A}}_j]\right)=\mathbf{A}_i,
\end{equation}
\begin{equation}
\label{4.21}
-\sum_{\ell=1}^s\left(J_{i\ell}^{(0)}[\boldsymbol{\mathcal{B}}_{ij},f_\ell^{(0)}]+
J_{i\ell}^{(0)}[f_i^{(0)},\boldsymbol{\mathcal{B}}_{\ell j}]\right)
=\mathbf{B}_{ij}+n_j \frac{\partial \zeta^{(0)}}{\partial n_j}
\boldsymbol{\mathcal{A}}_i,
\end{equation}
\begin{equation}
\label{4.22}
-\sum_{j=1}^s\left(J_{ij}^{(0)}[\mathcal{C}_{i,\lambda\beta},f_j^{(0)}]+
J_{ij}^{(0)}[f_i^{(0)},\mathcal{C}_{j,\lambda\beta}]\right)=C_{i,\lambda\beta}.
\end{equation}
%\begin{equation}
%\label{4.23}
%-\zeta^{(1,1)}T \frac{\partial f_i^{(0)}}{\partial T}-\sum_{j=1}^s\left(J_{ij}^{(0)}[\mathcal{D}_{i},f_j^{(0)}]+
%J_{ij}^{(0)}[f_i^{(0)},\mathcal{D}_{j}]\right)=D_{i}'.
%\end{equation}
%\end{widetext}

\section{Sonine polynomial approximation}
\label{sec5}

As usual, obtaining the explicit forms of the kinetic transport coefficients requires resorting to the leading terms in a Sonine polynomial expansion of the unknowns
$\boldsymbol{\mathcal{A}}_i$, $\boldsymbol{\mathcal{B}}_{ij}$, $\mathcal{C}_{i,\lambda \beta}$, and $\mathcal{D}_i$. Additionally, to obtain the collision contributions to the shear and bulk viscosities, the integrals $I_{\eta_c}$, $I_{\eta_b}'$ and $I_{\eta_b}''$ and the first-order contributions $\varpi_i$ to the partial temperatures $T_i$ must be estimated. %These quantities are given in terms of the unknowns $\mathcal{D}_i$ [see Eq. \ \eqref{b4}].
However, since determining the coefficients $\varpi_i$ is a quite difficult task that is beyond the goal of this paper, we will neglect the contribution of $\varpi_i$ to
$\eta_b$ and hence $\eta_b\simeq \eta_b^{(\text{I})}$. This approximation is justified, as previous results obtained for the IHS model \cite{GGG19b} have shown that the influence of the coefficients $\varpi_i$ on $\eta_b$ is in general quite small. We expect this feature to be present in the $\Delta$-model as well.

To estimate the integral $I_{\eta_b}'$ appearing in the expression of $\eta_b^{(\text{I})}$, one takes the Maxwellian approximation \eqref{4.19.1} to the scaled distributions $\varphi_i(\mathbf{c})$. This integral can be performed by the change of variables $\mathbf{g}_{12}^*=\mathbf{c}_1-\mathbf{c}_2$ and $\overline{\mathbf{G}}_{12}^*=\theta_i\mathbf{c}_1+\theta_j\mathbf{c}_2$, where the Jacobian is $(\theta_i+\theta_j)^{-d}$. The result in the two-dimensional case ($d=2$) is
\beq
\label{5.1}
I_{\eta_b}'=\frac{\sqrt{\pi}}{2}\Bigg(\frac{\theta_i+\theta_j}{\theta_i\theta_j}
\Bigg)^{1/2},
\eeq
while the integral $I_{\eta_c}$ appearing in the expression of $\eta_c$ for $d=2$ is
\beq
\label{5.1.1}
I_{\eta_c}=\frac{3}{32}\sqrt{\pi}\Bigg(\frac{\theta_i+\theta_j}{\theta_i\theta_j}\Bigg)^{3/2}.
\eeq
%and
%\beq
%I_{\eta_b}'' = \frac{\sqrt{\pi}}%%{4}\Bigg(\frac{\theta_j/\theta_i}%{\theta_i+\theta_j}\Bigg)^{1/2}
%\eeq
%\beq
%\label{5.2}
%I_{\eta_b}=\frac{\sqrt{\pi}}{2}\Bigg(\frac{\theta_i+\theta_j}{\theta_i\theta_j}
%\Bigg)^{1/2}.
%\eeq
With these expressions, $\eta_{b}^{(\text{I})}$ and $\eta_c$ for $d=2$ are given, respectively, by
\beqa
\label{5.3}
\eta_{b}^{(\text{I})}&=&\frac{\sqrt{\pi}}{4}\sum_{i=1}^s\sum_{j=1}^sn_i n_j \sigma_{ij}^{3}\chi_{ij} m_{ij}v_\text{th}\Bigg[\Bigg(\frac{\theta_i+\theta_j}{\theta_i\theta_j}
\Bigg)^{1/2}\nonumber\\
& & \times (1+\al_{ij})+\sqrt{\pi}\Delta_{ij}^*\Bigg],
\eeqa
\beqa
\label{5.4}
\eta_c&=&\frac{\pi}{4}\sum_{i=1}^s\sum_{j=1}^s n_j \sigma_{ij}^{2}\chi_{ij} \Bigg[\mu_{ji}(1+\al_{ij})+\frac{2}{\sqrt{\pi}}\frac{m_{ij}m_i}{\overline{m}^2\gamma_i^2}\nonumber\\
& & \times
\Bigg(\frac{\theta_j}{\theta_i^3(\theta_i+\theta_j)}
\Bigg)^{1/2}\Delta_{ij}^*\Bigg]\eta_{i}^k+\frac{1}{2}\eta_b^{(\text{I})}.
\eeqa
Equations \eqref{5.3} and \eqref{5.4} agree with the results derived in
Ref.\ \onlinecite{GBS18} for mechanically equivalent particles and those obtained in the conventional IHS model. \cite{GDH07,GHD07}

\subsection{Diffusion transport coefficients}

As mentioned before, the diffusion transport coefficients $D_i^T$ and $D_{ij}$ have only kinetic contributions. They are defined by Eqs.\ \eqref{4.8} and \eqref{4.9}, respectively. To estimate them, as usual, we consider the lowest Sonine approximations for $\boldsymbol{\mathcal{A}}_i$ and $\boldsymbol{\mathcal{B}}_{ij}$. Since they are vectorial quantities, then their leading order Sonine polynomial is proportional to $\mathbf{V}$, namely,
\beq
\label{5.5}
\boldsymbol{\mathcal{A}}_{i}(\mathbf{V})\rightarrow -\frac{\rho }{n_{i}T_{i}^{(0)}}
D_{i}^{T}f_{i,\text{M}}(\mathbf{V})\mathbf{V},
\eeq
\beq
\label{5.6}
\boldsymbol{\mathcal{B}}_{ij}(\mathbf{V})\rightarrow -\frac{ m_{i}\rho_{j}}{\rho n_{i}T_{i}^{(0)}}D_{ij}f_{i,\text{M}}(\mathbf{V})\mathbf{V},
\eeq
where $f_{i,\text{M}}(\mathbf{V})$ is defined by Eq.\ \eqref{3.20}. Multiplication of Eqs.\ \eqref{4.20} and \eqref{4.21} by $m_i\mathbf{V}$ and integration over velocity leads to the algebraic equations obeying the diffusion transport coefficients. They are given by
\begin{widetext}
\beqa
\label{5.7}
\sum_{j=1}^s\Big(\nu_{ij}+\frac{1}{2}\nu\Delta^*\frac{\partial \zeta_0^*}{\partial \Delta^*}\delta_{ij}\Big)
D_{j}^{T}
&=&-\frac{p\rho_i}{\rho^2}\Bigg(1-\frac{1}{2}\Delta^*\frac{\partial \ln p^*}{\partial \Delta^*}\Bigg)+
\frac{n_i T_i^{(0)}}{\rho}\Bigg(1-\frac{1}{2}\Delta^*\frac{\partial \ln \gamma_i}{\partial \Delta^*}\Bigg)
\nonumber\\
& & +\frac{\pi^{d/2}}{d\Gamma\left(\frac{d}{2}\right)}\frac{n_i T}{\rho}
\sum_{j=1}^s n_j \sigma_{ij}^d\chi_{ij}\Bigg[(1+\al_{ij})\mu_{ij}\gamma_j+\frac{2}{\sqrt{\pi}}\frac{m_{ij}}{\overline{m}}
\left(\frac{\theta_i}
{\theta_j(\theta_i+\theta_j)}\right)^{1/2}
\Delta_{ij}^*\Bigg]\nonumber\\
& & \times \Bigg(1-\frac{1}{2}\Delta^*\frac{\partial \ln \gamma_j}{\partial \Delta^*}\Bigg),
\eeqa
\beqa
\label{5.8}
\sum_{\ell=1}^s \nu_{i\ell} m_\ell D_{\ell j}&=&\frac{\rho^2}{m_j}\frac{\partial \zeta^{(0)}}{\partial n_j}D_i^T+\frac{\rho T}{m_j}\Big(\gamma_i \delta_{ij}+n_i\frac{\partial \gamma_i}{\partial n_j}\Big)-\frac{\rho_i}{m_j}\frac{\partial p}{\partial n_j}\nonumber\\
& & +\frac{\pi^{d/2}}{d\Gamma\left(\frac{d}{2}\right)}\frac{\rho T \rho_i}{m_j}
\sum_{\ell=1}^s \sigma_{i\ell}^d\chi_{i\ell}
\mu_{\ell i}\Bigg\{\Bigg[(1+\al_{i\ell})\left(\frac{\gamma_i}{m_i}+\frac{\gamma_\ell}{m_\ell}\right)+\frac{4}{\sqrt{\pi}}
\frac{\Delta_{i\ell}^*}{\overline{m}}\left(\frac{\theta_i+\theta_\ell}
{\theta_i\theta_\ell}\right)^{1/2}\Bigg]\Bigg[\delta_{j\ell}
+\frac{1}{2}\frac{n_\ell}{n_j}\nonumber\\
& &\times
\left(n_j \frac{\partial \ln \chi_{i\ell}}{\partial n_j}+I_{i\ell j}\right)\Bigg]+
n_\ell \frac{\partial \gamma_\ell}{\partial n_j}\Bigg[\frac{1+\al_{i\ell}}{m_\ell}+\frac{2}{\sqrt{\pi}}\frac{\Delta_{i\ell}^*}{\overline{m}\gamma_\ell}
\left(\frac{\theta_i}
{\theta_\ell(\theta_i+\theta_\ell)}\right)^{1/2}\Bigg]\Bigg\}.
\nonumber\\
\eeqa
\end{widetext}
In Eqs.\ \eqref{5.7} and \eqref{5.8},  the quantities $I_{i\ell j}$ are provided in the Appendix \ref{appD} for a granular binary mixture ($s=2$). Moreover, we have introduced the collision frequencies
\beq
\label{5.9}
\nu_{ii}=-\frac{1}{dn_iT_i^{(0)}}\sum_{j\neq i}^s\; \int \mathrm{d}\mathbf{v}\, m_i \mathbf{V}\cdot J_{ij}^{(0)}[f_{i,\text{M}}\mathbf{V},f_j^{(0)}],
\eeq
\beq
\label{5.10}
\nu_{ij}=-\frac{1}{dn_jT_j^{(0)}} \int \mathrm{d}\mathbf{v} m_i \mathbf{V}\cdot J_{ij}^{(0)}[f_i^{(0)},f_{j,\text{M}}\mathbf{V}], \quad i\neq j.
\eeq

Upon obtaining Eqs.\ \eqref{5.7} and \eqref{5.8}, use has been made of the results
\begin{widetext}
\beqa
\label{5.11}
\int \mathrm{d}\mathbf{V}\; m_i \mathbf{V}\cdot \boldsymbol{\mathcal{K}}_{ij}\Bigg[T\frac{\partial f_j^{(0)}}{\partial T}\Bigg]&=&\frac{\pi^{d/2}}{\Gamma\left(\frac{d}{2}\right)}n_i n_j \sigma_{ij}^d\chi_{ij}T\Bigg[(1+\al_{ij})\mu_{ij}\gamma_j+\frac{2}{\sqrt{\pi}}\frac{m_{ij}}{\overline{m}}
\Bigg(\frac{\theta_i}{\theta_j(\theta_i+\theta_j)}\Bigg)^{1/2}
\Delta_{ij}^*\Bigg]\nonumber\\
& & \times \Bigg(1-\frac{1}{2}\Delta^*\frac{\partial \ln \gamma_j}{\partial \Delta^*}\Bigg),
\eeqa
\beq
\label{5.12}
\int \mathrm{d}\mathbf{V}\; m_i \mathbf{V}\cdot \boldsymbol{\mathcal{K}}_{ij}\Big[f_j^{(0)}\Big]=\frac{\pi^{d/2}}{\Gamma\left(\frac{d}{2}\right)}n_i n_j\sigma_{ij}^d\chi_{ij}m_{ij}T \Bigg[(1+\al_{ij})\left(\frac{\gamma_i}{m_i}+\frac{\gamma_j}{m_j}\right)+\frac{4}{\sqrt{\pi}}
\frac{\Delta_{ij}^*}{\overline{m}}
\Bigg(\frac{\theta_i+\theta_j}{\theta_i\theta_j}\Bigg)^{1/2}\Bigg],
\eeq
\beqa
\label{5.13}
\int \mathrm{d}\mathbf{V}\; m_i \mathbf{V}\cdot \boldsymbol{\mathcal{K}}_{i\ell}\Bigg[n_j \frac{\partial f_\ell^{(0)}}{\partial n_j}\Bigg]&=&
\frac{\pi^{d/2}}{\Gamma\left(\frac{d}{2}\right)}n_i n_j \sigma_{i\ell}^d\chi_{i\ell}m_{i\ell}T\Bigg\{\delta_{j\ell}\Bigg[(1+\al_{i\ell})\left(\frac{\gamma_i}{m_i}+\frac{\gamma_\ell}{m_\ell}\right)
+\frac{4}{\sqrt{\pi}}\frac{\Delta_{i\ell}^*}{\overline{m}}
\left(\frac{\theta_i+\theta_\ell}{\theta_i\theta_\ell}\right)^{1/2}\Bigg]
\nonumber\\
& &
+n_\ell \frac{\partial \gamma_\ell}{\partial n_j}\Bigg[\frac{(1+\al_{i\ell})}{m_\ell}+
\frac{2}{\sqrt{\pi}}\frac{\Delta_{i\ell}^*}{\overline{m}\gamma_\ell}\Bigg(\frac{\theta_i}
{\theta_\ell(\theta_i+\theta_\ell)}\Bigg)^{1/2}\Bigg]\Bigg\}.
\eeqa
\end{widetext}
As in the case of $I_{\eta_b}'$, the collision integrals \eqref{5.11}--\eqref{5.13} have been evaluated by approaching $f_i^{(0)}(\mathbf{V})$ by its Maxwellian form $f_{i,\text{M}}(\mathbf{V})$ and by following similar steps as those made for the conventional IHS model. \cite{GHD07,G19} In the Maxwellian approximation,
one may use the identity
\beq
\label{5.13.1}
T\frac{\partial f_{j,\text{M}}}{\partial T}=-\frac{1}{2}\frac{\partial}{\partial \mathbf{V}}\cdot \Big(\mathbf{V}f_{j,\text{M}}\Big)\Bigg(1-\frac{1}{2}\Delta^*\frac{\partial \ln \gamma_j}{\partial \Delta^*}\Bigg).
\eeq
Similarly, explicit expressions of $\nu_{ii}$ and $\nu_{ij}$ can be obtained by replacing $f_i^{(0)}\to f_{i,\text{M}}$. They are given by
\beqa
\label{5.14}
\nu_{ii}&=&\frac{2\pi^{(d-1)/2}}{d\Gamma\left(\frac{d}{2}\right)}v_\text{th}\sum_{j\neq i}^s\;n_j \sigma_{ij}^{d-1}\chi_{ij}\mu_{ji}\Bigg[\Bigg(\frac{\theta_i+\theta_j}
{\theta_i\theta_j}\Bigg)^{1/2}\nonumber\\
& & \times (1+\al_{ij})+\sqrt{\pi}\Delta_{ij}^*\Bigg],
\eeqa
\beqa
\label{5.15}
\nu_{ij}&=&-\frac{2\pi^{(d-1)/2}}{d\Gamma\left(\frac{d}{2}\right)}v_\text{th}n_i \sigma_{ij}^{d-1}\chi_{ij}\mu_{ij}\Bigg[\Bigg(\frac{\theta_i+\theta_j}
{\theta_i\theta_j}\Bigg)^{1/2}\nonumber\\
& & \times (1+\al_{ij})+\sqrt{\pi}\Delta_{ij}^*\Bigg], \quad i\neq j.
\eeqa
When $\Delta_{ij}^*=0$, Eqs.\ \eqref{5.11}--\eqref{5.15} agree with previous expressions obtained in the IHS model.\cite{GDH07,GHD07} Additionally, in the tracer limit ($x_1\to 0$), Eqs.\ \eqref{5.11}--\eqref{5.15} are also consistent with the results derived  within the $\Delta$-model for moderate densities.\cite{GGBS24}

\subsection{Kinetic shear viscosity coefficient}

The kinetic contribution $\eta_k$ to the shear viscosity coefficient is defined in terms of the partial kinetic contributions $\eta_{i}^k$ by Eq.\ \eqref{4.12}. As in the case of the diffusion coefficients,
to obtain the kinetic coefficients $\eta_{i}^k$, one takes the leading Sonine approximation to $\mathcal{C}_{i,\lambda \beta}(\mathbf{V})$:
\beq
\label{5.16}
\mathcal{C}_{i,\lambda \beta}(\mathbf{V})\to -f_{i,\text{M}}(\mathbf{V})\frac{\eta_{i}^k}{n_i T_i^{(0)2}}R_{i,\lambda \beta}(\mathbf{V}).
\eeq
Multiplication of Eq.\ \eqref{4.22} by $R_{i,\lambda \beta}(\mathbf{V})$ and integration over $\mathbf{v}$ allows us to get the algebraic equations obeying the kinetic coefficients $\eta_{i}^k$. After some algebra, one achieves the result
\beq
\label{5.16.1}
\sum_{j=1}^s \tau_{ij} \eta_j^k=\Omega_i,
\eeq
where
\begin{widetext}
\beqa
\label{5.17}
\Omega_i&=&
n_i T_i^{(0)}+\frac{\pi^{d/2}}{d(d+2) \Gamma\left(\frac{d}{2}\right)}\frac{n_i T}{\overline{m}}\sum_{j=1}^sn_j \sigma_{ij}^d m_{ij}\chi_{ij}\Bigg\{(1+\al_{ij})
\Bigg[\left(\frac{\theta_i+\theta_j}{\theta_i \theta_j}\right)\mu_{ji}(3\al_{ij}-1)-4\frac{\overline{m}\left(\gamma_i-\gamma_j\right)}{m_i+m_j}\Bigg]\nonumber\\
& & -4\Delta_{ij}^*\Bigg[\left(\frac{\theta_i+\theta_j}{\pi \theta_i \theta_j}\right)^{1/2}\left[\frac{2\theta_j}{\theta_i+\theta_j}-4 \mu_{ji}(1+\al_{ij})\right]-2\mu_{ji}\Delta_{ij}^*\Bigg]\Bigg\}.
\eeqa
To obtain Eq.\ \eqref{5.17}, use has been made of the result
\beqa
\label{5.20}
\int \mathrm{d}\mathbf{V}\; R_{i,\lambda \beta}(\mathbf{V}) \mathcal{K}_{ij,\lambda}\left[\frac{\partial f_j^{(0)}}{\partial V_\beta}\right]&=&-2^{d-1}(d-1)x_j \left(\frac{\sigma_{ij}}{\sigma_i}\right)^d \frac{m_{ij}}{\overline{m}}\chi_{ij}\phi_i (1+\al_{ij})
\Bigg[\left(\frac{\theta_i+\theta_j}{\theta_i \theta_j}\right)\mu_{ji}(3\al_{ij}-1)\nonumber\\
& & -4\frac{\overline{m}}{m_i+m_j}\left(\gamma_i-\gamma_j\right)\Bigg]n T+2^{d+1}(d-1)x_j \left(\frac{\sigma_{ij}}{\sigma_i}\right)^d \frac{m_{ij}}{\overline{m}}\chi_{ij}\phi_i \Delta_{ij}^*\nonumber\\
& & \times
\Bigg\{\left(\frac{\theta_i+\theta_j}{\pi \theta_i \theta_j}\right)^{1/2}\Bigg[\frac{2\theta_j}{\theta_i+\theta_j}-4 \mu_{ji}(1+\al_{ij})\Bigg]-2\mu_{ji}\Delta_{ij}^*\Bigg\}n T,
\eeqa
where we have made the replacement $f_j^{(0)}\to f_{j,\text{M}}$ and have introduced the partial volume fraction of species $i$ as
\beq
\label{5.21}
\phi_i=\frac{\pi^{d/2}}{2^{d-1}d \Gamma\left(\frac{d}{2}\right)}n_i\sigma_i^d.
\eeq
In Eq.\ \eqref{5.16.1}, we have introduced the following collision frequencies:
\beq
\label{5.18}
\tau_{ii}=-\frac{1}{(d-1)(d+2)}\frac{1}{n_i T_i^{(0)2}}\Bigg(\sum_{j=1}^s \int \mathrm{d}\mathbf{v}R_{i,\lambda\beta}J_{ij}^{(0)}[f_{i,\text{M}}R_{i,\lambda\beta},f_j^{(0)}]+
\int \mathrm{d}\mathbf{v}R_{i,\lambda\beta}J_{ii}^{(0)}[f_i^{(0)},f_{i,\text{M}}R_{i,\lambda\beta}]\Bigg),
\eeq
\beq
\label{5.19}
\tau_{ij}=-\frac{1}{(d-1)(d+2)}\frac{1}{n_j T_j^{(0)2}} \int \mathrm{d}\mathbf{v}R_{i,\lambda\beta}J_{ij}^{(0)}[f_i^{(0)},f_{j,\text{M}}R_{j,\lambda\beta}], \quad i\neq j.
\eeq
A good estimate of $\tau_{ii}$ and $\tau_{ij}$ can be obtained by replacing the true distributions $f_i^{(0)}$ with their Maxwellian forms $f_{i,\text{M}}$. The Appendix \ref{appC} contains the explicit expressions of these collision frequencies for the sake of completeness.
\end{widetext}

\subsection{Mechanically equivalent particles}

Clearly, the expressions for the coefficients $D_{ij}$, $D_i^T$, $\eta$, and $\eta_b$ displayed in the previous subsections exhibit a quite complex dependence on the parameter space of the mixture. Thus, it is convenient to consider some specific limiting cases to more clearly see the impact of each parameter on the transport coefficients. One of the simplest situations is a mixture constituted by mechanically equivalent particles ($m_i=m$, $\sigma_i=\sigma$, $\Delta_{ij}=\Delta$, and $\al_{ij}=\al$). In this limiting case, since $\chi_{ij}=\chi$ and $\gamma_i=1$, careful analysis of Eq.\ \eqref{5.7} shows that its right hand side vanishes, so the thermal diffusion coefficient $D_i^T=0$, as expected. In the case of the coefficients $D_{ij}$, let us consider a binary mixture for the sake of simplicity. In this case, since $\mathbf{j}_1^{(1)}=-\mathbf{j}_2^{(1)}$, then $D_{21}=-D_{11}$ and $D_{22}=-D_{12}$. Additionally, as expected, according to Eq.\ \eqref{5.8} $D_{12}=-(n_1/n_2)D_{11}$, and the constitutive equation for the mass flux becomes
\begin{equation}
\label{5.22}
\mathbf{j}_1^{(1)}=-D_{\text{self}} \nabla x_1,
\end{equation}
where $x_1=n_1/n$ is the mole fraction or concentration of species $1$ and the self-diffusion coefficient $D_{\text{self}}$ in the two-dimensional case is
\beq
\label{5.23}
D_{\text{self}}=\frac{\sqrt{mT/\pi}}{\sigma \chi}\Bigg(1+\al+
\sqrt{\frac{\pi}{2}}\Delta^*\Bigg)^{-1}.
\eeq
The expression \eqref{5.23} agrees with the results derived in the tracer limit when the intruder and granular gas particles are mechanically equivalent.\cite{GGBS24}

We consider the shear viscosity coefficient $\eta$. For mechanically equivalent particles, $\varpi_i=0$ and so, $\eta_{b}^{(\text{II})}=0$. Furthermore, the collision integral
\eqref{5.20} reduces to
%\footnote{A misprint in the calculation of the collision integral \eqref{5.20} for monocomponent granular gases was found in Ref.\ \onlinecite{GBS18} while the present paper was written. The expression \eqref{5.21.1} corrects this result.}
\begin{widetext}
\beq
\label{5.21.1}
\int \mathrm{d}\mathbf{V}\; R_{i,\lambda \beta}(\mathbf{V}) \mathcal{K}_{ij,\lambda}\left[\frac{\partial f_j^{(0)}}{\partial V_\beta}\right]=2^d (d-1)x_i x_j \chi \phi n T\Bigg\{\frac{1}{4}(1+\al)(1-3\al)-\Delta^*\Bigg[\sqrt{\frac{2}{\pi}}(1+2\al)+\Delta^*\Bigg]
\Bigg\},
\eeq
\end{widetext}
where $\phi=\sum_i \phi_i$ is the total volume fraction of the mixture. For the sake of simplicity, let us consider the bidimensional case. In this case, to display the expression of $\eta$ for $d=2$, it is convenient to define it in dimensionless form, namely, $\eta^*=\eta/\eta_0$ where
\beq
\label{5.24}
\eta_0=\frac{\sqrt{m T/\pi}}{2\sigma}
\eeq
is the low-density value of the shear viscosity in the elastic limit. The expression of $\eta^*$ is
\beq
\label{5.25}
\eta^*=\left[1+\frac{1}{2}\phi \chi \left(1+\alpha+\sqrt{\frac{2}{\pi}}\Delta^*\right)\right]\eta_k^*
+\frac{1}{2}\eta_b^*,
\eeq
where\cite{Note1}
%\footnote{A misprint in the calculation of the collision integral \eqref{5.20} for monocomponent granular gases was found in Ref.\ \onlinecite{GBS18} while the present paper was written. The expression \eqref{5.21.1} corrects this result. For this reason, the expression \eqref{5.26} for $\eta_k^*$ differs slightly from the one derived in Ref.\ \onlinecite{GBS18}.}
\beqa
\label{5.26}
\eta_k^*&=&\nu_\eta^{^*-1}\Bigg\{1-\frac{1}{4}\phi \chi\Bigg[(1+\al)
(1-3 \alpha) -4\sqrt{\frac{2}{\pi}}\nonumber\\
& & \times(1+2\al)\Delta^*-4\Delta^{*2}\Bigg]\Bigg\},
\eeqa
\beq
\label{5.27}
\eta_b^*=\frac{8}{\pi}\phi^2 \chi \left(1+\alpha+\sqrt{\frac{\pi}{2}}\Delta^*\right).
\eeq
In Eqs.\ \eqref{5.25}--\eqref{5.27}, the (reduced) collision frequency $\nu_\eta^*$ for $d=2$ is given by
\beq
\label{5.28}
\nu_\eta^*=\frac{3}{8}\chi \Bigg[\left(\frac{7}{3}-\alpha\right)(1+\alpha)
+\frac{2\sqrt{2\pi}}{3}(1-\al)\Delta^*-\frac{2}{3}\Delta^{*2}\Bigg].
\eeq

\begin{figure}
%[hbtp]
\begin{center}
\begin{tabular}{lr}
\resizebox{6.5cm}{!}{\includegraphics{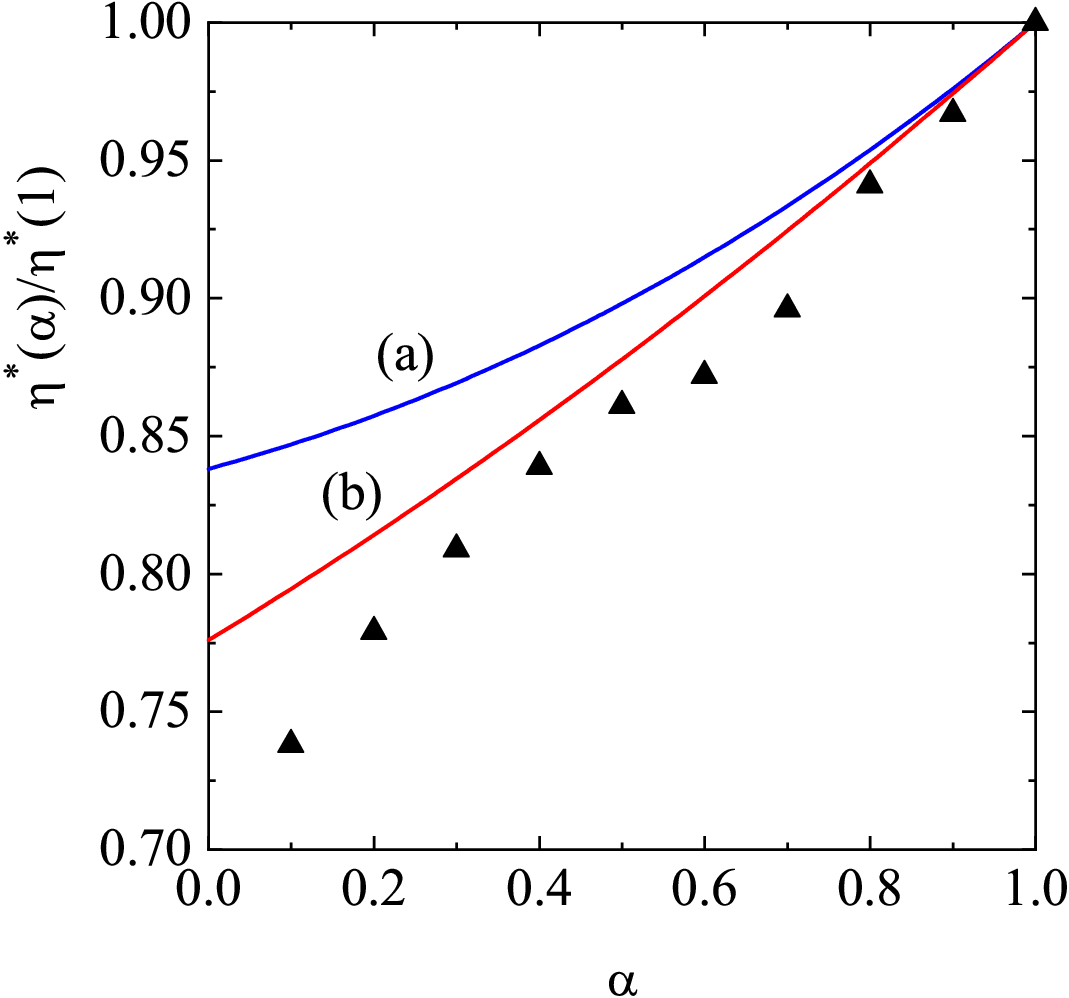}}
%&\resizebox{6.5cm}{!}{\includegraphics{Dpdil.pdf}}
\end{tabular}
\end{center}
\caption{Plot of the (scaled) shear viscosity coefficient $\eta^*(\al)/\eta^*(1)$ as a function of the coefficient of restitution $\al$ for $d=2$ and two different values of the solid volume fraction $\phi$: $\phi=0$ (a) and $\phi=0.314$ (b). The solid lines correspond to the results obtained from Eq.\ \eqref{5.25}. Symbols refer to MD simulations carried out in Ref.\ \onlinecite{SRB14} for $\phi=0.314$.
\label{fig1}}
\end{figure}

\vicente{The (reduced) cooling rate $\zeta^*$ for mechanically equivalent particles is
\beq
\label{5.28.1}
\zeta^*=\frac{\sqrt{2}\pi^{\frac{d-1}{2}}}{d\Gamma\left(\frac{d}{2}\right)}
\chi \left(1-\al^2-2\Delta^{*2}-\sqrt{2\pi}\al \Delta^*\right).
\eeq
In the steady state, the cooling rate vanishes. Thus, according to Eq.\ \eqref{5.28.1}, the condition $\zeta^*=0$ yields a quadratic equation in $\Delta^*$ whose physical solution (i.e., $\Delta^*=0$ if $\al=1$) provides the $\al$-dependence of $\Delta^*$. It is given by
\beq
\label{5.28.2}
\Delta^*(\al)=\frac{1}{2}\sqrt{\frac{\pi}{2}}\al\left[\sqrt{1+
\frac{4(1-\al^2)}{\pi \al^2}}-1\right].
\eeq
For identical particles, we have that  $\Delta^*=\Delta/\sqrt{2T/m}$. Thus, at given values of $\al$ and $\Delta$, Eq.\ \eqref{3.18} gives the value of the stationary temperature.  It should be noted that the relationship \eqref{5.28.2} has been tested against MD simulations, showing excellent agreement with deviations smaller than 2 \%, except for small values of the coefficient of restitution and/or high densities.\cite{BRS13} }

Figure \ref{fig1} plots the dependence of the (scaled) shear viscosity $\eta^*(\al)/\eta^*(1)$ on the coefficient of restitution $\al$ for two different values of the packing fraction, $\phi$: $\phi=0$ (dilute granular gas) and $\phi=0.314$ (moderately dense granular gas). Here, $\eta^*(1)$ refers to the shear viscosity for elastic collisions. The theoretical results obtained for the dense case are compared with those obtained by performing MD simulations.\cite{SRB14} For a given density, we observe that shear viscosity decreases with increasing inelasticity. Furthermore, for a given value of $\al$, the shear viscosity (scaled with respect to its elastic value) decreases with density. Regarding the comparison with MD simulations, we can conclude that the kinetic theory results qualitatively reproduce the trends observed in the simulations, despite the relatively high density. At a more quantitative level, the discrepancies between the two become more significant as inelasticity increases (let us say when $\al \lesssim 0.7$).

\section{Granular binary mixtures}
\label{sec6}

The results displayed in section \ref{sec5} apply for a mixture with an arbitrary number of species. To illustrate more clearly the dependence of both the diffusion transport coefficients and the shear viscosity on the parameters of the mixture we consider in this section a binary system ($s=2$). In addition, for the sake of simplicity, we will assume that $\Delta_{11}^*=\Delta_{22}^*=\Delta_{12}^*\equiv \Delta^*$.
This means that the
effective mechanism to transfer the kinetic energy injected by vibration in the vertical direction to the horizontal degrees of freedom of grains is the same for all the species.

As said before, in the case of a binary mixture (since $\mathbf{j}_1^{(1)}=-\mathbf{j}_2^{(1)}$),  one has the relations
\beq
\label{6.0}
D_{21}=-\frac{m_1}{m_2} D_{11}, \quad D_{22}=-\frac{m_1}{m_2} D_{12}, \quad D_1^T=-D_2^T.
\eeq
The expressions of the dimensionless transport coefficients $D_{11}^*$, $D_{12}^*$, and $D_1^{*T}$ are displayed in the Appendix \ref{appD} where
\beq
\label{6.1}
D_{ij}^*=\frac{m_i m_j \nu}{\rho T}D_{ij}, \quad D_{i}^{*T}=\frac{\rho \nu}{n T}D_{i}{^T}.
\eeq

Before examining the dependence of the transport coefficients on the system's parameter space in the dense regime, it is helpful to consider the low-density limiting case  ($\phi_i\to 0$).
\begin{figure}
%[hbtp]
\begin{center}
\begin{tabular}{lr}
\resizebox{6.5cm}{!}{\includegraphics{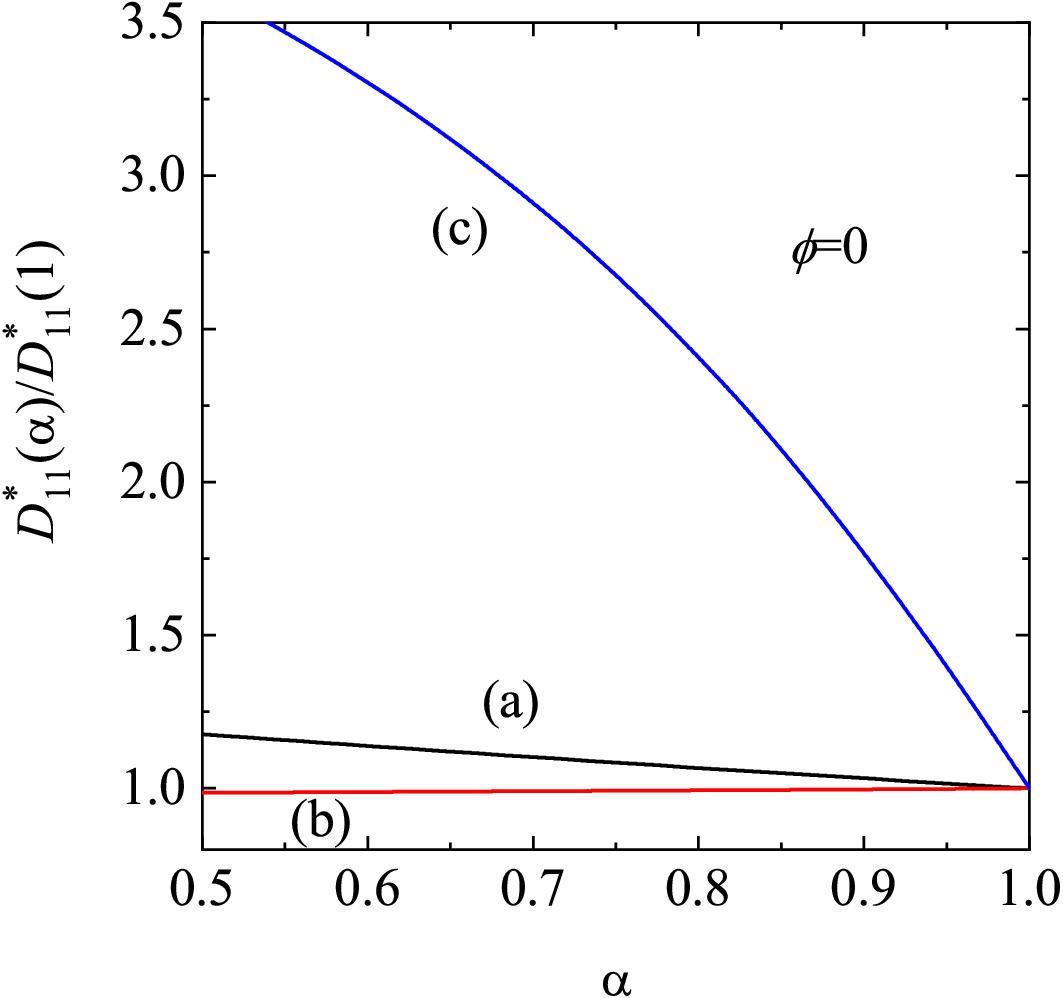}}
%&\resizebox{6.5cm}{!}{\includegraphics{Dpdil.pdf}}
\end{tabular}
\begin{tabular}{lr}
\resizebox{6.5cm}{!}{\includegraphics{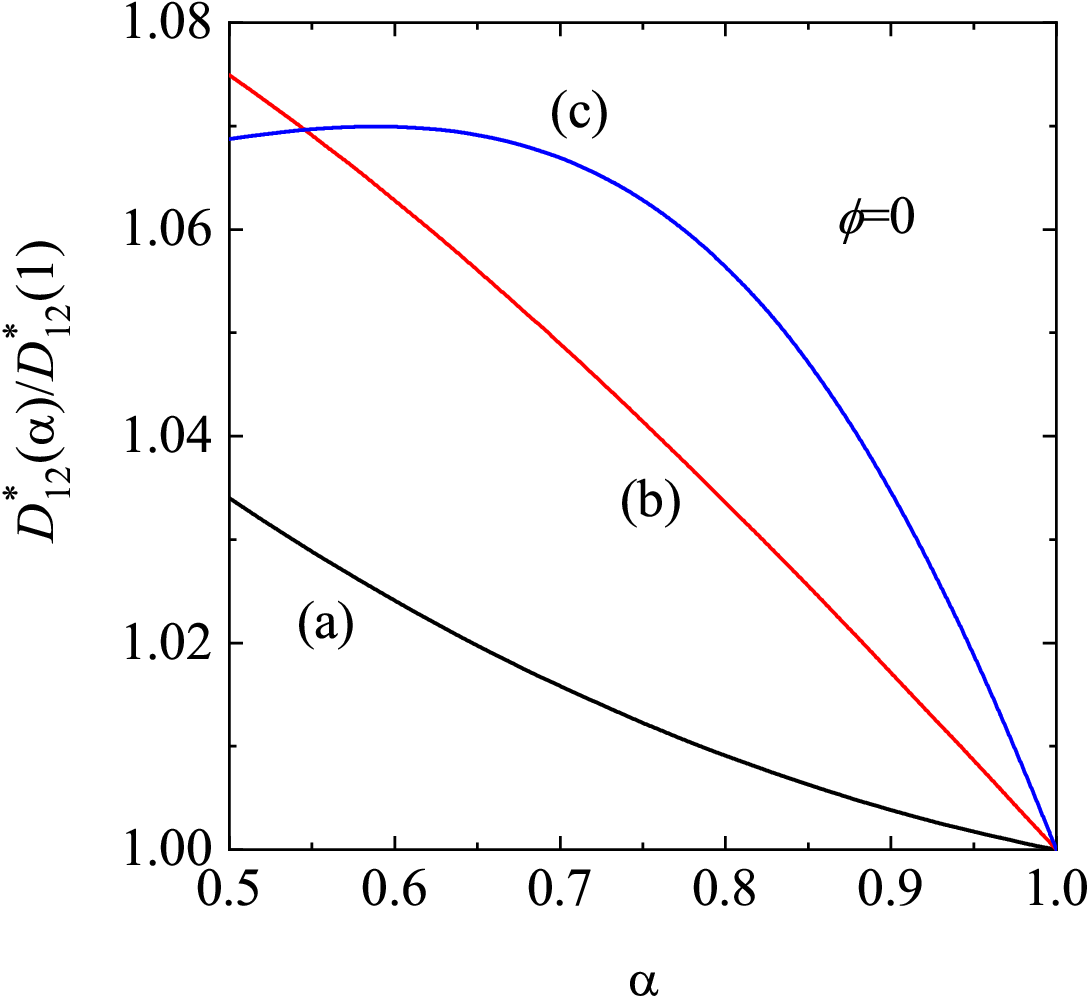}}
\end{tabular}
\begin{tabular}{lr}
\resizebox{6.5cm}{!}{\includegraphics{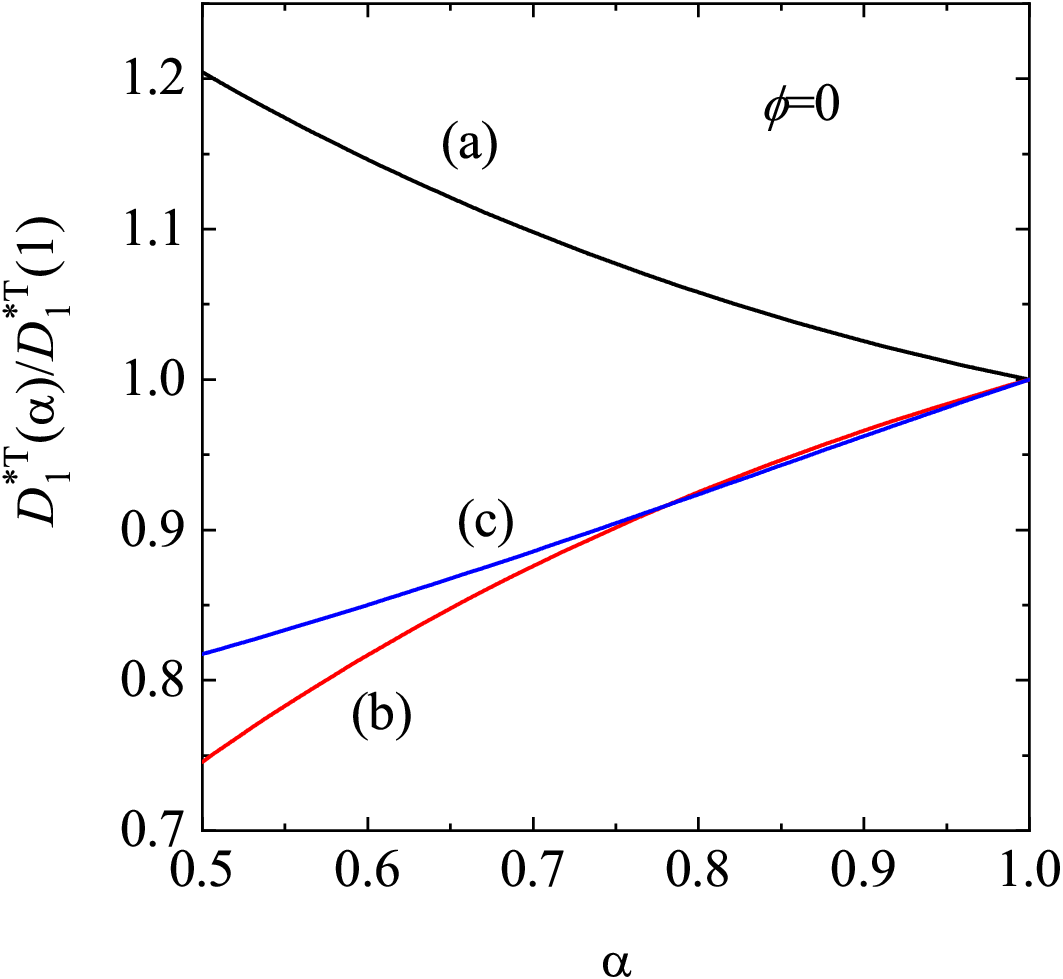}}
\end{tabular}
\end{center}
\caption{Plot of the (dimensionless) diffusion coefficients $D_{11}^*(\alpha)/D_{11}^*(1)$, $D_{12}^*(\alpha)/D_{12}^*(1)$, and $D_1^{*T}(\al)/D_1^{*T}(1)$ vs the (common) coefficient of restitution $\al_{ij}=\al$ for $d=2$, $x_1=0.5$, $\phi_i=0$, and three different binary mixtures: $\sigma_1/\sigma_2=2$, $m_1/m_2=2$ (a); $\sigma_1/\sigma_2=2$, $m_1/m_2=0.5$ (b); and $\sigma_1/\sigma_2=3$, $m_1/m_2=9$ (c). Here, $D_{11}^*(1)$, $D_{12}^*(1)$, and $D_1^{*T}(1)$ refer to the values of the diffusion coefficients for elastic collisions ($\al=1$).
\label{fig2}}
\end{figure}

\subsection{Diffusion transport coefficients. Low-density limit}

In the dilute regime, Eqs.\ \eqref{d3}, \eqref{d5} and \eqref{d6} reduce to
\beq
\label{6.2}
D_1^{*T}= \frac{x_1\left(\gamma_1-\frac{1}{2}\Delta^*\frac{\partial \gamma_1}{\partial \Delta^*}\right)-\frac{\rho_1}{\rho}}{\nu_D^*+\frac{1}{2}\Delta^*\frac{\partial \zeta_0^*}{\partial \Delta^*}},
\eeq
\beq
\label{6.3}
D_{11}^*=\frac{x_2 \frac{\partial \zeta_0^*}{\partial x_1}D_1^{*T}+\gamma_1+x_1x_2 \frac{\partial \gamma_1}{\partial x_1}-\frac{\rho_1}{\rho}}{\nu_D^*},
\eeq
\beq
\label{6.4}
D_{12}^*=-\frac{x_1 \frac{\partial \zeta_0^*}{\partial x_1}D_1^{*T}+x_1^2 \frac{\partial \gamma_1}{\partial x_1}+\frac{\rho_1}{\rho}}{\nu_D^*},
\eeq
where $\nu_D^*$ is given by Eq.\ \eqref{d4} with $\chi_ {12}=1$. To compare with the expressions derived in Ref.\ \onlinecite{GBS21}, one has first to express the mass flux $\mathbf{j}_1^{(1)}$ in terms of the spatial gradients $\nabla x_1$, $\nabla p$, and $\nabla T$. In this representation, the mass flux is written as\cite{GBS21}
\beq
\label{6.5}
\mathbf{j}_1^{(1)}=-\frac{m_1m_2n}{\rho} D\nabla x_1-\frac{\rho}{p}D_p \nabla p-\frac{\rho}{T}D_T \nabla T,
\eeq
where for a dilute granular mixture $p=nT$. In dimensionless form, the relationship between the coefficients $D_{11}^*$, $D_{12}^*$, and $D_1^{*T}$ and the coefficients $D^*=(m_1m_2\nu/\rho T)D$, $D_p^*=(\rho \nu/nT)D_p$, and $D_T^*=(\rho \nu/nT)D_T$ is
\beq
\label{6.6}
D^*=D_{11}^*-D_{12}^*, \quad D_p^*=x_1 D_{11}^*+x_2 D_{12}^*,
\eeq
\beq
\label{6.7}
D_T^*=D_1^{*T}-x_1 D_{11}^*-x_2 D_{12}^*.
\eeq
Substitution of Eqs.\ \eqref{6.2}--\eqref{6.4} into the relationships \eqref{6.6}--\eqref{6.7} yields Eqs.\ (102)--(104) of Ref.\ \onlinecite{GBS21}. This shows the consistency between the results derived here for the diffusion coefficients of dense granular mixtures with those derived in the low-density limit.\cite{GBS21}

\begin{figure}
%[hbtp]
\begin{center}
\begin{tabular}{lr}
\resizebox{6.5cm}{!}{\includegraphics{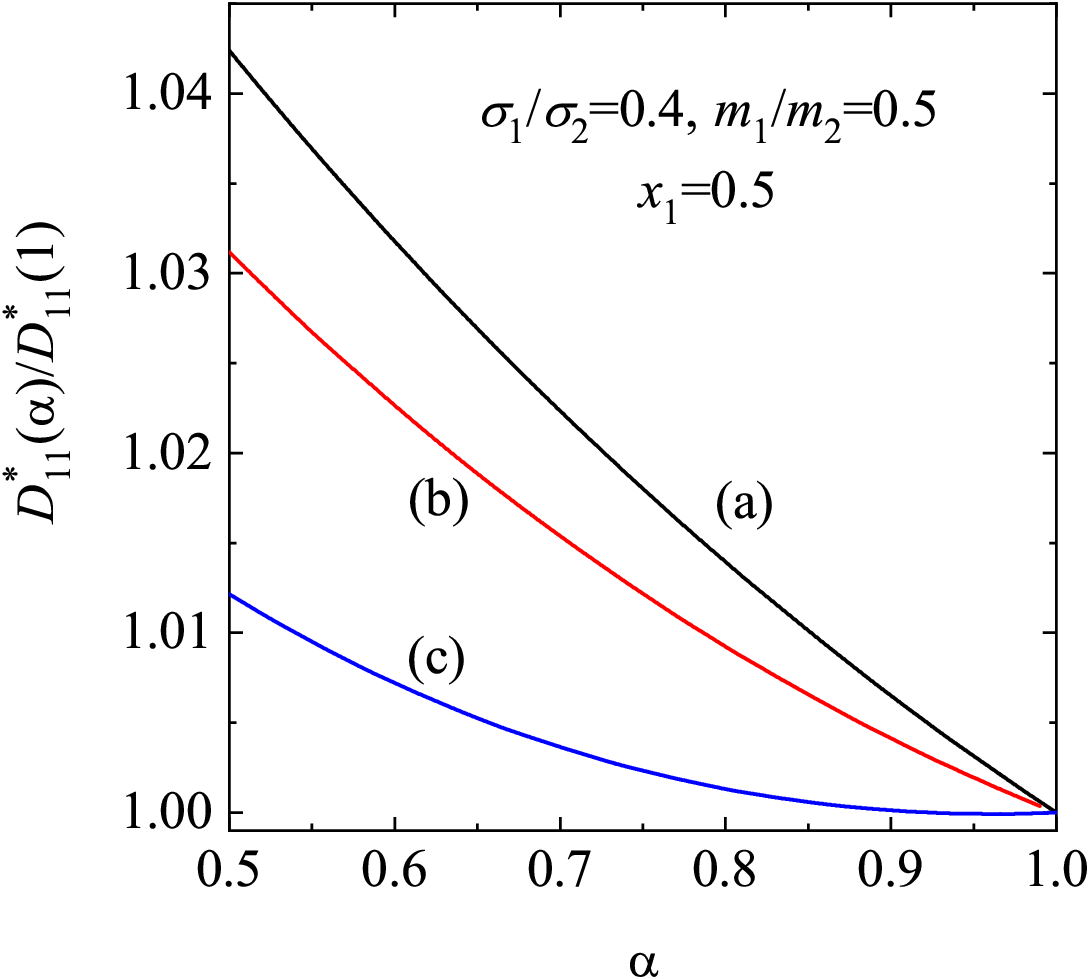}}
%&\resizebox{6.5cm}{!}{\includegraphics{Dpdil.pdf}}
\end{tabular}
\begin{tabular}{lr}
\resizebox{6.5cm}{!}{\includegraphics{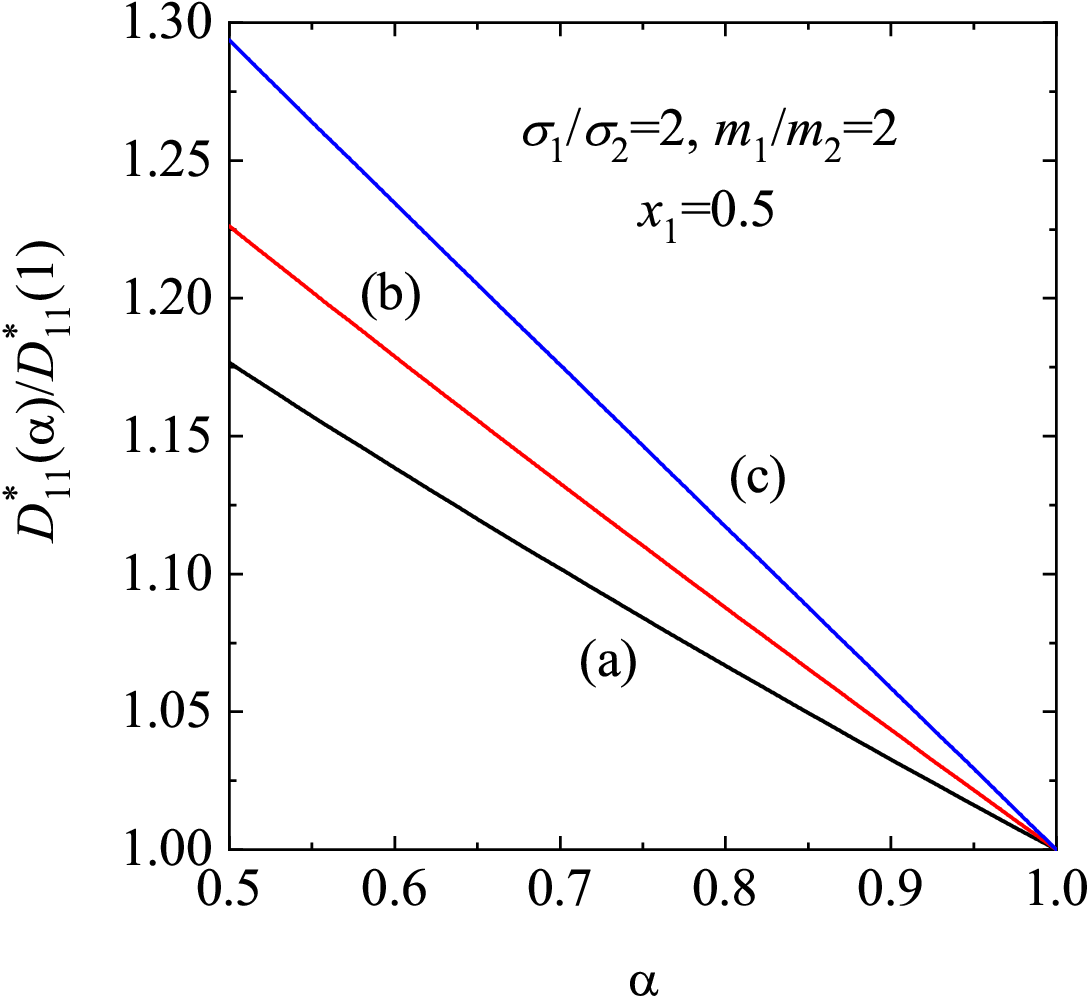}}
\end{tabular}
%\begin{tabular}{lr}
%\resizebox{6.5cm}{!}{\includegraphics{DTdilute.pdf}}
%\end{tabular}
\end{center}
\caption{Plot of the (dimensionless) diffusion coefficient $D_{11}^*(\alpha)/D_{11}^*(1)$ vs the (common) coefficient of restitution $\al_{ij}=\al$ for $d=2$, $x_1=0.5$, and two different mixtures: $\sigma_1/\sigma_2=0.5$, $m_1/m_2=0.4$, and $\sigma_1/\sigma_2=2$, $m_1/m_2=2$. Three different values of the solid volume fraction $\phi$ have been considered: $\phi=0$ (a), $\phi=0.1$ (b), and $\phi=0.2$ (c).
\label{fig3}}
\end{figure}
\begin{figure}
%[hbtp]
\begin{center}
\begin{tabular}{lr}
\resizebox{6.5cm}{!}{\includegraphics{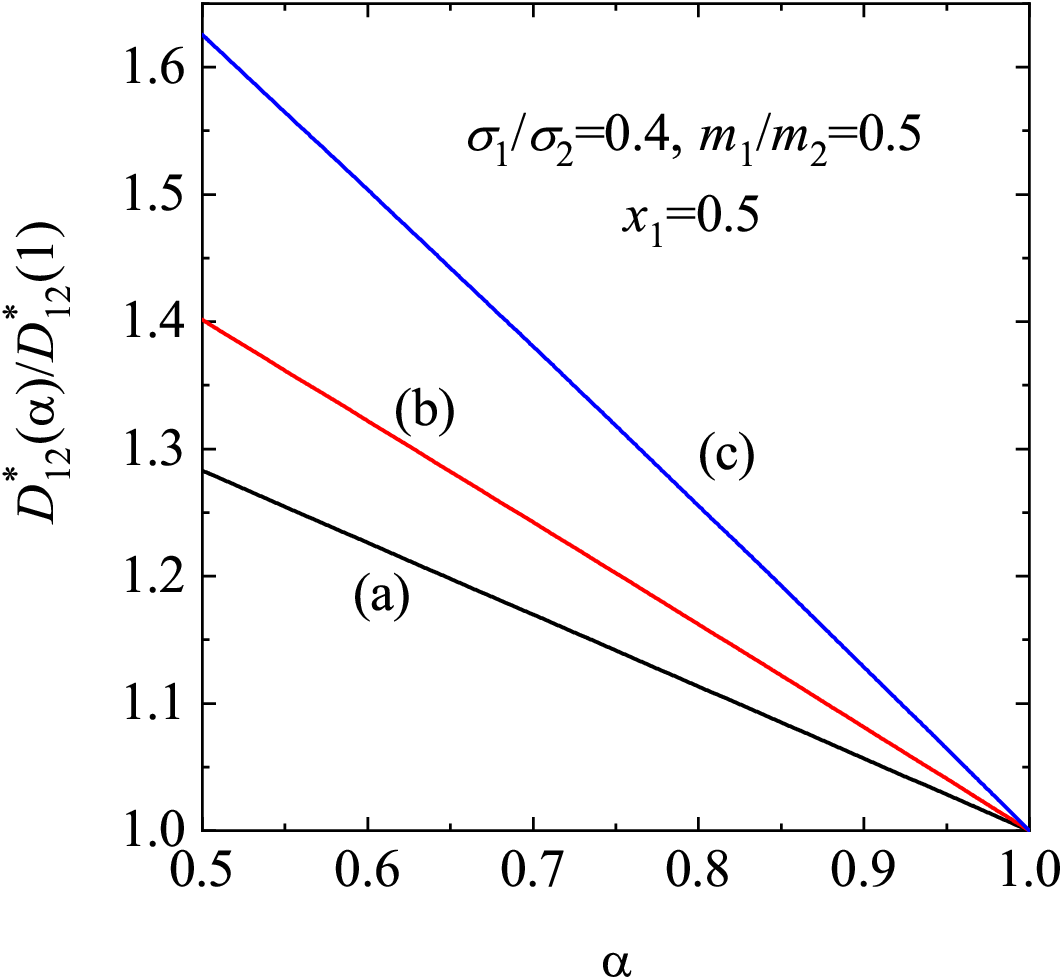}}
%&\resizebox{6.5cm}{!}{\includegraphics{Dpdil.pdf}}
\end{tabular}
\begin{tabular}{lr}
\resizebox{6.5cm}{!}{\includegraphics{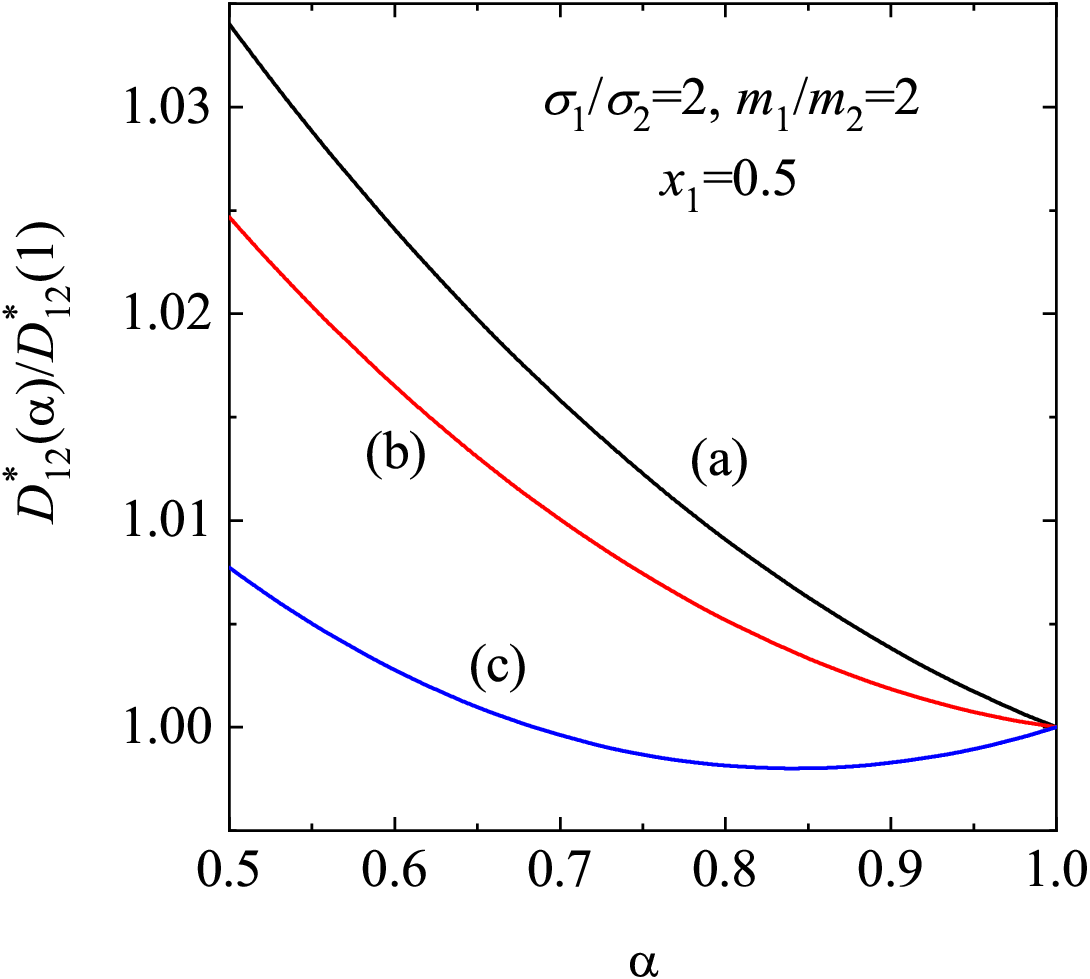}}
\end{tabular}
%\begin{tabular}{lr}
%\resizebox{6.5cm}{!}{\includegraphics{DTdilute.pdf}}
%\end{tabular}
\end{center}
\caption{Plot of the (dimensionless) diffusion coefficient $D_{12}^*(\alpha)/D_{12}^*(1)$ vs the (common) coefficient of restitution $\al_{ij}=\al$ for $d=2$, $x_1=0.5$, and two different mixtures: $\sigma_1/\sigma_2=0.5$, $m_1/m_2=0.4$, and $\sigma_1/\sigma_2=2$, $m_1/m_2=2$. Three different values of the solid volume fraction $\phi$ have been considered: $\phi=0$ (a), $\phi=0.1$ (b), and $\phi=0.2$ (c).
\label{fig4}}
\end{figure}
\begin{figure}
%[hbtp]
\begin{center}
\begin{tabular}{lr}
\resizebox{6.5cm}{!}{\includegraphics{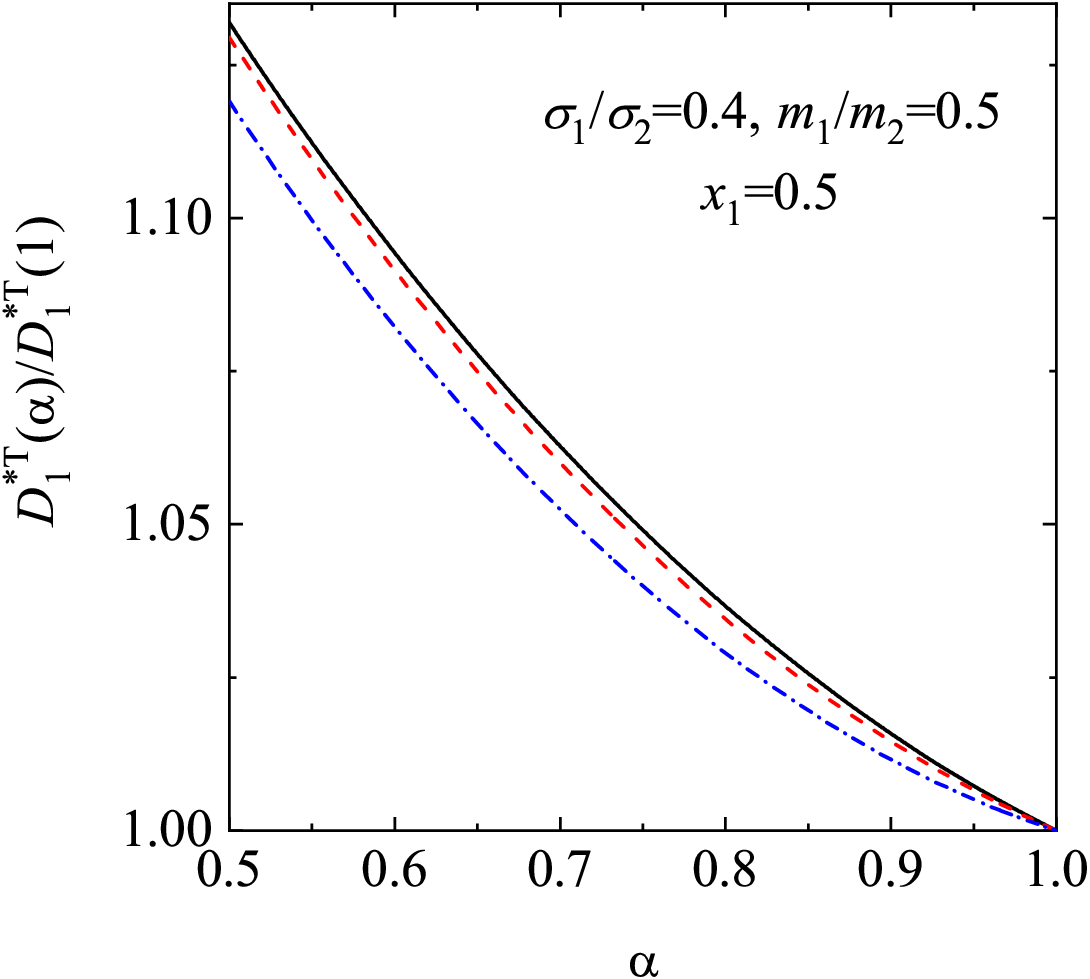}}
%&\resizebox{6.5cm}{!}{\includegraphics{Dpdil.pdf}}
\end{tabular}
\begin{tabular}{lr}
\resizebox{6.5cm}{!}{\includegraphics{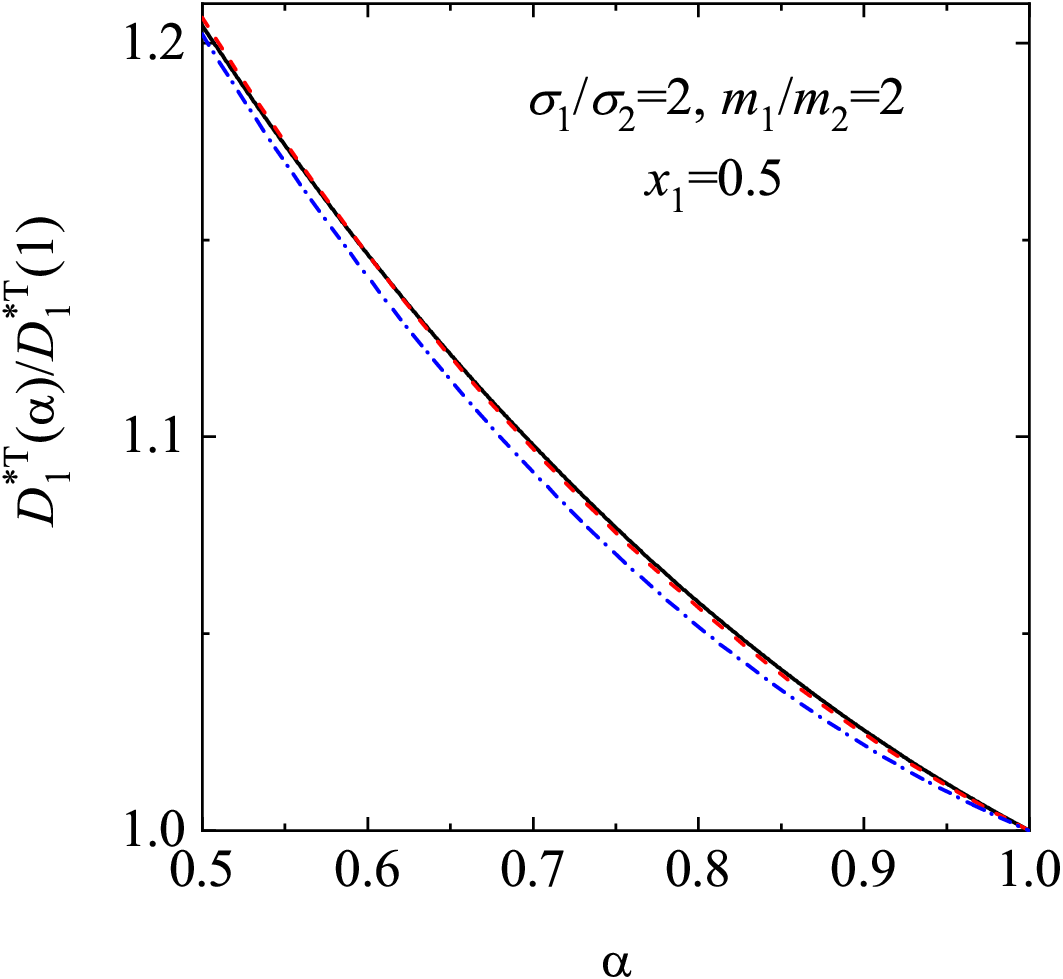}}
\end{tabular}
%\begin{tabular}{lr}
%\resizebox{6.5cm}{!}{\includegraphics{DTdilute.pdf}}
%\end{tabular}
\end{center}
\caption{Plot of the (dimensionless) diffusion coefficient $D_{1}^{*T}(\alpha)/D_{1}^{*T}(1)$ vs the (common) coefficient of restitution $\al_{ij}=\al$ for $d=2$, $x_1=0.5$, and two different mixtures: $\sigma_1/\sigma_2=0.5$, $m_1/m_2=0.4$, and $\sigma_1/\sigma_2=2$, $m_1/m_2=2$. Three different values of the solid volume fraction $\phi$ have been considered: $\phi=0$ (solid line), $\phi=0.1$ (dashed line), and $\phi=0.2$ (dash-dotted line).
\label{fig5}}
\end{figure}

\vicente{In contrast to the monocomponent limiting case, for which the dependence of $\Delta^*$ on $\al$ is explicitly given by Eq.\ \eqref{5.28.2}, the dependence of $\Delta^*$ and $T_1^{(0)}/T_2^{(0)}$ on the parameters of the mixture is determined by numerically solving the set of coupled equations $\zeta_1^*=0$ and $\zeta_2^*=0$. An estimate of the partial cooling rates $\zeta_i^*$ by assuming Maxwellian distributions for the zeroth-order solutions $f_i^{(0)}(\mathbf{V})$ are given by Eq.\  \eqref{3.21}. As previously mentioned, the forms \eqref{3.21} yield good agreement with computer simulations for both $\Delta^*$ and $T_1^{(0)}/T_2^{(0)}$.\cite{BSG20}}

Figure \ref{fig2} shows the (reduced) diffusion transport coefficients $D_{11}^*(\al)/D_{11}^*(1)$, $D_{12}^*(\al)/D_{12}^*(1)$, and $D_1^{*T}(\al)/D_1^{*T}(1)$ as functions of the (common) coefficient of restitution $\al_{ij}\equiv \al$ for $d=2$ and $x_1=0.5$. Three different mixtures have been considered in the low-density regime ($\phi=0$). For hard disks ($d=2$), the solid volume fraction $\phi$ is
\beq
\label{6.8.1}
\phi=\frac{\pi}{4}\left(n_1\sigma_1^2+n_2\sigma_2^2\right).
\eeq
Here, $D_{11}^*(1)$, $D_{12}^*(1)$, and $D_1^{*T}(1)$ refer to the values of the diffusion coefficients for elastic collisions ($\al=1$). \vicente{It must be recalled that $\Delta^*$ and $T_1^{(0)}/T_2^{(0)}$ change in Fig.\ \ref{fig2} with the parameters of the mixture.}
Although the deviations of the diffusion transport coefficients from their forms for elastic collisions in the $\Delta$-model are generally smaller than those in the conventional IHS model,\cite{GD02,GMD06,GM07} we observe that the effect of inelasticity on mass transport in two-dimensional, confined granular mixtures can be significant, especially for mixtures of particles with the same mass density (the case $\sigma_1/\sigma_2=3$ and $m_1/m_2=9$). Regarding the dependence on the mass ratio, we find that the (dimensionless) coefficients $D_{11}^*$ and $D_{12}^*$ increase monotonically with respect to their elastic forms with decreasing $\al$ (i.e, as the inelasticity in collisions increases), except for $D_{12}^*$ when $m_1/m_2=(\sigma_1/\sigma_2)^2=9$, where this coefficient exhibits weak non-monotonic dependence on $\al$ for relatively high inelasticity.

For the thermal diffusion coefficient, $D_1^{*T}$, it is apparent that, for a given diameter ratio, the coefficient increases with inelasticity when the mass ratio is greater than 1 and decreases when the mass ratio is less than 1. Conversely, for $m_1/m_2=9$ (where both species have the same mass density), the ratio of $D_1^{*T}(\al)/D_1^{*T}(1)$ decreases monotonically with decreasing $\al$. Figure \ref{fig2} also shows that all diffusion coefficients are positive.

\subsection{Diffusion transport coefficients. Moderately dense mixtures}

Now, we want to assess how density affects the diffusion transport coefficients. As before, we will scale these coefficients with respect to their values for elastic collisions. Figures \ref{fig3}--\ref{fig5} show the dependence of the ratios $D_{11}^*(\alpha)/D_{11}^*(1)$, $D_{12}^*(\alpha)/D_{12}^*(1)$, and $D_1^{*T}(\al)/D_1^{*T}(1)$ on the (common) coefficient of restitution $\al_{ij}\equiv \al$ for three different values of the solid volume fraction $\phi$: $\phi=0$ (low-density mixture), $\phi=0.1$ (moderately low density), and $\phi=0.2$ (moderately high density).
Two different mixtures are considered.

As expected, the deviations of the diffusion coefficients from their elastic forms are generally less significant than those obtained in the IHS model (see for instance, Figs. 5.5, 5.6, and 5.7 of Ref.\ \onlinecite{G19}). In general, we observe that the three dimensionless diffusion coefficients increase monotonically with $\al$, except for the coefficient $D_{12}^*$  at high density ($\phi=0.2$) when $\sigma_1>\sigma_2$ and $m_1>m_2$. The increase of the diffusion coefficients with increasing inelasticity is consistent with the trends observed in the IHS model.\cite{G19} Regarding the influence of density on diffusion transport coefficients, the effect of density on $D_{11}^*$ ($D_{12}^*$) is more significant when the mass and diameter ratios are greater than (smaller) than 1.
The effect of density on the thermal diffusion coefficient, $D_1^{*T},$ is generally weaker than on the other two diffusion coefficients. Additionally, for a given coefficient of restitution, we observe that $D_{11}^*$ increases with increasing density when the mass and diameter ratios are greater than 1 and decreases when they are smaller than 1. However, the opposite occurs for $D_{12}^*$.

\begin{figure}
%[hbtp]
\begin{center}
\begin{tabular}{lr}
\resizebox{6.5cm}{!}{\includegraphics{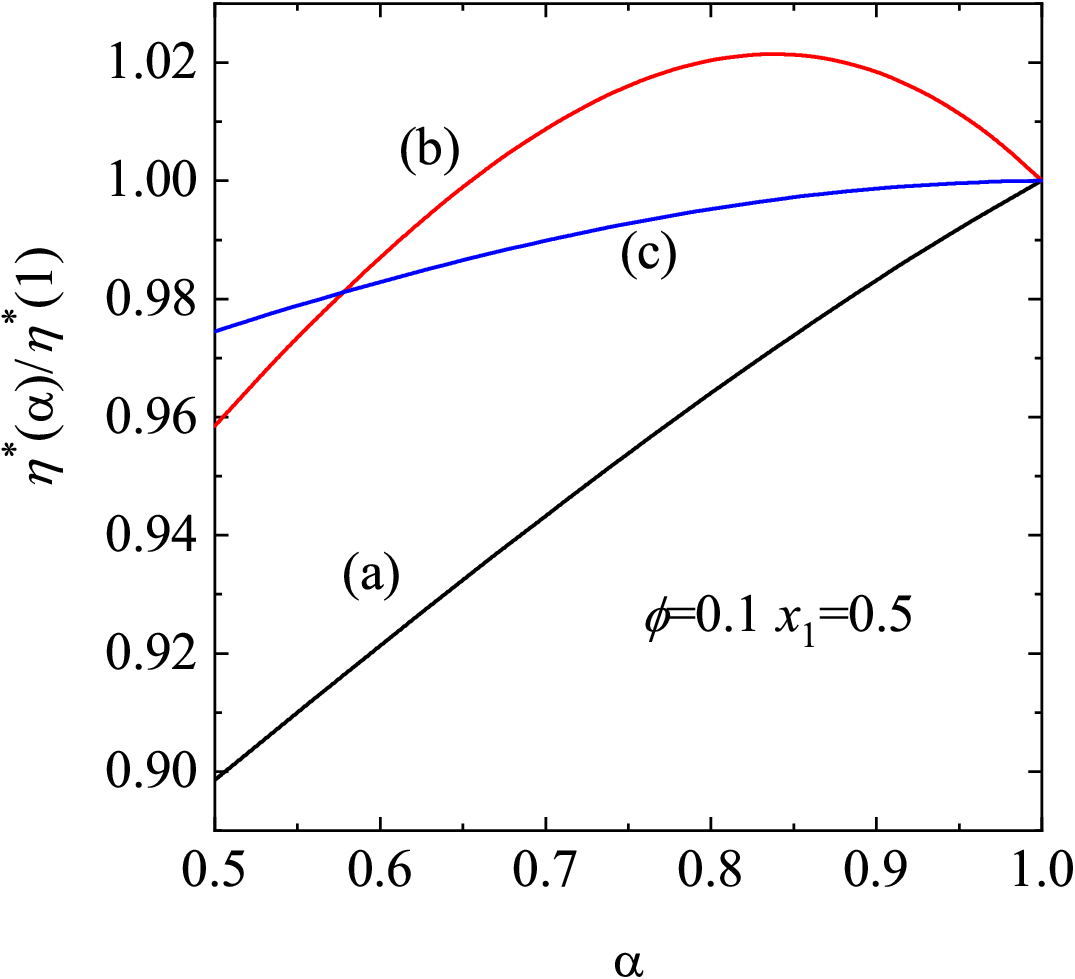}}
%&\resizebox{6.5cm}{!}{\includegraphics{Dpdil.pdf}}
\end{tabular}
\end{center}
\caption{Plot of the (scaled) shear viscosity coefficient $\eta^*(\al)/\eta^*(1)$ as a function of the (common) coefficient of restitution $\al$ for $d=2$, $x_1=0.5$, $\phi=0.1$, and three different mixtures: $\sigma_1/\sigma_2=2$, $m_1/m_2=2$ (a); $\sigma_1/\sigma_2=2$, $m_1/m_2=4$ (b); and $\sigma_1/\sigma_2=2$, $m_1/m_2=0.5$ (c).
\label{fig6}}
\end{figure}
\begin{figure}
%[hbtp]
\begin{center}
\begin{tabular}{lr}
\resizebox{6.5cm}{!}{\includegraphics{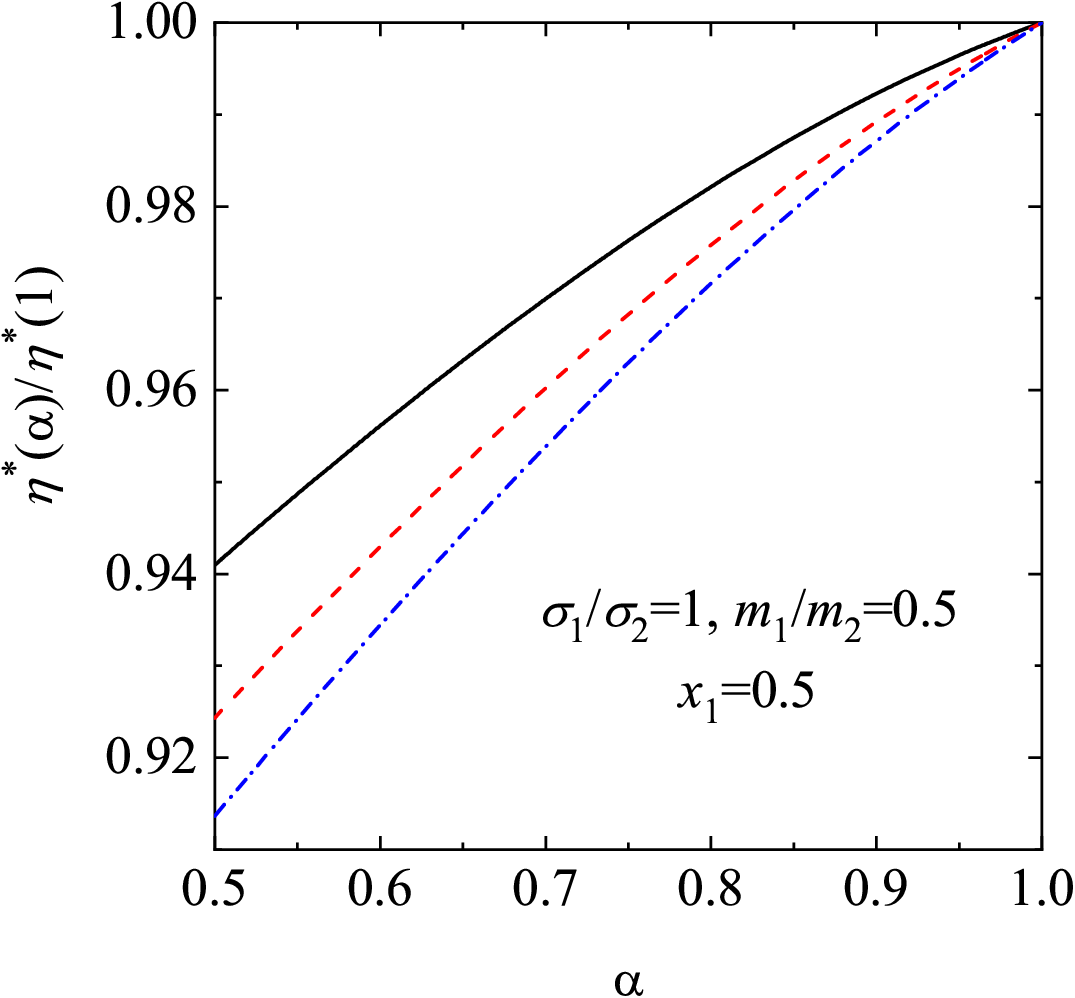}}
%&\resizebox{6.5cm}{!}{\includegraphics{Dpdil.pdf}}
\end{tabular}
\begin{tabular}{lr}
\resizebox{6.5cm}{!}{\includegraphics{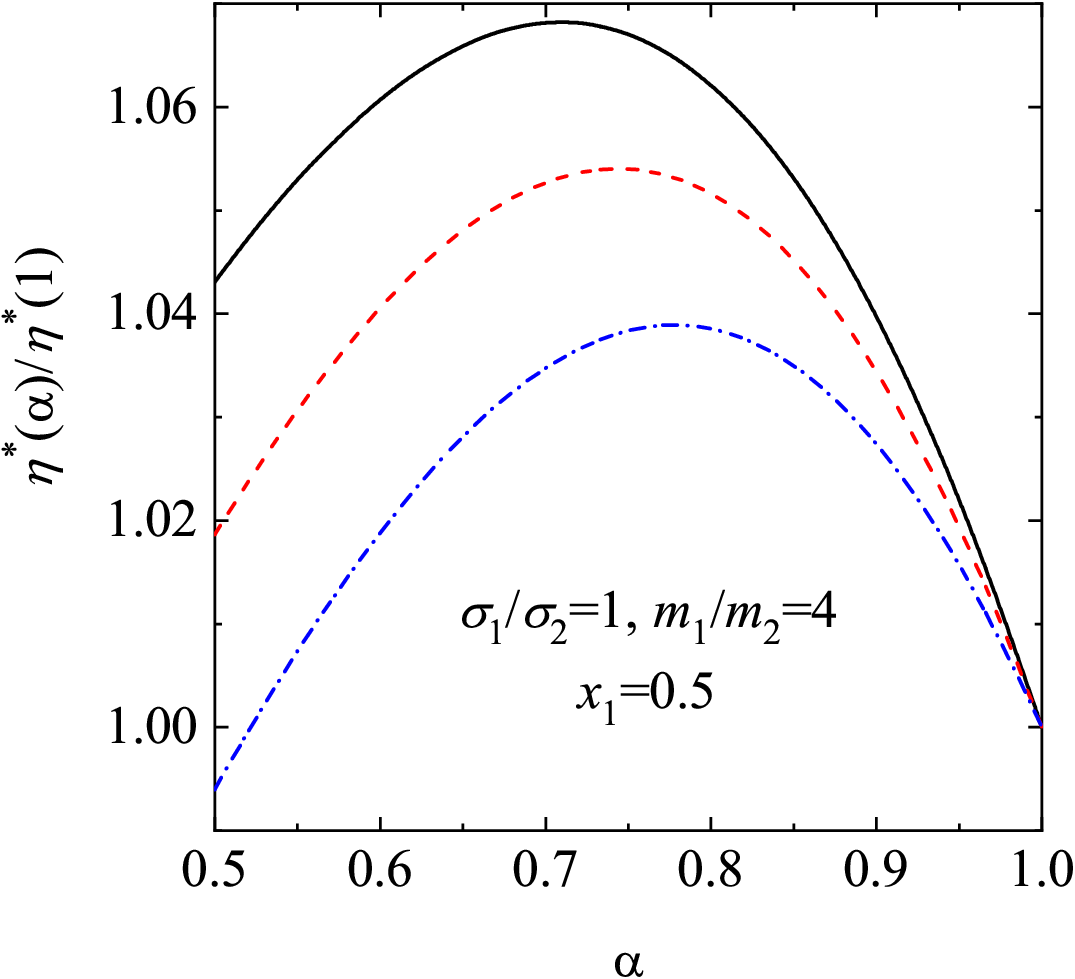}}
\end{tabular}
%\begin{tabular}{lr}
%\resizebox{6.5cm}{!}%{\includegraphics{DTdilute.pdf}}
%\end{tabular}
\end{center}
\caption{Plot of the (scaled) diffusion coefficient shear viscosity coefficient $\eta^*(\al)/\eta^*(1)$ vs the (common) coefficient of restitution $\al_{ij}=\al$ for $d=2$, $x_1=0.5$, and two different mixtures: $\sigma_1/\sigma_2=1$, $m_1/m_2=0.5$, and $\sigma_1/\sigma_2=1$, $m_1/m_2=4$. Three different values of the solid volume fraction $\phi$ have been considered: $\phi=0$ (solid line), $\phi=0.1$ (dashed line), and $\phi=0.2$ (dash-dotted line).
\label{fig7}}
\end{figure}

\subsection{Shear and bulk viscosities}

In the case of a binary mixture, the shear viscosity $\eta=\eta_1^k+\eta_2^k+\eta_c$ where
\beq
\label{6.8.n1}
\eta_1^k=\frac{\tau_{22}\Omega_1-\tau_{12}\Omega_2}{\tau_{11}\tau_{22}-\tau_{12}\tau_{21}}, \quad
\eta_2^k=\frac{\tau_{11}\Omega_2-\tau_{21}\Omega_1}{\tau_{11}\tau_{22}-\tau_{12}\tau_{21}}.
\eeq
Here, $\Omega_i$ is given by Eq.\ \eqref{5.17} and the expressions of the collision frequencies $\tau_{ij}$ are displayed in the Appendix \ref{appC}. For a two-dimensional system, the collisional shear viscosity $\eta_c$ is given by Eq.\ \eqref{5.4}.

The dependence of the (scaled) shear viscosity $\eta^*(\al)/\eta^*(1)$ on the (common) coefficient of restitution $\al$ is shown in Fig.\ \ref{fig6} for $d=2$, $x_1=0.5$, $\phi=0.1$, and three different binary mixtures. Here, $\eta^*=(\nu/nT)\eta$ and $\eta^*(1)$ refers to the value of $\eta^*$ for elastic collisions. Depending on the mass and/or diameter ratios, the ratio of shear viscosities $\eta^*(\al)/\eta^*(1)$ decreases with increasing inelasticity or shows non-monotonic dependence on $\al$.
To complement Fig.\ \ref{fig6}, Fig.\ \ref{fig7} shows the influence of the density on the (dimensionless) shear viscosity. We plot $\eta^*(\al)/\eta^*(1)$ versus $\al$ for two different mixtures and three values of the solid volume fraction $\phi$. As with confined single-component granular gases,\cite{GBS18} it is quite apparent that the scaled shear viscosity coefficient in the $\Delta$ model exhibits weaker density dependence than the conventional IHS model.\cite{GMD06,G19}
We also observe that, when $m_1<m_2$, the ratio $\eta^*(\al)/\eta^*(1)$ decreases with decreasing $\al$, whereas, when $m_1>m_2$, the ratio exhibits a non-monotonic dependence with inelasticity. In the latter case, the shear viscosity of the confined granular mixture is in general larger than its corresponding value for elastic collisions. Additionally, the effect of inelasticity on the shear viscosity of a confined granular mixture is much less significant than that found in previous works in the IHS model (see, for example, Figs.\ 5.8 and 5.9 of Ref.\ \onlinecite{G19}).
\begin{figure}
%[hbtp]
\begin{center}
\begin{tabular}{lr}
\resizebox{6.5cm}{!}{\includegraphics{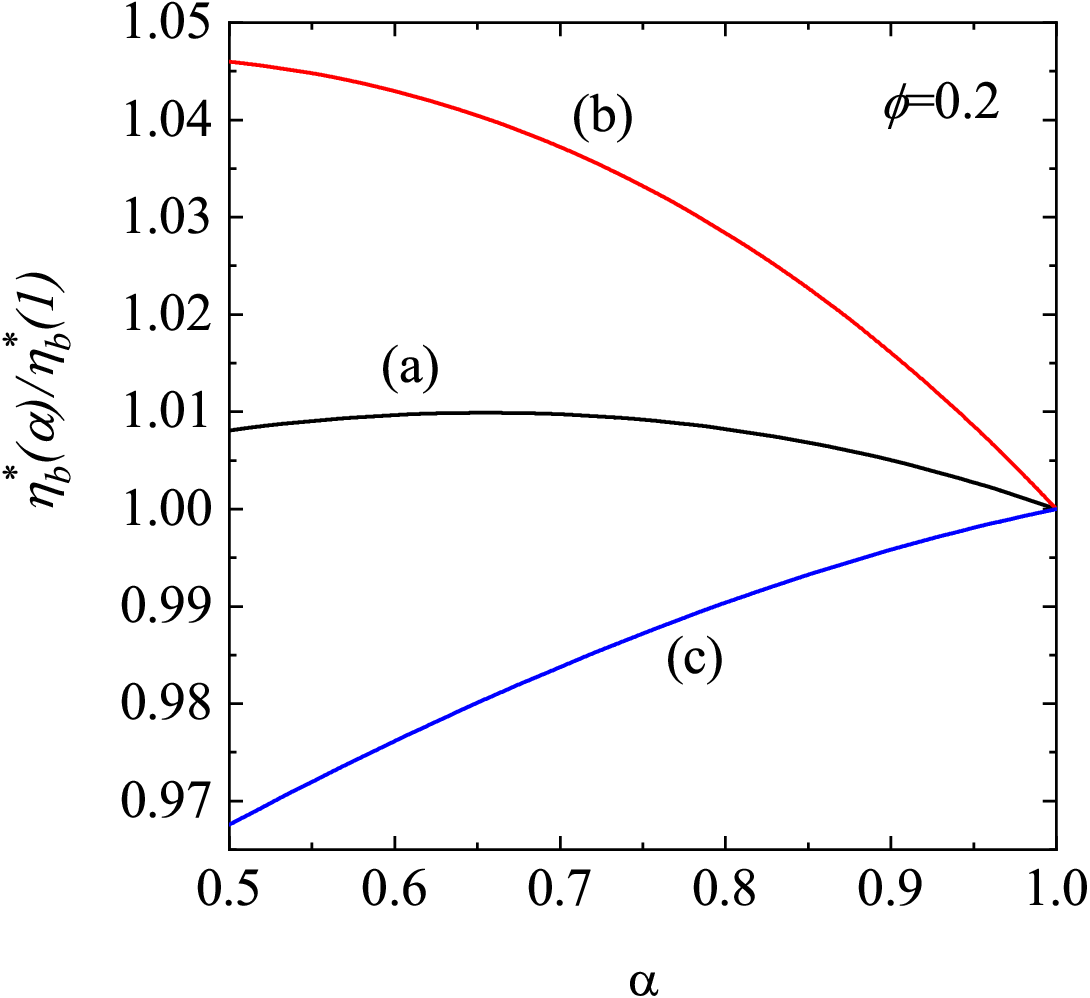}}
%&\resizebox{6.5cm}{!}{\includegraphics{Dpdil.pdf}}
\end{tabular}
\end{center}
\caption{Plot of the (scaled) bulk viscosity coefficient $\eta_b^*(\al)/\eta_b^*(1)$ as a function of the (common) coefficient of restitution $\al$ for $d=2$, $x_1=0.5$, $\phi=0.2$, and three different mixtures: $\sigma_1/\sigma_2=2$, $m_1/m_2=2$ (a); $\sigma_1/\sigma_2=2$, $m_1/m_2=4$ (b); and $\sigma_1/\sigma_2=2$, $m_1/m_2=0.5$ (c).
\label{fig8}}
\end{figure}

From Eqs.\ \eqref{4.14.1}--\eqref{4.16}, it is clear that evaluating the bulk viscosity $\eta_b$ requires knowing the first-order contributions $\varpi_i$ to the partial temperatures. However, determining the above coefficients is beyond the scope of this paper, as it involves lengthy and complex calculations. On the other hand, as mentioned in section \ref{sec5}, according to the previous results derived from the IHS model\cite{GGG19b} and for low-density confined granular mixtures\cite{GBS18} the influence of $\varpi_i$ on the value of the bulk viscosity $\eta_b$ is generally very small. Thus, for practical purposes, the bulk viscosity $\eta_b$ can be well estimated by $\eta_b^{(\text{I})}$.
The ratio $\eta_b^*(\al)/\eta_b^*(1)$ is plotted in Fig.\ \ref{fig8} as a function of the (common) coefficient of restitution $\al$ for $d=2$, $x_1=0.5$, $\phi=0.2$ and three different mixtures. Here, $\eta_b^*=(\nu/nT)\eta_b$ where $\eta_b^*(1)$ is the (dimensionless) bulk viscosity for elastic collisions. As with the case of the shear viscosity, we observe that collisional dissipation has a weaker influence on the bulk viscosity than in the IHS model.\cite{GGG19b}

\section{An application: Thermal diffusion segregation in a confined granular dense mixture}
\label{sec7}

Knowing the complete set of diffusion transport coefficients allows us to apply our theoretical results to one of the most interesting problems in multicomponent systems: the segregation and mixing of dissimilar species or components in a binary mixture. In the context of granular systems, the segregation problem is relevant not only from a fundamental point of view, but also from a practical one. This problem has led to significant experimental, computational, and theoretical research in granular media, particularly when the system is fluidized by vibrating walls. In our model, segregation is induced by the combined effects of gravity and a thermal gradient. The objective here is to extend our previous results (which were derived for dilute systems with arbitrary concentration\cite{GBS24,GBS24a} and for moderate densities in the tracer limit\cite{GGBS24}) to arbitrary concentrations and moderate densities. Thus, our present study encompasses the previous two works,\cite{GBS24,GGBS24} whose results are recovered when appropriate limiting cases are taken.

\begin{figure}
%[hbtp]
\begin{center}
\begin{tabular}{lr}
\resizebox{7.5cm}{!}{\includegraphics{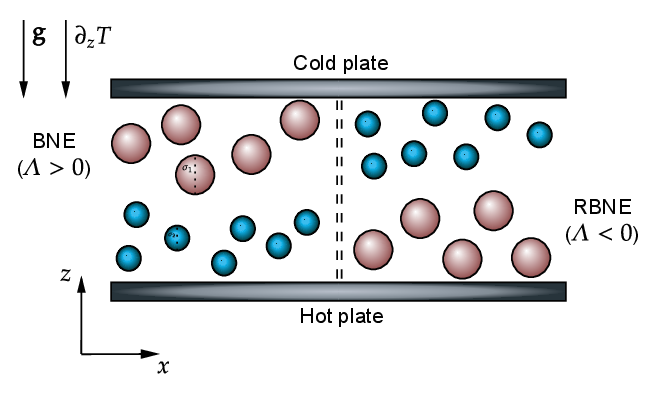}}
%\resizebox{7.5cm}{!}{\includegraphics{BNE_RBNE_new.pdf}}
%&\resizebox{6.5cm}{!}{\includegraphics{Dpdil.pdf}}
\end{tabular}
\end{center}
\caption{Illustration of the segregation process behavior in a granular binary mixture. The BNE effect ($\Lambda>0$) occurs when the large (brown) particles tend to accumulate near the cold (top) plate of the system. The RBNE effect ($\Lambda<0$) occurs when the large (brown) particles tend to accumulate near the hot (bottom) plate of the system.
\label{BNE}}
\end{figure}

Thermal diffusion is well-known to be caused by the relative motion of species within a mixture due to a thermal gradient. The motion of the species of the mixture gives rise to concentration gradients that lead to diffusion processes. A steady state is reached where the segregation effect from thermal diffusion is balanced by the mixing effect of ordinary diffusion.\cite{KCL87} The amount of segregation parallel to the thermal gradient can be measured by the thermal diffusion factor $\Lambda$. This quantity is defined in an inhomogeneous non-convecting ($\mathbf{U}=\mathbf{0}$) steady state with zero mass flux ($\mathbf{j}_i=\mathbf{0}$) as
\beq
\label{7.1}
-\Lambda \frac{\partial \ln T}{\partial z}=\frac{\partial}{\partial z}\ln \left(\frac{n_1}{n_2}\right).
\eeq
A binary mixture has been considered here, and, for simplicity, it is assumed that, in two-dimensional systems, gradients occur only along the $z$-axis. Furthermore, without loss of generality, we assume that $\sigma_1>\sigma_2$ and that the gravitational field is parallel to the thermal gradient, so $\mathbf{g}=-g \hat{e}_z$ where $\hat{e}_z$ is the unit vector in the positive direction of the $z$-axis This means that the bottom plate is hotter than the top plate, so $\partial_z T<0$ \vicente{(see for instance the sketch of Fig.\ \ref{BNE})}.

According to Eq.\ \eqref{7.1}, when $\Lambda>0$, the larger particles 1 rise to the top (cold) plate, since $\partial_z \ln (n_1/n_2)>0$. \vicente{We recall that this effect is usually known as the Brazil nut effect (BNE)}.
Conversely, when $\Lambda<0$, the larger particles 1 sink to the bottom (hot) plate, since $\partial_z \ln (n_1/n_2)<0$. \vicente{This effect is known as the reverse Brazil nut effect (RBNE)}.
 Now, we write the thermal diffusion factor $\Lambda$ in terms of the (dimensionless) diffusion transport coefficients $D_{11}^*$, $D_{12}^*$, and $D_{1}^{*T}$. First, the momentum balance equation \eqref{2.11} leads to
\beq
\label{7.2}
\frac{\partial p}{\partial z}=-\rho g,
\eeq
where we have accounted for that $P_{ij}=p\delta_{ij}$. According to Eq.\ \eqref{3.19.1}, the hydrostatic pressure $p=nT p^*$ where the dimensionless pressure $p^*$ depends on $z$ through its dependence on $x_1$, $\phi$, and $\Delta^*=\Delta/v_\text{th}(T(z))$. Thus, in dimensionless form, Eq.\ \eqref{7.2} can be rewritten as
\beq
\label{7.3}
x_1 \xi_1 \Lambda_1+x_2 \xi_2 \Lambda_2-\frac{1}{2}\Delta^* \frac{\partial p^*}{\partial \Delta^*}=-\left(p^*+g^*\right),
\eeq
where
\beq
\label{7.4}
\Lambda_1=\frac{\partial_z \ln n_1}{\partial_z \ln T}, \quad \Lambda_2=\frac{\partial_z \ln n_2}{\partial_z \ln T},
\eeq
\vspace{0.1mm}
\beq
\label{7.5}
\xi_1=T^{-1}\frac{\partial p}{\partial n_1}=p^*+\frac{\phi_1}{x_1}\frac{\partial p^*}{\partial \phi}+x_2
\frac{\partial p^*}{\partial x_1},
\eeq
\beq
\label{7.6}
\xi_2=T^{-1}\frac{\partial p}{\partial n_2}=p^*+\frac{\phi_2}{x_2}\frac{\partial p^*}{\partial \phi}-x_1
\frac{\partial p^*}{\partial x_1},
\eeq
and
\beq
\label{7.7}
g^*=\frac{\rho g}{n\left(\frac{\partial T}{\partial z}\right)}<0
\eeq
is a dimensionless parameter measuring the competing effect between gravity and thermal gradient on segregation. In addition, according to the constitutive equation \eqref{4.6} for a binary mixture, the steady state condition $\mathbf{j}_1^{(0)}=\mathbf{0}$  yields the relationship
\beq
\label{7.8}
-D_1^{*T}=x_1 \Lambda_1 D_{11}^*+x_2 \Lambda_2 D_{12}^*.
\eeq
The solution to the set of linear equations \eqref{7.3} and \eqref{7.8} for $\Lambda_1$ and $\Lambda_2$ is
\beq
\label{7.9}
\Lambda_1=\frac{(p^*+g^*-\frac{1}{2}\Delta^* \frac{\partial p^*}{\partial \Delta^*})D_{12}^*-\xi_2D_1^{T*}}{x_1(\xi_2 D_{11}^*-\xi_1 D_{12}^*)},
\eeq
\vspace{0.05mm}
\beq
\label{7.9.1}
\quad
\Lambda_2=\frac{\xi_1 D_1^{T*}-(p^*+g^*-\frac{1}{2}\Delta^* \frac{\partial p^*}{\partial \Delta^*})D_{11}^*}{x_2(\xi_2 D_{11}^*-\xi_1 D_{12}^*)}.
\eeq
According to Eqs.\ \eqref{7.1} and \eqref{7.4}, $\Lambda=\Lambda_1-\Lambda_2$ and hence the thermal diffusion factor $\Lambda$ can be finally written as
\begin{widetext}
\begin{equation}
\label{7.10}
\Lambda=\frac{D_1^{T*}(x_1\xi_1+x_2\xi_2)-(p^*+g^*-\frac{1}{2}\Delta^* \frac{\partial p^*}{\partial \Delta^*})(x_1D_{11}^*+x_2D_{12}^*)}{x_1x_2(\xi_2D_{11}^*-\xi_1D_{12}^*)}.
\end{equation}
\end{widetext}
The explicit form of $\Lambda$ on the parameters of the mixture can be obtained when one substitutes Eqs.\ \eqref{d3}, \eqref{d5} and \eqref{d6} for the diffusion transport coefficients $D_1^{T*}$, $D_{11}^*$, and $D_{12}^*$, respectively,  and Eq.\ \eqref{3.19.1} for $p^*$ (and its corresponding derivatives $\xi_i$) into Eq.\ \eqref{7.10}. This yields the dependence of the thermal diffusion factor on the parameter space of the problem, including mass and size ratios, mole fraction, scaled gravity, solid volume fraction, and coefficients of restitution. In particular, inspecting the dependence of the denominator $x_1x_2(\xi_2D_{11}^*-\xi_1D_{12}^*)$ of Eq.\ \eqref{7.10} on the parameters of the mixture shows that it is usually positive.
Thus, the condition of setting the thermal diffusion factor to zero yields the curves delineating the regimes between \vicente{the segregation toward the cold and the hot wall (BNE/RBNE transition)}. This yields the segregation criterion
\begin{widetext}
\beq
\label{7.11}
(x_1\xi_1+x_2\xi_2)D_1^{T*}=\left(p^*+g^*-\frac{1}{2}\Delta^* \frac{\partial p^*}{\partial \Delta^*}\right)\left(x_1D_{11}^*+x_2D_{12}^*\right).
\eeq
\end{widetext}
Since criterion \eqref{7.11} involves many parameters, it is helpful to first consider some limiting cases to understand the different competing mechanisms that appear in the segregation problem.

\begin{figure}
%[hbtp]
\begin{center}
\begin{tabular}{lr}
\resizebox{7.0cm}{!}{\includegraphics{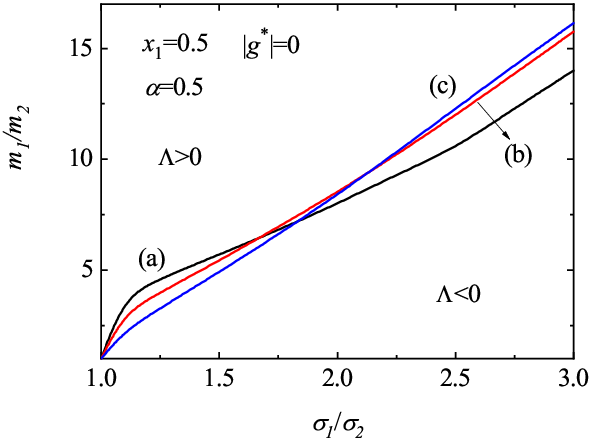}}
%&\resizebox{6.5cm}{!}{\includegraphics{Dpdil.pdf}}
\end{tabular}
\end{center}
\caption{Phase diagram for the marginal segregation curve ($\Lambda=0$) in the $(\sigma_1/\sigma_2, m_1/m_2)$-plane for a two-dimensional system with $x_1=0.5$, $|g^*|=0$, a (common) coefficient of restitution $\al=0.5$, and three different values of the solid volume fraction: $\phi=0$ (a); $\phi=0.1$ (b); and $\phi=0.2$ (c).
\label{fig9}}
\end{figure}

\subsection{Mechanically equivalent particles}

In this case, $D_1^{T*}=0$, $D_{11}^*=-(x_2/x_1)D_{12}^*$ and hence, Eq.\ \eqref{7.11} holds for any value of the coefficients of restitution, masses, diameters, solid volume fraction, and $\Delta^*$. As expected, therefore, no segregation appears in the mixture.

\subsection{Low-density regime}

For dilute granular mixtures, $p^*=1$, $\xi_1=\xi_2=1$, and the diffusion transport coefficients are given by Eqs.\ \eqref{6.2}--\eqref{6.4}. In this regime, Eq.\ \eqref{7.11} becomes
\beq
\label{7.12}
D_1^{T^*}=\left(1+g^*\right) \left(x_1D_{11}^*+x_2D_{12}^*\right).
\eeq
Equation \eqref{7.12} agrees with the segregation criterion found in Ref.\ \onlinecite{GBS24} when one takes into account the relations \eqref{6.6} and \eqref{6.7}.

\subsection{Tracer limit for moderate densities}

In the tracer limit ($x_1\to 0$) and moderate densities, $x_1\xi_1+x_2\xi_2\simeq \xi=p^*+\phi \partial_\phi p^*$ and Eq.\  \eqref{7.11} leads to
\beq
\label{7.13}
\xi \overline{D}_1^{T}=\left(p^*+g^*-\frac{1}{2}\Delta^* \frac{\partial p^*}{\partial \Delta^*}\right)\left(D_{11}^*+\overline{D}_{12}\right),
\eeq
where $\overline{D}_1^{T}=x_1^{-1}D_1^{*T}$ and $\overline{D}_{12}=x_1^{-1}D_{12}^*$. Equation \eqref{7.13} is consistent with the results obtained in Ref.\ \onlinecite{GGBS24} for the segregation of an intruder in a granular confined dense gas.

\begin{figure}
%[hbtp]
\begin{center}
\begin{tabular}{lr}
\resizebox{7.0cm}{!}{\includegraphics{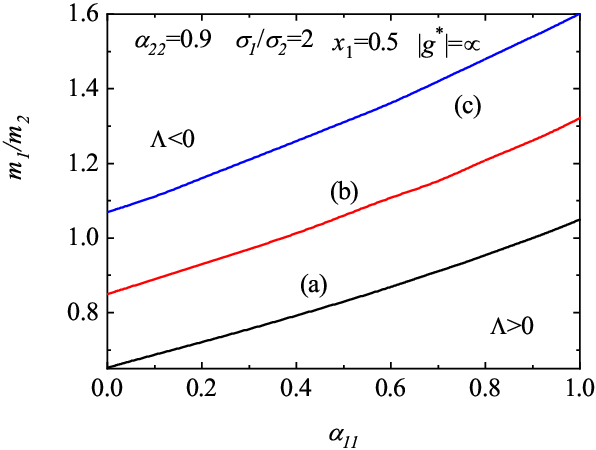}}
%&\resizebox{6.5cm}{!}{\includegraphics{Dpdil.pdf}}
\end{tabular}
\end{center}
\caption{Plot of the dependence of the marginal segregation curve ($\Lambda=0$) on the coefficient of restitution $\al_{11}$ in for a two-dimensional system with $x_1=0.5$, $\sigma_1/\sigma_2=2$, $\al_{22}=0.9$, $\al_{12}=(\al_{11}+\al_{22})/2$,
and three different values of the solid volume fraction: $\phi=0$ (a); $\phi=0.1$ (b); and $\phi=0.2$ (c). The limiting case $|g^*|\to \infty$ is considered.
\label{fig10}}
\end{figure}
\begin{figure}
%[hbtp]
\begin{center}
\begin{tabular}{lr}
\resizebox{7.0cm}{!}{\includegraphics{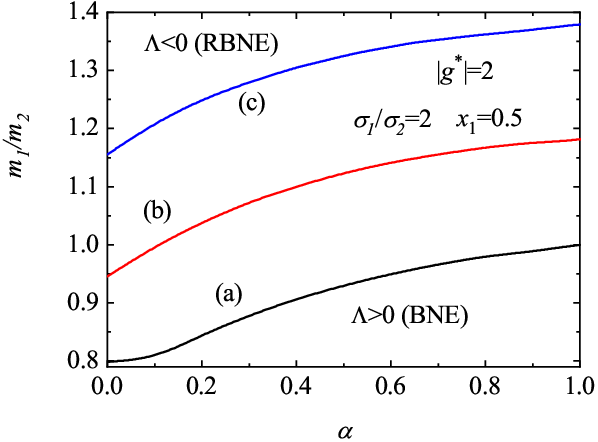}}
%&\resizebox{6.5cm}{!}{\includegraphics{Dpdil.pdf}}
\end{tabular}
\end{center}
\caption{Plot of the dependence of the marginal segregation curve ($\Lambda=0$) on the (common) coefficient of restitution $\al_{ij}\equiv \al$ for a two-dimensional system with $x_1=0.5$, $\sigma_1/\sigma_2=2$, $|g^*|=2$,
and three different values of the solid volume fraction: $\phi=0$ (a); $\phi=0.1$ (b); and $\phi=0.2$ (c).
\label{fig11}}
\end{figure}

\subsection{Moderately dense regime in a confined granular binary mixture}

We now consider granular binary mixtures with an arbitrary concentration at moderate densities. As in Ref.\ \onlinecite{GBS24}, although gravity and the thermal gradient are of the same order of magnitude in our analysis, it is interesting for illustrative purposes to separate the influence of each term in Eq. \eqref{7.11} on segregation. Thus, we first consider cases where gravity is absent ($g=0$ but $\partial_z T\neq 0$) or \textit{thermalized} systems where the effect of gravity on segregation is much more important than that of thermal gradient  ($g\neq 0$ but $\partial_z T\to 0$).

\subsubsection{Absence of gravity ($|g^*|=0$)}

We study here a situation where gravity is absent ($|g^*|=0$) and segregation is induced by the presence of a thermal gradient. In this limiting case, the condition \eqref{7.11} for obtaining the marginal segregation curve ($\Lambda=0$) reduces to
\beq
\label{7.14}
(x_1\xi_1+x_2\xi_2)D_1^{T*}=\left(p^*-\frac{1}{2}\Delta^* \frac{\partial p^*}{\partial \Delta^*}\right)(x_1D_{11}^*+x_2D_{12}^*).
\eeq
A phase diagram \vicente{delineating the regimes between BNE and RBNE} in the $(\sigma_1/\sigma_2; m_1/m_2)$-plane for a two-dimensional system is plotted in Fig.\ \ref{fig9}. The lines are determined from the condition $\Lambda=0$. We have considered mixtures with a concentration
$x_1=0.5$, a (common) coefficient of restitution $\al=0.5$, and three different values of the solid volume fraction: a \emph{dilute} mixture ($\phi=0$) and two mixtures with moderate densities ($\phi=0.1$ and $\phi=0.2$). In general, we observe that density primarily increases weakly the size of the \vicente{RBNE region ($\Lambda<0$)}. This means that larger particles tend to accumulate near the hot plate as the system becomes denser. Additionally, at a given diameter ratio, the region with larger particles attempting to move toward the cold plate \vicente{(i.e., the BNE region where $\Lambda>0$)} appears essentially when larger particles are much heavier than the other species. Comparing these results with those obtained in the conventional IHS model shows qualitative agreement between both models (see for instance, Fig. 5 \ of Ref.\ \onlinecite{G11}).

\subsubsection{Thermalized systems ($\partial_z T\to 0$)}

We consider now a problem where the segregation is only driven by the gravitational force since inhomogeneities in the temperature are neglected. This limiting situation (gravity dominates the temperature gradient) can be achieved in the shaken or sheared systems employed in numerical simulations and physical experiments. \cite{HQL01,BEKR03,SBKR05} When $|g^*|\to \infty$, Eq.\ \eqref{7.11} gives $\Lambda\approx |g^*|(x_1 D_{11}^*+x_2 D_{12}^*)/[(x_1\xi_1+x_2\xi_2)D_1^{T*}]$ and hence, the marginal segregation curve ($\Lambda=0$) is obtained from the condition
\beq
\label{7.15}
x_1 D_{11}^*+x_2 D_{12}^*=0.
\eeq
To illustrate this limiting segregation case ($|g^*|\to \infty$), Fig.\ \ref{fig10} shows the dependence of the marginal segregation curve on the coefficient of restitution $\al_{11}$ for a two-dimensional system. The system has the following parameters: $x_1=0.5$, $\sigma_1/\sigma_2=2$, $\al_{22}=0.9$, $\al_{12}=(\al_{11}+\al_{22})/2$, and three values of the solid volume fraction ($\phi=0$, 0.1 and 0.2). For a given value of the coefficient of restitution $\al_{11}$, we observe that the thermal diffusion factor $\Lambda$ is always negative \vicente{(RBNE effect)} when the larger particles are much heavier than the smaller ones. In this case, the larger particles accumulate near the hot plate. The effect of density on the marginal segregation curve is also apparent, as the region where the thermal diffusion factor becomes positive \vicente{(and larger particles move toward the cold plate, BNE effect)} increases with increasing density.

\subsubsection{General case}

Finally, we consider the general case for finite values of reduced gravity, $|g^*|$. To illustrate this situation, Fig.\ \eqref{fig11} plots the marginal segregation curve ($\Lambda=0$) versus the common coefficient of restitution $\al_{ij}\equiv \al$ for a two-dimensional system with $x_1=0.5$, $\sigma_1/\sigma_2=2$, and three values of the solid volume fraction ($\phi=0$, 0.1 and 0.2). As can be seen, the behavior of the marginal segregation curve with inelasticity is quite similar to that found when $|g^*|\to\infty$. At a given value of $\al$, \vicente{the size of BNE region ($\Lambda>0$)} increases as the density of the system increases.

To the best of our knowledge, there are no computer simulations available in the granular literature for comparison with the theoretical results reported here for the \vicente{BNE/RBNE transition} in the context of the $\Delta$-model for mixtures. We expect the current results to encourage the development of such simulations, which will help assess the reliability of the theoretical results derived for the thermal diffusion factor $\Lambda$.

\section{Discussion}
\label{sec8}

In kinetic theory, it is well known that the most accurate description of multicomponent molecular mixtures is based on the revised Enskog kinetic equation for hard spheres.\cite{BE73a,BE73b,BE73c}
More than 40 years ago,\cite{LCK83} this kinetic equation was solved by applying the Chapman--Enskog method\cite{CC70} to first order in gradients. The Chapman--Enskog method also allowed the identification of the expressions of all the parameters involved in Navier--Stokes hydrodynamics, including the equation of state and transport coefficients.

This previous work\cite{LCK83} was then extended to \textit{granular} mixtures, namely a mixture of hard spheres with inelastic collisions.\cite{GDH07,GHD07} It is important to note that modifying the collisions to account for inelasticity yields significant differences from molecular (elastic) mixtures, but the formal structure of the modified Chapman-Enskog expansion remains the same. As for elastic collisions,\cite{CC70,FK72} several approximations (based on the truncation of a series of Sonine polynomials) are required for practical evaluations.

\vicente{
The objective of the present study is to extend the results derived in Refs.\ \onlinecite{GDH07,GHD07} for moderately dense granular mixtures in the context of the conventional IHS model to the $\Delta$-model.\cite{BRS13}  The $\Delta$-model is a coarse-grained collisional model that
attempts to incorporate collisional energy injection into the dynamics of granular particles in a minimal way, particularly in situations that mimic vertically vibrated and confined systems. As illustrated in Fig.\ \ref{fig0}, in this geometry} the system is confined in a box in which the $z$-direction is slight larger than one particle diameter. The box is vertically vibrated, causing the particles to gain energy through collisions with the walls. This energy gained is then transferred to the horizontal degrees of freedom of grains when collisions between particles take place.

\vicente{Due to the technical difficulties associated with describing this type of system using kinetic theory, the $\Delta$-model is usually considered as the starting point to determine the dynamic properties of the system. In the $\Delta$-model}, when particles collide, part of the gained energy (due to their collisions with the vibrating walls) is released into their horizontal degrees of freedom. The model accounts for this effect by introducing positive factors, denoted by $\Delta_{ij}$, which are added to the relative motion of colliding spheres. The quantities $\Delta_{ij}$ can be related to the intensity of vertical vibrations in experiments.\cite{MGB19a}

In the context of the $\Delta$-model for granular mixtures, two different works\cite{GBS18,GGBS24} have recently analyzed transport. However, both studies considered some special limiting situations. One study\cite{GBS18} used the first-order Chapman--Enskog solution to determine the Navier--Stokes transport coefficients of a \textit{dilute} binary mixture, and the other\cite{GGBS24} considered binary mixtures at moderate densities in the tracer limit (namely, when the concentration of one of the species is negligible). The goal now is to develop a kinetic theory that is valid for moderate densities and arbitrary values of the concentration of each species.

Unlike dilute granular mixtures,\cite{GBS18} the transport coefficients at moderate densities generally have kinetic and collisional contributions. These contributions are expressed in terms of zeroth- and first-order distribution functions, obtained by solving the Enskog equation using the Chapman–Enskog method up to the first order of spatial gradients. Although the exact forms of the zeroth-order distributions, $f_i^{(0)}(\mathbf{V})$, are not yet known, previous work\cite{BSG20} has clearly shown that Maxwellian distributions at the zeroth-order partial temperatures, $T_i^{(0)}$, are a good approximation for them. The first-order distributions, $f_i^{(1)}(\mathbf{V})$, are defined in terms of the quantities $\boldsymbol{\mathcal A}_i(\mathbf{V})$, $\boldsymbol{\mathcal B}_{ij}(\mathbf{V})$, $\mathcal{C}_{ij}(\mathbf{V})$, and $\mathcal{D}_i(\mathbf{V})$, which obey the set of coupled linear integral equations \eqref{4.2}--\eqref{4.5}, respectively. Solving these integral equations provides the kinetic contributions to the transport coefficients. However, since the evaluation of the complete set of Navier-Stokes transport coefficients for the mixture is quite lengthy and cumbersome, this work addresses the determination of the diffusion transport coefficients, as well as the shear and bulk viscosities. We plan to obtain the heat flux coefficients in a subsequent paper.

The constitutive equation of the mass flux is given by Eq.\ \eqref{4.6} where the diffusion transport coefficients $D_i^T$ and $D_{ij}$ are defined by Eqs.\ \eqref{4.8} and \eqref{4.9}, respectively. These coefficients have only kinetic contributions. The  constitutive equation of the pressure tensor is given by Eq.\ \eqref{4.10}. The kinetic contribution to the shear viscosity $\eta$ is provided by Eqs.\ \eqref{4.12} and \eqref{4.13}, while its collisional contribution is given by Eq.\ \eqref{4.18}. As expected, the bulk viscosity $\eta_b$ has only collisional contributions, which are given by Eqs.\ \eqref{4.14.1}--\eqref{4.16}.

To achieve analytical expressions for the transport coefficients, the relevant state of a confined mixture with a stationary temperature has been considered. \vicente{Additionally, although the analytical results have been derived in this paper for a $d$-dimensional system ($d=2$ for hard disks and $d=3$ for hard sphere), the case $d=2$ has been specifically considered. This is because the $\Delta$-model was primarily proposed to reproduce the results obtained in confined, quasi-two-dimensional setups. We expect the main conclusions reported here for $d=2$ to be similar to those for $d=3$.}

The explicit forms for the diffusion coefficients $D_{ij}$ and $D_i^T$ and the kinetic shear viscosity $\eta_k$ have been obtained by taking the leading terms in a Sonine polynomial expansion of the first-order distribution functions. This is the standard procedure for determining these coefficients for elastic\cite{CC70,FK72} and inelastic\cite{G19} mixtures.
In the case of the bulk viscosity $\eta_b$, the contributions coming from the first-order partial temperatures $\varpi_i$ have been neglected. This approximation is based on the results obtained in the IHS model\cite{GGG19b} where the influence of $\varpi_i$ on $\eta_b$ has been in general shown quite small. We expect that this approximation is accurate in the $\Delta$-model as well.
Under the above approximations, for binary mixtures, the diffusion coefficients are given by Eqs.\ \eqref{d3}, \eqref{d5}, and \eqref{d6}, the shear viscosity by Eqs.\ \eqref{5.4} and \eqref{6.8.n1}, and the bulk viscosity by Eq.\ \eqref{5.3}.

To illustrate how the above transport coefficients depend on inelasticity in a binary mixture, for simplicity, we have assumed a common coefficient of restitution ($\al\equiv \al_{11}=\al_{22}=\al_{12}$) and the case $\Delta_{11}=\Delta_{22}=\Delta_{12}$. Figures \ref{fig3}--\ref{fig8} highlight the influence of the coefficient of restitution $\al$  on the mass and momentum transport for mixtures with different densities. As expected from the results derived in the low-density regime, the impact of $\al$ on transport is in general smaller than that of the conventional IHS model.\cite{GDH07,GHD07,G19}  Regarding the influence of density on transport, the effect of the solid volume fraction $\phi$ on diffusion transport coefficients and shear viscosity is also weaker than that found for the IHS model.

We have also analyzed thermal diffusion segregation induced by both gravity and a thermal gradient as an application of the derived kinetic theory. In this situation, the so-called thermal diffusion factor $\Lambda$ [defined by Eq.\ \eqref{7.1}] provides a segregation criterion. An explicit expression of $\Lambda$ in terms of the (dimensionless) diffusion coefficients $D_{11}^*$, $D_{12}^*$, and $D_1^{T*}$, the (reduced) pressure $p^*$ and the (reduced) gravity $g^*$ is given by Eq.\ \eqref{7.10}. Assuming the lower plate of the container is hotter than the top plate, when the thermal diffusion factor is positive ($\Lambda > 0$, BNE effect), the larger particles tend to accumulate near the cold plate against gravity, while the smaller particles sink to the bottom plate. When the thermal diffusion factor is negative ($\Lambda<0$, RBNE effect), the opposite effect occurs. The condition $\Lambda=0$ provides the focus line between the two opposite behaviors. Comparing these results with those obtained in the low-density regime \cite{GBS24,GBS24a} shows that the effect of density on the phase diagrams for the marginal segregation curve ($\Lambda=0$) is generally significant. Specifically, the RBNE region ($\Lambda<0$) is increased as the system becomes more dense in the absence of gravity (see Fig.\ \ref{fig9}), but decreases in the presence of gravity (see Figs.\ \ref{fig10} and \ref{fig11}).

\vicente{Since the results derived in the paper have been obtained from the (inelastic) Enskog kinetic equation, one could speculate about the range of validity for reproducing the results obtained from MD simulations. As for molecular (elastic) mixtures, the Enskog equation provides a semiquantitative description of the hard-sphere system that neglects velocity correlations between the particles that are about to collide (molecular chaos hypothesis). Although the existence of these velocity correlations restricts the range of validity of the (inelastic) Enskog theory, the latter can be still considered as a good approximation (especially at the level of the transport properties) for both moderate inelasticities and/or densities. In particular, the Enskog results have been shown in general to compare quite well with MD simulations\cite{LBD02,MDCPH11,ChS13,MGH14} and even with real experiments\cite{YHCMW02,HYCMW04} for moderately high densities such as $n\sigma^3\lesssim 0.25$ for $d=3$. In the case of the $\Delta$-model, Fig.\ \ref{fig1} highlights the good agreement for the shear viscosity over the entire range of the coefficient of normal restitution, even at relatively high density ($\phi=0.314$). Thus, within the context of the $\Delta$-model, we expect that the range of validity of the present results to be similar to that found in previous works for the conventional IHS model.}

\vicente{It is important to recall that our theory applies to relatively dilute granular materials, in which grains mostly interact through collisions rather than enduring contacts. This means that since grains move quickly, inertial effects become important. To measure the influence of inertia on the dynamic properties of grains, the inertial number $P$ is usually introduced. For a monocomponent granular gas, $P=\dot{\gamma}\sigma/\sqrt{p/\rho_m}$, where $\dot{\gamma}$ is the shear rate and $\rho_m$ is the particle density. Since our results are restricted to the Navier--Stokes domain (where $\dot{\gamma}$ is very small), the transport coefficients are independent of the shear rate  or the inertia number. An interesting problem is extending our results to shearing granular flows to determine the nonlinear dependence of the rheological properties on the shear rate, or equivalently, on the inertia number. This will enable us to compare our results with those of previous studies on polyatomic molecules\cite{GL98,EO01} and with those obtained using micropolar generalized Navier--Stokes equations.\cite{Eringen66}}

As said before, another future challenging problem  within the $\Delta$-model is to determine the heat flux transport coefficients for dense mixtures. Knowing the complete set of Navier-Stokes transport coefficients will enable us to perform a linear stability analysis of the homogeneous steady state, among other applications. In particular, since the homogeneous steady state is stable in the dilute limit \cite{GBS21}, we want to see if density corrections to the transport coefficients can modify the stability of the homogeneous state. Additionally, the reliability of the theoretical results derived here, which were obtained under certain approximations, should be assessed against computer simulations. One possible project is to carry out computer simulations to measure shear viscosity. As for dry granular mixtures, \cite{MG03,GM03,ChG23} we plan to perform simulations on a granular mixture subjected to simple shear flow, where the cooling effects associated with collision dissipation are precisely offset by the presence of the $\Delta$ parameters in the collisional rules. Work along these lines is in progress.

\acknowledgments

We acknowledge financial support from grant no. PID2024-156352NB-I00 funded by MCIU/ AEI/10.13039/501100011033/FEDER, UE and from grant no. GR24022 funded by Junta de Extremadura (Spain) and by European Regional Development Fund (ERDF) ``A way of making Europe''. The research of David Gonz\'alez M\'endez has been supported by the predoctoral fellowship FPU24/01056 from the Spanish Government.

\vspace{0.5cm}

\noindent \textbf{AUTHOR DECLARATIONS}\\

\noindent \textbf{Conflict of Interest}\\

The authors have no conflicts to disclose.

\vspace{0.5cm}

\noindent \textbf{DATA AVAILABILITY}\\

The data that support the findings of this study are available from the corresponding author upon reasonable request.

\appendix

\section{First-order approximation}
\label{appA}

\begin{widetext}

Given that the application of the Chapman--Enskog method \cite{CC70} to granular mixtures has been carried out in several previous papers (see for instance, Ref.\ \onlinecite{G19}), we provide in this Appendix some intermediate steps in the $\Delta$-model to achieve the final linear integral equations defining the Navier--Stokes transport coefficients. To first order, the Enskog kinetic equation for the one-particle distribution function $f_i^{(1)}$ of species $i$ is
\beq
\label{a1}
\partial_t^{(0)}f_i^{(1)}-\sum_{j=1}^s\;J_{ij}^{(1)}[f_i,f_j]=-\left(\text{D}_t^{(1)}+\mathbf{V}\cdot \nabla+\mathbf{g}\cdot \frac{\partial}{\partial \mathbf{v}} \right)f_i^{(0)},
\eeq
where $\text{D}_t^{(1)}=\partial_t^{(1)}+\mathbf{U}\cdot \nabla$ and the first order contribution $\sum_{j=1}^s\; J_{ij}^{(1)}[f_i,f_j]$ to the Enskog collision operator is given by \cite{G19}
\begin{eqnarray}
\label{a2}
\sum_{j=1}^s\; J_{ij}^{(1)}[f_i,f_j]&\to& -\sum_{j=1}^s\;\sum_{\ell=1}^s \left\{\boldsymbol{\mathcal{K}}_{i\ell}\left[n_j
\frac{\partial f_\ell ^{(0)}}{\partial n_j}\right]+\frac{1}{2}\left(n_j \frac{\partial \ln \chi_{i\ell}}{\partial n_j}+I_{i\ell j}\right)\boldsymbol{\mathcal{K}}_{i\ell}\left[f_\ell^{(0)}\right] \right\}\cdot \nabla \ln n_j \nonumber\\
& &
-\sum_{j=1}^s\boldsymbol{\mathcal{K}}_{ij}\left[T\frac{\partial f_j^{(0)}}{\partial T}\right]\cdot \nabla \ln T+\frac{1}{2} \sum_{j=1}^s\mathcal{K}_{ij,\lambda}
\left[\frac{\partial f_j^{(0)}}{\partial V_\beta}\right]\left(\partial_\lambda U_\beta+\partial_\beta U_\lambda-\frac{2}{d}\delta_{\lambda\beta}\nabla \cdot \mathbf{U}\right)\nonumber\\
& &  +\frac{1}{d} \sum_{j=1}^s\mathcal{K}_{ij,\beta}\left[\frac{\partial f_j^{(0)}}{\partial V_\beta}\right]\nabla \cdot \mathbf{U}-\left({\cal L}f^{(1)}\right)_i.
\end{eqnarray}
The linear operator $\left({\cal L}f^{(1)}\right)_i$ is defined by
\beq
\label{a3}
\left({\cal L}X\right)_i=-\sum_{j=1}^s\left(J_{ij}^{(0)}[X_i,f_j^{(0)}]+
J_{ij}^{(0)}[f_i^{(0)},X_j]\right),
\eeq
where $J_{ij}^{(0)}$ is given by Eq.\ \eqref{3.12} and the operator $\boldsymbol{\mathcal{K}}_{ij}[X_j]$ for a multicomponent mixture in the $\Delta$-model is
\beqa
\label{a4}
\boldsymbol{\mathcal{K}}_{ij}\Big[X\Big]&=&-\sigma_{ij}^{d}\chi_{ij} \int \mathrm{d}{\bf v}_{2}\int \mathrm{d}\widehat{\boldsymbol{\sigma}}
\Theta (-\widehat{{\boldsymbol {\sigma }}}\cdot {\bf g}_{12}-2\Delta_{ij})(-\widehat{{\boldsymbol {\sigma }}}\cdot {\bf g}_{12}-2\Delta_{ij})
\widehat{\boldsymbol{\sigma}}\al_{ij}^{-2}f_i^{(0)}(\mathbf{V}_1'')X(\mathbf{V}_2'')\nonumber\\
& &
+\sigma_{ij}^{d}\chi_{ij} \int \mathrm{d}{\bf v}_{2}\int \mathrm{d}\widehat{\boldsymbol{\sigma}}
\Theta (\widehat{{\boldsymbol {\sigma}}}\cdot {\bf g}_{12})(\widehat{{\boldsymbol {\sigma}}}\cdot {\bf g}_{12})
\widehat{\boldsymbol{\sigma}}f_i^{(0)}(\mathbf{V}_1) X(\mathbf{V}_2).
\eeqa
In addition, the quantities $I_{i\ell j}$ are defined in terms of the functional derivative of the (local) pair distribution function $\chi_{ij}$ with respect to the (local) partial densities $n_\ell$. These quantities are the origin of the primary difference between the so-called Standard Enskog Theory and the Revised Enskog Theory for elastic collisions. \cite{BE73a,BE73b,BE73c,BE79} Given the mathematical difficulties involved in the determination of the above functional derivatives, these parameters are chosen here to recover the results derived for molecular fluid mixtures. Their explicit forms will be displayed in the Appendix \ref{appD} for a binary mixture.

The balance equations at this order provide the actions of the operators $\text{D}_t^{(1)}$ over the hydrodynamic fields as
\beq
\label{a5}
\text{D}_t^{(1)}n_i=-n_i \nabla \cdot \mathbf{U}, \quad \text{D}_t^{(1)} \mathbf{U}=-\rho^{-1}\nabla p+\mathbf{g},
\eeq
\beq
\label{a6}
\frac{d}{2}n\text{D}_t^{(1)}T=-p \nabla \cdot \mathbf{U}-\frac{d}{2}n T \zeta_U \nabla \cdot \mathbf{U},
\eeq
As usual, upon writing Eq.\ \eqref{a6} we have taken into account that since the cooling rate is a scalar, corrections to first order in gradients can only arise from the divergence of the flow velocity: $\zeta^{(1)}=\zeta_U \nabla \cdot \mathbf{U}$. The right-hand side of Eq.\ \eqref{a1} can be more explicitly evaluated by taking into account Eqs.\ \eqref{a5} and \eqref{a6}. The result is
\beqa
\label{a7}
& & \left(\text{D}_t^{(1)}+\mathbf{V}\cdot \nabla+\mathbf{g}\cdot \frac{\partial}{\partial \mathbf{v}} \right)f_i^{(0)}=\sum_{j=1}^s
\Bigg(\mathbf{V} n_j \frac{\partial f_i^{(0)}}{\partial n_j}+\frac{n_j}{\rho}
\frac{\partial p}{\partial n_j}
\frac{\partial f_i^{(0)}}{\partial \mathbf{V}}\Bigg)\cdot \nabla \ln n_j \nonumber\\
& & +\Bigg[\frac{p}{\rho}\Big(1-\frac{1}{2}\sum_{\ell = 1}^s\sum_{j=1}^s\Delta_{\ell j}^*\frac{\partial \ln p^*}{\partial \Delta_{\ell j}^*}\Big)\frac{\partial f_i^{(0)}}{\partial \mathbf{V}}+T \frac{\partial f_i^{(0)}}{\partial T}\mathbf{V}\Bigg]\cdot \nabla \ln T
-V_\lambda \frac{\partial f_i^{(0)}}{\partial V_\beta}\frac{1}{2}\left(\partial_\beta U_\lambda+\partial_\lambda U_\beta-\frac{2}{d}\delta_{\lambda\beta}\nabla \cdot \mathbf{U}\right)\nonumber\\
& &
-\Bigg[\frac{1}{d}\mathbf{V}\cdot \frac{\partial f_i^{(0)}}{\partial \mathbf{V}}
+\left(\zeta_U+\frac{2}{d}\frac{p}{nT}\right)T\frac{\partial f_i^{(0)}}{\partial T}
+\sum_{j=1}^s n_j \frac{\partial f_i^{(0)}}{\partial n_j}\Bigg]\nabla
\cdot \mathbf{U}.
\eeqa

With this result, the kinetic equation for the first-order distribution function $f_i^{(1)}(\mathbf{V})$ can be written as
\beq
\label{a8}
\partial_t^{(0)}f_i^{(1)}+\left({\cal L}f^{(1)}\right)_i=\mathbf{A}_i\cdot \nabla \ln T+
\sum_{j=1}^s\mathbf{B}_{ij}\cdot \nabla \ln n_j+C_{i,\lambda \beta}\frac{1}{2}\left(\partial_\beta U_\lambda+\partial_\lambda U_\beta-\frac{2}{d}\delta_{\lambda\beta}\nabla \cdot \mathbf{U}\right)+
D_i\nabla \cdot \mathbf{U},
\eeq
where the coefficients of the field gradients on the right hand side are given by
\begin{equation}
\label{a9}
\mathbf{A}_i\left(\mathbf{V}\right)=-T \frac{\partial f_i^{(0)}}{\partial T}\mathbf{V}-\frac{p}{\rho}\left(1-\frac{1}{2}\sum_{\ell = 1}^s\sum_{j=1}^s\Delta_{\ell j}^*\frac{\partial \ln p^*}{\partial \Delta_{\ell j}^*}\right)
\frac{\partial f_i^{(0)}}{\partial \mathbf{V}}
-\sum_{j=1}^s\boldsymbol{\mathcal{K}}_{ij}\Bigg[T \frac{\partial f_j^{(0)}}{\partial T}\Bigg],
\end{equation}
\beq
\label{a11}
\mathbf{B}_{ij}\left(\mathbf{V}\right)= -\mathbf{V} n_j \frac{\partial f_i^{(0)}}{\partial n_j}-\frac{n_j}{\rho}
\frac{\partial p}{\partial n_j}
\frac{\partial f_i^{(0)}}{\partial \mathbf{V}}-\sum_{\ell=1}^s \left\{\boldsymbol{\mathcal{K}}_{i\ell}\left[n_j
\frac{\partial f_\ell ^{(0)}}{\partial n_j}\right]+\frac{1}{2}\left(n_j \frac{\partial \ln \chi_{i\ell}}{\partial n_j}+I_{i\ell j}\right)\boldsymbol{\mathcal{K}}_{i\ell}\left[f_\ell^{(0)}\right] \right\},
\eeq
\begin{equation}
\label{a12}
C_{i,\lambda\beta}\left(\mathbf{V}\right)=V_\lambda\frac{\partial f_i^{(0)}}{\partial V_\beta}+\sum_{j=1}^s\mathcal{K}_{ij,\lambda}\left[\frac{\partial f_j^{(0)}}{\partial V_\beta}
\right],
\end{equation}
\begin{equation}
\label{a13}
D_i\left(\mathbf{V}\right)=\frac{1}{d}\mathbf{V}\cdot \frac{\partial f_i^{(0)}}{\partial \mathbf{V}}
+\left(\zeta_U+\frac{2}{d}\frac{p}{nT}\right)T\frac{\partial f_i^{(0)}}{\partial T}+\sum_{j=1}^s \left\{n_j\frac{\partial f_i^{(0)}}{\partial n_j}+
\frac{1}{d}\mathcal{K}_{ij,\beta}\left[\frac{\partial f_j^{(0)}}{\partial V_\beta}\right]\right\}.
\end{equation}

The first-order contribution to the cooling rate $\zeta_U$ can be written as
\beq
\label{a14}
\zeta_U=\zeta^{(1,0)}+\zeta^{(1,1)},
\eeq
where
\beqa
\label{a15}
\zeta^{(1,0)}&=&\frac{4\pi^{d/2}}{d^2\Gamma\left(\frac{d}{2}\right)}\sum_{i,j}\chi_{ij}\frac{m_{ij}}{\overline{m}}x_ix_j n\sigma_{ij}^d\Delta_{ij}^{*2}+\frac{4\pi^{(d-1)/2}}{d^2\Gamma\left(\frac{d+1}{2}\right)}\frac{v_{\text{th}}}{nT}\sum_{i,j}\chi_{ij}
m_{ij}\sigma_{ij}^d \al_{ij}\Delta_{ij}^*\int \mathrm{d}\mathbf{v}_1 \int \mathrm{d}\mathbf{v}_2\; g_{12} f_i^{(0)}(\mathbf{V}_1)f_j^{(0)}(\mathbf{V}_2)\nonumber\\
& & -\frac{3\pi^{d/2}}{d^2\Gamma\left(\frac{d}{2}\right)}\sum_{i,j}\chi_{ij}\mu_{ji}x_i x_j n\sigma_{ij}^d\gamma_i (1-\al_{ij}^2),
\eeqa
\beq
\label{a16}
\zeta^{(1,1)}=-\frac{4\pi^{(d-1)/2}}{d n T}\sum_{i,j}\chi_{ij}m_{ij}\sigma_{ij}^{d-1}\int \mathrm{d}\mathbf{v}_1 \int \mathrm{d}\mathbf{v}_2\;
f_i^{(0)}(\mathbf{V}_1)\mathcal{D}_j(\mathbf{V}_2)\Bigg[\frac{\Delta_{ij}^2}{\Gamma\left(\frac{d+1}{2}\right)}g_{12}+\frac{\sqrt{\pi}}
{d\Gamma\left(\frac{d}{2}\right)}\al_{ij}\Delta_{ij}g_{12}^2-
\frac{1-\al_{ij}^2}{4\Gamma\left(\frac{d+3}{2}\right)}g_{12}^3\Bigg].
\eeq
In Eq.\ \eqref{a15} we recall that $\Delta_{ij}^*=\Delta_{ij}/v_\text{th}$. The second term in the right-hand side of Eq.\ \eqref{a15} can be explicitly evaluated when one takes the Maxwellian approximation \eqref{3.20} for the distributions $ f_i^{(0)}(\mathbf{V})$ and $ f_j^{(0)}(\mathbf{V})$. In this approximation, one gets the result
\beq
\label{a17}
\int \mathrm{d}\mathbf{v}_1 \int \mathrm{d}\mathbf{v}_2\; g_{12} f_i^{(0)}(\mathbf{V}_1)f_j^{(0)}(\mathbf{V}_2)=\frac{\Gamma\left(\frac{d+1}{2}\right)}{\Gamma\left(\frac{d}{2}\right)}n_in_j \left(\frac{\theta_i+\theta_j}{\theta_i\theta_j}\right)^{1/2}v_{\text{th}}.
\eeq
Thus, the expression of $\zeta^{(1,0)}$ can be more explicitly written when one takes into account the result
\eqref{a17}:
\beq
\label{a18}
\zeta^{(1,0)}=\frac{4 \pi^{d/2}}{d^2\Gamma\left(\frac{d}{2}\right)}\sum_{i,j}\chi_{ij}x_i x_j n \sigma_{ij}^{d}\Bigg\{\frac{m_{ij}}{\overline{m}}\Delta_{ij}^*\Bigg[\Delta_{ij}^*+\frac{2}{\sqrt{\pi}}\al_{ij}
\left(\frac{\theta_i+\theta_j}{\theta_i\theta_j}\right)^{1/2}\Bigg]-\frac{3}{4}\mu_{ji}\gamma_i(1-\al_{ij}^2)\Bigg\}.
\eeq

The solution to Eq.\ \eqref{a8} is
\beq
\label{a19}
f_i^{(1)}(\mathbf{V})=\boldsymbol{\mathcal{A}}_i\left(\mathbf{V}\right)\cdot  \nabla \ln
T+\sum_{j=1}^s\boldsymbol{\mathcal{B}}_{ij}\left(\mathbf{V}\right) \cdot \nabla \ln n_j
+\mathcal{C}_{i,\lambda \beta}(\mathbf{V})\frac{1}{2}\left(\partial_\beta U_\lambda+\partial_\lambda U_\beta-\frac{2}{d}\delta_{\lambda\beta}\nabla \cdot \mathbf{U}\right)+\mathcal{D}_i
\left(\mathbf{V}\right) \nabla \cdot \mathbf{U},
\eeq
where the integral equations obeying the unknown functions $\boldsymbol{\mathcal{A}}_i$, $\boldsymbol{\mathcal{B}}_{ij}$, $\mathcal{C}_{i,\lambda \beta}$, and $\mathcal{D}_i$ can be obtained by substituting the expression \eqref{a19} of $f_i^{(1)}(\mathbf{V})$ into Eq.\ \eqref{a8} and identifying the coefficients of the independent gradients. To obtain them, one has to take into account the contributions from the operator $\partial_t^{(0)}$ acting on the temperature gradient. These contributions are given by
\beqa
\label{a20}
\partial_t^{(0)}\nabla \ln T&=&-\nabla \zeta^{(0)}=-T\frac{\partial \zeta^{(0)}}{\partial T}\nabla \ln T-\sum_{j=1}^s\; n_j \frac{\partial \zeta^{(0)}}{\partial n_j}\nabla \ln n_j\nonumber\\
&=&-\frac{1}{2}\zeta^{(0)}\Bigg(1-\sum_{\ell = 1}^s\sum_{j=1}^s\;\Delta_{\ell j}^*\frac{\partial \ln \zeta_0^*}{\partial \Delta_{\ell j}^*}\Bigg)\nabla \ln T-\sum_{j=1}^s\; n_j \frac{\partial \zeta^{(0)}}{\partial n_j}\nabla \ln n_j,
\eeqa
where we recall that $\zeta_0^*=\zeta^{(0)}/\nu$. The corresponding integral equations for the unknowns can be finally achieved when one identifies the coefficients of the independent gradients. These integral equations are given by Eqs.\ \eqref{4.2}--\eqref{4.5}, where
\begin{equation}
\label{a21}
D_i'\left(\mathbf{V}\right)=\frac{1}{d}\mathbf{V}\cdot \frac{\partial f_i^{(0)}}{\partial \mathbf{V}}
+\left(\zeta^{(1,0)}+\frac{2}{d}\frac{p}{nT}\right)T\frac{\partial f_i^{(0)}}{\partial T}+\sum_{j=1}^s n_j \left\{\frac{\partial f_i^{(0)}}{\partial n_j}+
\frac{1}{d}\mathcal{K}_{ij,\beta}\left[\frac{\partial f_j^{(0)}}{\partial V_\beta}\right]\right\}.
\end{equation}

\section{Collisional contributions to the pressure tensor}
\label{appB}

The first-order contributions to the collisional pressure tensor $P_{\lambda\beta}^c$ are obtained in this Appendix. To first order in spatial gradients, $P_{\lambda\beta}^c$ can be split as
\beq
\label{b1}
P_{\lambda\beta}^{c}=P_{\lambda\beta}^{(\Delta=0)c}+P_{\lambda\beta}^{(\Delta\neq 0)c},
\eeq
where $P_{\lambda\beta}^{(\Delta=0)c}$ denotes the contributions to $P_{\lambda\beta}^c$ obtained in the conventional IHS (i.e., when $\Delta_{ij}=0$) while $P_{\lambda\beta}^{(\Delta\neq 0)c}$ denotes the contributions involving terms proportional to $\Delta_{ij}$. Given that $P_{\lambda\beta}^{(\Delta=0)c}$ was obtained in previous works, \cite{GDH07,GHD07,GGG19b} we focus here on the evaluation of $P_{\lambda\beta}^{(\Delta\neq 0)c}$.

To first order in spatial gradients, $P_{\lambda\beta}^{(\Delta\neq 0)c}$ is given by
\beqa
\label{b2}
P_{\lambda\beta}^{(\Delta\neq 0)c}&=&2\sum_{i,j}\chi_{ij} m_{ij} \sigma_{ij}^d \Delta_{ij}\int
d \mathbf{v}_1\int \mathrm{d} \mathbf{v}_2
\int \mathrm{d}\widehat{\boldsymbol {\sigma}}\,\Theta (\widehat{{\boldsymbol {\sigma}}}
\cdot \mathbf{g}_{12})(\widehat{\boldsymbol {\sigma }}\cdot {\bf g}_{12})\widehat{\sigma}_\lambda \widehat{\sigma}_\beta\nonumber\\
& &\times \Bigg[f_i^{(1)}(\mathbf{V}_1)f_j^{(0)}(\mathbf{V}_2)
+\frac{1}{2}f_i^{(0)}(\mathbf{V}_1)\frac{\partial f_j^{(0)}(\mathbf{V}_2)}{\partial \mathbf{r}}\cdot {\boldsymbol {\sigma}}_{ij}\Bigg]\nonumber\\
&=&\frac{2B_1}{d+1} \sum_{i,j}\chi_{ij} m_{ij} \sigma_{ij}^d \Delta_{ij}
\int \mathrm{d} \mathbf{v}_1\int \mathrm{d} \mathbf{v}_2
g_{12}^{-1}\left(g_{12,\lambda} g_{12,\beta}+g_{12}^2 \delta_{\lambda\beta}\right)f_i^{(1)}(\mathbf{V}_1)f_j^{(0)}(\mathbf{V}_2)\nonumber\\
& & -\frac{B_2}{d+2} \sum_{i,j}\chi_{ij} m_{ij} \sigma_{ij}^{d+1} \Delta_{ij}\partial_\nu U_\mu
\int \mathrm{d} \mathbf{v}_1\int \mathrm{d}\mathbf{v}_2 \left(g_{12,\nu} \delta_{\lambda\beta}+g_{12,\lambda} \delta_{\nu\beta}+g_{12,\beta} \delta_{\lambda\nu}\right)
f_i^{(0)}(\mathbf{V}_1)\frac{\partial f_j^{(0)}(\mathbf{V}_2)}{\partial V_{2\mu}}\nonumber\\
&=&\frac{2B_1}{d+1} \sum_{i,j}\chi_{ij} m_{ij} \sigma_{ij}^d \Delta_{ij}\int \mathrm{d} \mathbf{v}_1\int \mathrm{d} \mathbf{v}_2
g_{12}^{-1}\left(g_{12,\lambda} g_{12,\beta}+g_{12}^2 \delta_{\lambda\beta}\right)\nonumber\\
& & \times \Bigg[\mathcal{C}_{i,\mu \nu}(\mathbf{V}_1)\frac{1}{2}\left(\partial_\mu U_\nu+\partial_\nu U_\mu-\frac{2}{d}\delta_{\mu\nu}\nabla \cdot \mathbf{U}\right)+\mathcal{D}_i
\left(\mathbf{V}_1\right) \nabla \cdot \mathbf{U}\Bigg]f_j^{(0)}(\mathbf{V}_2)\nonumber\\
& & -\frac{B_2}{d+2} \sum_{i,j}\chi_{ij} m_{ij} n_i n_j \sigma_{ij}^{d+1} \Delta_{ij}\Bigg[\left(\partial_\beta U_\lambda+\partial_\lambda U_\beta-\frac{2}{d}\delta_{\lambda\beta}\nabla \cdot \mathbf{U}\right)+\frac{d+2}{d}\delta_{\lambda\beta}\nabla \cdot \mathbf{U}\Bigg],
\nonumber\\
\eeqa
where the coefficients $B_k$ have been defined in Eq.\ \eqref{3.15}. In addition, upon deriving Eq.\ \eqref{b2}, we have taken into account that by symmetry reasons the only contributions to $P_{\lambda\beta}^{(\Delta\neq 0)c}$ coming from $f_i^{(1)}$ only involve the unknowns $\mathcal{C}_{i,\mu\nu}$ and $\mathcal{D}_{i}$.
To compute the first term on the right hand side of Eq.\ \eqref{b2}, we take the leading Sonine approximations to $\boldsymbol{\mathcal{C}}_i$ and $\mathcal{D}_i$. In the case of $\boldsymbol{\mathcal{C}}_i$ is given by Eq. \eqref{5.16}   while in the case of $\mathcal{D}_i$ is
\beq
\label{b3}
\mathcal{D}_i(\mathbf{V})\to f_{i,\text{M}}(\mathbf{V}) W_i(\mathbf{V}) \frac{\varpi_i}{T_i^{(0)}}, \quad
W_i(\mathbf{V})=\frac{m_i V^2}{2T_i^{(0)}}-\frac{d}{2},
\eeq
where $T_i^{(1)}=\varpi_i \nabla \cdot \mathbf{U}$. Thus, taking into account the leading Sonine approximations to $\boldsymbol{\mathcal{C}}_i$ and $\mathcal{D}_i$, after some algebra one gets the expression
\beqa
\label{b4}
P_{\lambda\beta}^{(\Delta\neq 0)c}&=&-\frac{8\pi^{(d-1)/2}}{(d+1)\Gamma\left(\frac{d+1}{2}\right)} \sum_{i,j}\chi_{ij} \frac{m_{ij}m_i}{\overline{m}^2}\frac{n_j \sigma_{ij}^d}{\gamma_i^2} \left(\frac{\theta_j}{\theta_i+\theta_j}\right)^2\Delta_{ij}^*
\Big[\int \mathrm{d}\mathbf{c}_1\int \mathrm{d}\mathbf{c}_2
g_{12}^{*-1}g_{12,x}^{*2}g_{12,y}^{*2}\varphi_{i,\text{M}}(\mathbf{c}_1)\varphi_{j,\text{M}}(\mathbf{c}_2)\Big]\nonumber\\
& & \times \eta_i^k\left(\partial_\beta U_\lambda+\partial_\lambda U_\beta-\frac{2}{d}\delta_{\lambda\beta}\nabla \cdot \mathbf{U}\right)-\frac{\pi^{d/2}}{d(d+2)\Gamma\left(\frac{d}{2}\right)}  \sum_{i,j}\chi_{ij}n_in_j \sigma_{ij}^{d+1}m_{ij}v_\text{th}\Delta_{ij}^*\nonumber\\
& & \times \Bigg[\left(\partial_\beta U_\lambda+\partial_\lambda U_\beta-\frac{2}{d}\delta_{\lambda\beta}\nabla \cdot \mathbf{U}\right)+\frac{d+2}{d}\delta_{\lambda\beta}\nabla \cdot \mathbf{U}\Bigg]\nonumber\\
& & + \frac{4\pi^{(d-1)/2}}{d\Gamma\left(\frac{d}{2}\right)} \sum_{i,j}\chi_{ij} \mu_{ji} \theta_i n_i n_j \sigma_{ij}^d \Delta_{ij}^* \varpi_i \left[ \int \dd \mathbf{c}_1 \int \dd \mathbf{c}_2 \, g_{12}^* \left( \theta_i c_1^2 - \frac{d}{2} \right) \varphi_{i,M}(\textbf{c}_1)\varphi_{j}(\textbf{c}_2)  \right] \delta_{\lambda\beta} \nabla \cdot \mathbf{U},
% +\frac{2\pi^{(d-1)/2}}{d\Gamma\left(\frac{d}{2}\right)} \sum_{i,j}\chi_{ij}\mu_{ji}n_i n_j \sigma_{ij}^d \Delta_{ij}^*
% \left(\frac{\theta_i\theta_j}{\theta_i+\theta_j}\right)^{1/2}\delta_{\lambda\beta}\varpi_i \nabla \cdot \mathbf{U},
\eeqa
where we recall that $\mathbf{g}_{12}^*=\mathbf{g}_{12}/v_\text{th}$, $\mathbf{c}_{i}=\mathbf{v}_{i}/v_\text{th}$, and we have replaced the scaled distribution $\varphi_j(\mathbf{c}_2)$ by its Maxwellian form $\varphi_{j,\text{M}}(\mathbf{c}_2)$ in the first identity of Eq.\ \eqref{b4} for the sake of simplicity.

According to Eq.\ \eqref{b4}, the collisional transfer contributions to $\eta$ and $\eta_b$ involving terms proportional to $\Delta_{ij}$ are given by
\beqa
\label{b5}
\eta_c^{(\Delta\neq 0)}&=&\frac{8\pi^{(d-1)/2}}{(d+1)\Gamma\left(\frac{d+1}{2}\right)} \sum_{i,j}\chi_{ij} \frac{m_{ij}m_i}{\overline{m}^2}\frac{n_j \sigma_{ij}^d}{\gamma_i^2} \left(\frac{\theta_j}{\theta_i+\theta_j}\right)^2\Delta_{ij}^*\Big[\int \mathrm{d}\mathbf{c}_1\int \mathrm{d}\mathbf{c}_2
g_{12}^{*-1}g_{12,x}^{*2}g_{12,y}^{*2}\varphi_{i,\text{M}}(\mathbf{c}_1)\varphi_{j,\text{M}}(\mathbf{c}_2)\Big]\eta_i^k
\nonumber\\
& &
+\frac{d}{d+2}\eta_b^{(\Delta\neq 0)(\text{I})},
\eeqa
\beq
\label{b6}
\eta_b^{(\Delta\neq 0)}=\eta_b^{(\Delta\neq 0)(\text{I})}+\eta_b^{(\Delta\neq 0)(\text{II})},
\eeq
where
\beq
\label{b7}
\eta_b^{(\Delta\neq 0)(\text{I})}=\frac{\pi^{d/2}}{d^2\Gamma\left(\frac{d}{2}\right)}\sum_{i,j}\chi_{ij}n_in_j \sigma_{ij}^{d+1}m_{ij}v_\text{th}\Delta_{ij}^*,
\eeq
\beq
\label{b8}
\eta_b^{(\Delta \neq 0)(\text{II})} = -\frac{4\pi^{(d-1)/2}}{d\Gamma\left(\frac{d}{2}\right)} \sum_{i,j}\chi_{ij} \mu_{ji} \theta_i n_i n_j \sigma_{ij}^d \Delta_{ij}^*  \left[ \int \dd \mathbf{c}_1 \int \dd \mathbf{c}_2 \, g_{12}^* \left( \theta_i c_1^2 - \frac{d}{2} \right) \varphi_{i,M}(\textbf{c}_1)\varphi_{j}(\textbf{c}_2)  \right] \varpi_i
%\eta_b^{(\Delta\neq 0)''}=-\frac{2\pi^{(d-1)/2}}{d\Gamma\left(\frac{d}{2}\right)} \sum_{i,j}\chi_{ij}\mu_{ji}n_i n_j \sigma_{ij}^d \Delta_{ij}^*
%\left(\frac{\theta_i\theta_j}{\theta_i+\theta_j}\right)^{1/2}\varpi_i.
\eeq
The expressions \eqref{4.14.1}--\eqref{4.16} for $\eta_b$ and \eqref{4.18} for $\eta_c$ can be obtained from Eqs.\ \eqref{b5}--\eqref{b8} and those corresponding to the contributions to $P_{\lambda\beta}^{(\Delta=0)c}$.

\section{Expressions of $\tau_{ii}$ and $\tau_{ij}$}
\label{appC}

Although the expressions of the collision frequencies $\tau_{ii}$ and $\tau_{ij}$ were displayed for dilute granular mixtures, for the sake of completeness, it is convenient to provide them for moderate densities. To estimate them, as usual we make the replacement $f_i^{(0)}\to f_{i,\text{M}}$ in Eqs.\ \eqref{5.18} and \eqref{5.19}. In this approximation, $\tau_{ii}$ and $\tau_{ij}$ can be written as
\beq
\label{c1}
\tau_{ii}=\tau_{ii}^{(0)}+\tau_{ii}^{(1)}, \quad \tau_{ij}=\tau_{ij}^{(0)}+\tau_{ij}^{(1)},
\eeq
where
\begin{eqnarray}
 \label{c2}
\tau_{ii}^{(0)} &=&\frac{2\pi ^{(d-1)/2}}{d(d+2)\Gamma \left( \frac{d}{2}\right) } \upsilon_{\text{th}}\Bigg\{ n_{i}\sigma
_{i}^{d-1}\chi_{ii}(2\theta_{i})^{-1/2}(3+2d-3\alpha_{ii})(1+\alpha_{ii})+2\sum_{j\neq i}^s
n_{j}\sigma_{ij}^{d-1}\chi_{ij}\mu_{ji}(1+\alpha_{ij})\theta_{i}^{3/2}
\theta_{j}^{-1/2}\nonumber\\
& & \times \left[(d+3)(\mu_{ij}\theta_j-\mu_{ji}\theta_i)\theta_{i}^{-2}(\theta_{i}+\theta_{j})^{-1/2}+\frac{3+2d-3\alpha_{ij}}{2}\mu_{ji}
\theta_{i}^{-2}(\theta_{i}+\theta_{j})^{1/2}\right.\nonumber\\
& &\left.+\frac{2d(d+1)-4}{2(d-1)}\theta
_{i}^{-1}(\theta_{i}+\theta_{j})^{-1/2}\right] \Bigg\},
\end{eqnarray}
\beqa
\label{c3}
\tau_{ii}^{(1)}&=&\frac{\sqrt{2}\pi^{(d-1)/2}}{d(d+2)\Gamma\left(\frac{d}{2}\right)}\upsilon_{\text{th}}
n_i\sigma_i^{d-1}\chi_{ii}\Delta_{ii}^*
\left[\sqrt{2\pi}(d-2\al_{ii})-2\theta_i^{-1/2}\Delta_{ii}^*\right]
\nonumber\\
& &
-\frac{8\pi^{(d-1)/2}}{d(d+2)\Gamma\left(\frac{d}{2}\right)}\upsilon_{\text{th}}\sum_{j\neq i}^s n_j\sigma_{ij}^{d-1}\chi_{ij}
\mu_{ji}^2\theta_i^2 \Delta_{ij}^* \Big[\sqrt{\pi}\theta_j^{-2}(1+\al_{ij})+\left(\theta_i+\theta_j\right)^{-1/2}\theta_i^{1/2}\theta_j^{-3/2}\Delta_{ij}^*\Big],
\eeqa
\begin{eqnarray}
 \label{c4}
\tau_{ij}^{(0)}&=&\frac{4\pi ^{(d-1)/2}}{d(d+2)\Gamma \left( \frac{d}{2}\right) } \upsilon_{\text{th}} n_{i}
\sigma_{ij}^{d-1}\chi_{ij}\mu_{ij}\theta_{j}^{3/2}\theta_{i}^{-1/2}(1+\alpha_{ij})\Big[ (d+3)(\mu_{ij}\theta_j-\mu_{ji}\theta_i)\theta_{j}^{-2}(\theta_{i}
+\theta_{j})^{-1/2}\nonumber\\
& & +\frac{3+2d-3\alpha_{ij}}{2}\mu_{ji}\theta_{j}^{-2}(\theta_{i}+\theta_{j})^{1/2}
-\frac{2d(d+1)-4}{2(d-1)}\theta_{j}^{-1}(\theta_{i}+\theta
_{j})^{-1/2}\Big],
\end{eqnarray}
\beq
\label{c5}
\tau_{ij}^{(1)}=-\frac{4\pi^{(d-1)/2}}{d(d+2)\Gamma\left(\frac{d}{2}\right)}\upsilon_{\text{th}}n_i\sigma_{ij}^{d-1}\chi_{ij}
\mu_{ij}\theta_j^2 \Delta_{ij}^* \Big[2\sqrt{\pi}\mu_{ji}\theta_i^{-2}(1+\al_{ij})-(d+2)\sqrt{\pi}\theta_i^{-2}
+2\mu_{ji}\left(\theta_i+\theta_j\right)^{-1/2}\theta_j^{1/2}\theta_i^{-3/2}\Delta_{ij}^*\Big].
\eeq
In the case of mechanically equivalent particles, Eqs.\ \eqref{c2}--\eqref{c5} are consistent with previous results obtained from the $\Delta$-model for dilute monocomponent granular gases. \cite{BBMG15}

\section{Expressions of the diffusion coefficients for a granular binary mixture}
\label{appD}

In this Appendix, we provide the expressions of the diffusion transport coefficients for a granular binary mixture ($s=2$) in steady state conditions. In this case, since $\mathbf{j}_1^{(1)}=-\mathbf{j}_2^{(1)}$, the diffusion coefficients obey the relations
\beq
\label{d1}
D_{21}=-\frac{m_1}{m_2}D_{11}, \quad D_{22}=-\frac{m_1}{m_2}D_{12}, \quad D_2^T=-D_1^T.
\eeq
For the sake of simplicity, we consider the case where $\Delta_{11}=\Delta_{22}=\Delta_{12}$. We introduce the dimensionless coefficients
\beq
\label{d2}
D_{ij}^*=\frac{m_i m_j \nu}{\rho T}D_{ij}, \quad D_{i}^{*T}=\frac{\rho \nu}{n T}D_{i}{^T}.
\eeq

The dimensionless thermal diffusion coefficient $D_{1}^{*T}$ is
\beqa
\label{d3}
D_1^{*T}&=&\Big(\nu_{D}^*+\frac{1}{2}\Delta^*\frac{\partial \zeta_0^*}{\partial \Delta^*}\Big)^{-1}\Bigg\{-\frac{
\rho_1}{\rho}\Big(p^*-\frac{1}{2}\Delta^*\frac{\partial p^*}{\partial \Delta^*}\Big)+x_1\Big(\gamma_1-\frac{1}{2}\Delta^*\frac{\partial \gamma_1}{\partial \Delta^*}\Big) \nonumber\\
& & +\frac{\pi^{d/2}}{d\Gamma\left(\frac{d}{2}\right)}n\sigma_{12}^d x_1 \Bigg[x_1 \left(\frac{\sigma_1}{\sigma_{12}}\right)^d\chi_{11}\Bigg(\frac{1+\al_{11}}{2}\gamma_1+
\sqrt{\frac{2}{\pi \theta_1}}\mu_{12}\Delta^*\Bigg)
\Bigg(1-\frac{1}{2}\Delta^*\frac{\partial \ln \gamma_1}{\partial \Delta^*}\Bigg)
\nonumber\\
& & +x_2 \chi_{12}\Bigg((1+\al_{12})\mu_{12}\gamma_2+
\frac{4}{\sqrt{\pi}}\mu_{12}\mu_{21}\Delta^* \sqrt{\frac{\theta_1}{\theta_2(\theta_1+\theta_2)}}\Bigg)
\Bigg(1-\frac{1}{2}\Delta^*\frac{\partial \ln \gamma_2}{\partial \Delta^*}\Bigg)
\Bigg]\Bigg\},
\eeqa
where
\beq
\label{d4}
\nu_D^*=\nu_{11}^*-\nu_{12}^*=\frac{2\pi^{(d-1)/2}}{d\Gamma\left(\frac{d}{2}\right)}
\chi_{12}\left(x_2\mu_{21}+x_1\mu_{12}\right)\Bigg[\sqrt{\frac{\theta_1+\theta_2}{\theta_1\theta_2}}(1+\al_{12})
+\sqrt{\pi}\Delta^*\Bigg].
\eeq
Here, $\nu_{ij}^*=\nu_{ij}/\nu$.

The expression of the coefficient $D_{11}^*$ can be written as
\beqa
\label{d5}
\nu_{D}^*D_{11}^*&=&\frac{n_1}{x_1}\frac{\partial \zeta_0^*}{\partial n_1}D_1^{*T}+\gamma_1+n_1\frac{\partial \gamma_1}{\partial n_1}-\frac{\rho_1}{\rho}\Bigg(p^*+\frac{n_1}{x_1}\frac{\partial p^*}{\partial n_1}\Bigg)+\frac{\pi^{d/2}}{d\Gamma\left(\frac{d}{2}\right)}n\sigma_{1}^d x_1 \chi_{11}
\nonumber\\
& & \times\Bigg\{\Bigg[(1+\al_{11})\gamma_1+4
\mu_{12}\Delta^* \sqrt{\frac{2}{\pi\theta_1}}\Bigg]\Bigg(1+\frac{1}{2}n_1\frac{\partial \ln \chi_{11}}{\partial n_1}\Bigg)+\frac{1}{2}n_1\frac{\partial \gamma_{1}}{\partial n_1}\Bigg(1+\al_{11}+4
\frac{\mu_{12}}{\gamma_1}\Delta^* \sqrt{\frac{1}{2\pi \theta_1}}\Bigg)\Bigg\}
\nonumber\\
& &
+\frac{\pi^{d/2}}{d\Gamma\left(\frac{d}{2}\right)}n\sigma_{12}^d x_1 \chi_{12}\mu_{21}\Bigg\{
\Bigg[(1+\al_{12})\left(\gamma_1+\mu \gamma_2\right)+\frac{8}{\sqrt{\pi}}
\mu_{12}\Delta^* \sqrt{\frac{\theta_1+\theta_2}{\theta_1\theta_2}}\Bigg]
\nonumber\\
& &\times \Bigg[\frac{x_2}{2x_1}\left(n_1 \frac{\partial \ln \chi_{12}}{\partial n_1}+I_{121}\right)\Bigg]
+n_2\frac{\partial \gamma_{2}}{\partial n_1}\Bigg[(1+\al_{12})\mu+\frac{4}{\sqrt{\pi}}
\frac{\mu_{12}}{\gamma_2}\Delta^* \sqrt{\frac{\theta_1}{\theta_2(\theta_1+\theta_2)}}\Bigg]
\Bigg\},
\eeqa
where $\mu=m_1/m_2$ is the mass ratio. The coefficient $D_{12}^*$ is
\beqa
\label{d6}
\nu_{D}^*D_{12}^*&=&\frac{n_2}{x_2}\frac{\partial \zeta_0^*}{\partial n_2}D_1^{*T}+n_1\frac{\partial \gamma_1}{\partial n_2}-\frac{\rho_1}{\rho}\Bigg(p^*+\frac{n_2}{x_2}\frac{\partial p^*}{\partial n_2}\Bigg)+\frac{\pi^{d/2}}{d\Gamma\left(\frac{d}{2}\right)}n\sigma_{1}^d x_1 \chi_{11}
\nonumber\\
& & \times \Bigg\{\frac{1}{2}\frac{x_1}{x_2}\Bigg[(1+\al_{11})\gamma_1+4
\mu_{12}\Delta^* \sqrt{\frac{2}{\pi\theta_1}}\Bigg]n_2\frac{\partial \ln \chi_{11}}{\partial n_2}+\frac{1}{2}n_1\frac{\partial \gamma_{1}}{\partial n_2}\Bigg(1+\al_{11}+4
\frac{\mu_{12}}{\gamma_1}\Delta^* \sqrt{\frac{1}{2\pi\theta_1}}\Bigg)\Bigg\}\nonumber\\
& &
+\frac{\pi^{d/2}}{d\Gamma\left(\frac{d}{2}\right)}n\sigma_{12}^d x_1 \chi_{12}\mu_{21}\Bigg\{
\Bigg[(1+\al_{12})\left(\gamma_1+\mu \gamma_2\right)+\frac{8}{\sqrt{\pi}}
\mu_{12}\Delta^* \sqrt{\frac{\theta_1+\theta_2}{\theta_1\theta_2}}\Bigg]
\nonumber\\
& &\times \Bigg[1+\frac{1}{2}\left(n_2 \frac{\partial \ln \chi_{12}}{\partial n_2}+I_{122}\right)\Bigg]
+n_2\frac{\partial \gamma_{2}}{\partial n_2}\Bigg[(1+\al_{12})\mu+\frac{4}{\sqrt{\pi}}
\frac{\mu_{12}}{\gamma_2}\Delta^* \sqrt{\frac{\theta_1}{\theta_2(\theta_1+\theta_2)}}\Bigg]
\Bigg\}.
\eeqa
\end{widetext}

It is quite apparent that the explicit form of the transport coefficients $D_{11}^*$ and $D_{12}^*$ requires the knowledge of
the quantities $I_{i\ell j}$. These parameters are given in terms of the functional derivative of the (local) pair
distribution function $\chi_{ij}$ with respect to the (local) partial densities $n_\ell$
[see Eq.\ (C11) of Ref.\ \onlinecite{GDH07}]. The quantities $I_{i\ell j}$ are zero if
$i=\ell$, but otherwise are not zero. As said in the Appendix \ref{appA}, in granular mixtures they are usually chosen to recover the results derived by L\'opez de Haro {\em et al.} for elastic mixtures \cite{LCK83} (see Appendix C of Ref.\ \onlinecite{GHD07}).

In a binary mixture, the nonzero parameters $I_{121}$ and $I_{122}$ appearing in Eqs.\ \eqref{d5} and \eqref{d6} are given by \cite{GHD07,G19}
\beqa
\label{d7}
I_{121}&=&\frac{1}{TB_2n_2\sigma_{12}^d\chi_{12}}\left[n_1\left(\frac{\partial \mu_1}{\partial n_1}
\right)_{T,n_{2}}-T\right]-2\frac{n_1\sigma_1^d\chi_{11}}
{n_2\sigma_{12}^d\chi_{12}}\nonumber\\
& & -
\frac{n_1^2\sigma_1^d}{n_2\sigma_{12}^d\chi_{12}}\frac{\partial\chi_{11}}{\partial n_1}
-\frac{n_1}{\chi_{12}}\frac{\partial\chi_{12}}{\partial n_1},
\eeqa
\beqa
\label{d8}
I_{122}&=&\frac{1}{TB_2\sigma_{12}^d\chi_{12}}\left(\frac{\partial \mu_1}{\partial n_2}
\right)_{T,n_{1}}-2-
\frac{\sigma_1^d n_1}{\sigma_{12}^d\chi_{12}}\frac{\partial\chi_{11}}{\partial n_2}\nonumber\\
& &
-\frac{n_2}{\chi_{12}}\frac{\partial\chi_{12}}{\partial n_2},
\eeqa
where $\mu_1$ is the chemical potential of the species 1. It must remarked that since granular fluids lack a thermodynamic description, the concept of chemical potential could be questionable. On the other hand, as discussed in previous works, \cite{GGBS24} the presence of the chemical potential $\mu_i$ in the theory is essentially due to the choice of the quantities $I_{i\ell j}$. Given that the explicit form of the chemical potential must be known to evaluate the diffusion transport coefficients, for practical purposes, the expression considered here for $\mu_i$ is the same as the one obtained for molecular mixtures ($\alpha_{ij}=1$). Although this evaluation requires the use of thermodynamic relations that only apply for elastic systems, we expect that this approximation could be reliable for not too small values of the coefficients of restitution. More comparisons with computer simulations are needed to support the above expectation.

In the case of hard disks ($d=2$), a good approximation for the pair distribution function $\chi_{ij}$ is \cite{JM87}
\begin{equation}
\label{d9}
\chi_{ij}=\frac{1}{1-\phi}+\frac{9}{16}\frac{\phi}{(1-\phi)^2}\frac{\sigma_i\sigma_jM_1}{\sigma_{ij}M_2},
\end{equation}
where $\phi=\sum_i\; \pi n_i\sigma_i^2/4$ is the solid volume fraction for disks and
\begin{equation}
\label{d10}
M_n=\sum_{k=1}^2\; x_k \sigma_k^n.
\end{equation}
The expression of the chemical potential $\mu_i$ of the species $i$
consistent with the approximation (\ref{d9}) is \cite{S16}
\begin{eqnarray}
\label{d11}
\frac{\mu_i}{T}&=&\ln (\lambda_i^2n_i)-\ln (1-\phi)+\frac{M_1}{4M_2}\left[\frac{9\phi}{1-\phi}+\ln (1-\phi)\right]\sigma_i
\nonumber\\
& &
-\frac{1}{8}\Big[\frac{M_1^2}{M_2^2}\frac{\phi(1-10\phi)}{(1-\phi)^2}-
\frac{8}{M_2}\frac{\phi}{1-\phi}+\frac{M_1^2}{M_2^2}\ln (1-\phi)\Big]\nonumber\\
& &\times \sigma_i^2,
\end{eqnarray}
where $\lambda_i(T)$ is the (constant) de Broglie's thermal wavelength. \cite{RG73} Note that for mechanically equivalent particles ($m_1=m_2$, $\sigma_1=\sigma_2$), $I_{121}=I_{122}=0$, as expected since the SET and the RET lead to the same Navier-Stokes transport coefficients for a monocomponent granular gas. \cite{GD99a}

According to Eqs.\ \eqref{d3}, \eqref{d5} and \eqref{d6}, it is quite apparent that the diffusion transport coefficients are given in terms of several derivatives. In particular, the derivatives $(\partial \gamma_1/\partial \Delta^*)$, $(\partial \gamma_1/\partial x_1)$ and $(\partial \gamma_1/\partial \phi)$ are given by \cite{GBS21}
\beq
\label{d12}
\left(\frac{\partial \gamma_1}{\partial \Delta^*}\right)=\frac{\sqrt{Y^2-4X Z}-Y}{2X},
\eeq
where $X=N \Delta^*$,
\beq
\label{d13}
Y=M \Delta^*-2 N \gamma_{1}+\gamma_{1}
\left(\frac{\partial \zeta_1^*}{\partial \gamma_1}\right), \quad Z=\gamma_{1}\left(\frac{\partial \zeta_1^*}{\partial \Delta^*}\right)-2 M \gamma_{1}.
\eeq
%\beq
%\label{a11.2}
%Z=\gamma_{1}\left(\frac{\partial \zeta_1^*}{\partial \gamma_1}\right)-2 A \gamma_{1}.
%\eeq
Here,
\beq
\label{d14}
M=\frac{1}{2}\Bigg[x_1 \gamma_1 \Big(\frac{\partial \zeta_1^*}{\partial \Delta^*}\Big)_{\gamma_1}
+x_2 \gamma_2 \Big(\frac{\partial \zeta_2^*}{\partial \Delta^*}\Big)_{\gamma_1}\Bigg],
\eeq
\beq
\label{d15}
N=\frac{1}{2}\Bigg(x_1 \gamma_1 \frac{\partial \zeta_1^*}{\partial \gamma_1}
+x_2 \gamma_2 \frac{\partial \zeta_2^*}{\partial \gamma_1}\Bigg).
\eeq
In addition, in the above equations
\beq
\label{d16}
\left(\frac{\partial \zeta_i^*}{\partial \Delta^*}\right)=\left(\frac{\partial \zeta_i^*}{\partial \Delta^*}\right)_{\gamma_1}+\left(\frac{\partial \zeta_i^*}{\partial \gamma_1}\right)\left(\frac{\partial \gamma_1}{\partial \Delta^*}\right).
\eeq

The derivatives $\partial \gamma_1/\partial x_1$ and $(\partial \gamma_1/\partial \phi)$ can be written as
\begin{widetext}
\beq
\label{d17}
\frac{\partial \gamma_1}{\partial x_1}=-\frac{\gamma_{1}\frac{\partial \zeta_1^*}{\partial x_1}+\frac{1}{2}\left(x_1\gamma_{1}\frac{\partial \zeta_1^*}{\partial x_1}+x_2\gamma_{2}\frac{\partial \zeta_2^*}{\partial x_1}\right)\left[\Delta^* \left(\frac{\partial \gamma_1}{\partial \Delta^*}\right)-2\gamma_1\right]}
{\gamma_{1}\frac{\partial \zeta_1^*}{\partial \gamma_1}+\frac{1}{2}\left(x_1\gamma_{1}\frac{\partial \zeta_1^*}{\partial \gamma_1}+x_2\gamma_{2}\frac{\partial \zeta_2^*}{\partial \gamma_1}\right)\left[\Delta^* \left(\frac{\partial \gamma_1}{\partial \Delta^*}\right)-2\gamma_1\right]},
\eeq
\beq
\label{d18}
\frac{\partial \gamma_1}{\partial \phi}=-\frac{\gamma_{1}\frac{\partial \zeta_1^*}{\partial \phi}+\frac{1}{2}\left(x_1\gamma_{1}\frac{\partial \zeta_1^*}{\partial \phi}+x_2\gamma_{2}\frac{\partial \zeta_2^*}{\partial \phi}\right)\left[\Delta^* \left(\frac{\partial \gamma_1}{\partial \Delta^*}\right)-2\gamma_1\right]}
{\gamma_{1}\frac{\partial \zeta_1^*}{\partial \gamma_1}+\frac{1}{2}\left(x_1\gamma_{1}\frac{\partial \zeta_1^*}{\partial \gamma_1}+x_2\gamma_{2}\frac{\partial \zeta_2^*}{\partial \gamma_1}\right)\left[\Delta^* \left(\frac{\partial \gamma_1}{\partial \Delta^*}\right)-2\gamma_1\right]}.
\eeq
\end{widetext}

In Eqs.\ \eqref{d17} and \eqref{d18}, it is understood that the derivatives $\partial_{x_1}\zeta_i^*$ and $\partial_{\phi}\zeta_i^*$ are taken at $\gamma_1\equiv \text{const}$. Similarly to Eq.\ \eqref{d16}, we have the relations
\beq
\label{d19}
\left(\frac{\partial p^*}{\partial \Delta^*}\right)=\left(\frac{\partial p^*}{\partial \Delta^*}\right)_{\gamma_1}+\left(\frac{\partial p^*}{\partial \gamma_1}\right)\left(\frac{\partial \gamma_1}{\partial \Delta^*}\right),
\eeq
\beq
\label{d20}
\left(\frac{\partial \zeta_0^*}{\partial \Delta^*}\right)=\left(\frac{\partial \zeta_0^*}{\partial \Delta^*}\right)_{\gamma_1}+\left(\frac{\partial \zeta_0^*}{\partial \gamma_1}\right)\left(\frac{\partial \gamma_1}{\partial \Delta^*}\right).
\eeq
Finally, since the set $\Lambda\equiv \left\{\zeta_0, p^*, \gamma_1\right\}$ depends on the number densities $n_1$ and $n_2$ through its dependence on $x_1$ and $\phi$ we have the identities
\beq
\label{d21}
n_1 \frac{\partial \Lambda}{\partial n_1}=x_1 x_2 \frac{\partial \Lambda}{\partial x_1}+\phi_1 \frac{\partial \Lambda}{\partial \phi},
\eeq
\beq
\label{d22}
n_2 \frac{\partial \Lambda}{\partial n_2}=-x_1 x_2 \frac{\partial \Lambda}{\partial x_1}+\phi_2 \frac{\partial \Lambda}{\partial \phi},
\eeq
\beq
\label{d23}
n_2 \frac{\partial \Lambda}{\partial n_1}=x_2^2 \frac{\partial \Lambda}{\partial x_1}+\frac{x_2}{x_1}\phi_1 \frac{\partial \Lambda}{\partial \phi},
\eeq
\beq
\label{d24}
n_1 \frac{\partial \Lambda}{\partial n_2}=-x_1^2 \frac{\partial \Lambda}{\partial x_1}+\frac{x_1}{x_2}\phi_2 \frac{\partial \Lambda}{\partial \phi},
\eeq
where $\phi_i=\pi n_i\sigma_i^2/4$.

%\bibliography{diffusionDelta}

%\end{document}

%

\end{document}